\documentclass[11pt,dvips]{article}

\usepackage{amsmath,epsfig,ulem}
\usepackage{rotating}
\usepackage{cite}
\usepackage{hhline}
\usepackage{array}
\usepackage{amssymb}
\usepackage{color}
\usepackage{dsfont}
\usepackage{equations}

\begin{document}
\tolerance=100000
\interfootnotelinepenalty=10000

\newcommand{\mathbold}[1]{\mbox{\boldmath $\bf#1$}}

\def\tablename{\bf Table}%
\def\figurename{\bf Figure}%
\def\ifmath#1{\relax\ifmmode #1\else $#1$\fi}
\def\ls#1{\ifmath{_{\lower1.5pt\hbox{$\scriptstyle #1$}}}}

\newcommand\phm{\phantom{-}}
\newcommand{\sts}{\scriptstyle}
\newcommand{\ngs}{\!\!\!\!\!\!}
\newcommand{\rb}[2]{\raisebox{#1}[-#1]{#2}}
\newcommand{\CP}{${\cal CP}$~}
\newcommand{\sbomu}{\frac{\sin 2 \beta}{2 \mu}}
\newcommand{\kmol}{\frac{\kappa \mu}{\lambda}}
\newcommand{\s}{\\ \vspace*{-3.5mm}}
\newcommand{\lsim}{\raisebox{-0.13cm}{~\shortstack{$<$\\[-0.07cm] $\sim$}}~}
\newcommand{\gsim}{\raisebox{-0.13cm}{~\shortstack{$>$\\[-0.07cm] $\sim$}}~}
\newcommand{\kr}{\color{red}}

\renewcommand{\Re}{\thinspace{\rm Re\thinspace}}
\renewcommand{\Im}{\thinspace{\rm Im\thinspace}}
\newcommand{\imag}{\Im}
\newcommand{\real}{\Re}
\def\nicefrac#1#2{\hbox{$\frac{#1}{#2}$}}
\def\half{\nicefrac{1}{2}}

\renewcommand{\thefootnote}{\arabic{footnote}}
\renewcommand{\theequation}{{\rm \thesection.\arabic{equation}}}

\begin{titlepage}

\begin{flushright}
DESY 06-066\\
IFT-06/022\\
SCIPP-06/12\\
\today
\end{flushright}

\vspace{0.6cm}

\begin{center}
{\Large \bf The Neutralino Sector in the U(1)-Extended Supersymmetric Standard
            Model}\\[1cm]
{\large S.Y. Choi$^{1,2}$, H.E. Haber$^{3}$, J. Kalinowski$^{4}$ and
        P.M. Zerwas$^{2,3}$}\\[1cm]
{\it $^1$ Department of Physics and RIPC, Chonbuk National University, Jeonju
          561-756, Korea\\
     $^2$ Deutsches Elektronen-Synchrotron DESY, D-22603 Hamburg, Germany\\
     $^3$ SCIPP, University of California, Santa Cruz, CA 95064, USA\\
     $^4$ Institute of Theor. Physics, Warsaw University, 00681 Warsaw,
          Poland}
\end{center}

\vspace{1.8cm}

\begin{abstract}
\noindent
Motivated by grand unified theories and string theories we analyze the
general structure of the neutralino sector in the USSM, an extension of
the Minimal Supersymmetric Standard Model that involves a broken
extra U(1) gauge symmetry. This supersymmetric U(1)-extended model
includes an Abelian gauge superfield and a Higgs singlet superfield
in addition to the standard gauge and Higgs
superfields of the MSSM. The interactions between the MSSM fields
and the new fields are in general weak and the mixing is small, so
that the coupling of the two subsystems can be treated
perturbatively. As a result, the mass spectrum and mixing matrix in the
neutralino sector can be analyzed analytically and the structure of
this 6-state system is under good theoretical control. We describe
the decay modes of the new states and the impact of this extension on
decays of the original MSSM neutralinos, including radiative transitions
in cross-over zones. Production channels in cascade decays at the LHC and
pair production at $e^+e^-$ colliders are also discussed.
\end{abstract}

\end{titlepage}

\newpage

\section{Introduction}
\setcounter{equation}{0}

Adding an extra U(1)$_X$ broken gauge symmetry to the gauge symmetries
of the Standard Model is well motivated by grand unified theories
\cite{grand_unification}.  The corresponding supersymmetric
extension that generalizes the minimal supersymmetric Standard Model
(MSSM) often appears as the low energy effective theory of
superstring theories \cite{superstring}.  This U(1)$_X$ extended
supersymmetric gauge theory shall henceforth be denoted as the USSM.\s

The Higgs sector associated with the broken U(1)$_X$ gauge symmetry
provides an elegant solution to the $\mu$ problem in supersymmetric
theories \cite{u1ssm0,mu_problem}. An effective $\mu$ parameter is
generated by the vacuum expectation value of the new singlet Higgs
field $S$, which breaks the U(1)$_X$ gauge symmetry. This is the same
mechanism employed by the next-to-minimal supersymmetric standard model
(NMSSM) \cite{nmssm}. However, the USSM possesses an additional
advantage by avoiding the extra discrete symmetries of the NMSSM that,
in the canonical version, result in the existence of domain-walls
that are incompatible with the observed energy density of the universe.
Moreover, the upper bound on the mass of the lightest Higgs boson
of the MSSM is relaxed in the USSM
due to contributions from the new singlet-doublet Higgs
interactions and the U(1)$_X$ $D$-terms \cite{u1prime_D}.
Various scenarios of this type have been discussed in the literature, see e.g.
Refs.~\cite{erler,u1ssm3},
in which the U(1)$_X$ gauge symmetry is embedded in the
grand unification
group E$_6$ (or one of its rank-five subgroups).\s

Including the extra symmetry, the gauge group is extended to $G={\rm
SU}(3)_C\times{\rm SU}(2)_L\times{\rm U}(1)_Y\times{\rm U}(1)_X$ with
the couplings $g_3, g_2, g_Y, g_X$, respectively. The matter particle
content in the supersymmetric theory includes, potentially among
others, the left-handed chiral superfields $\hat{L}_i,\, \hat{E}^c_i;
\,\hat{Q}_i,\, \hat{U}^c_i,\, \hat{D}^c_i$, where the subscript $i$
denotes the generation index, and the Higgs superfields $\hat{H}_d,\,
\hat{H}_u, \,\hat{S}$. The usual MSSM Yukawa terms $\hat{W}_Y$ of the
MSSM superpotential ({\it i.e.} without the $\mu$ term)
are augmented by an additional term that couples
the iso-singlet to the two iso-doublet Higgs fields:
\begin{eqnarray}\label{eqi}
\hat{W}=\hat{W}_Y +\lambda \hat{S}\, (\hat{H}_u\hat{H}_d)\,.
\label{eq:superpotential}
\end{eqnarray}
The coupling $\lambda$ is dimensionless. Gauge invariance of the
superpotential $\hat{W}$ under U(1)$_X$ requires the U(1)$_X$ charges to satisfy
$Q_{H_d}+Q_{H_u}+Q_S=0$ and corresponding relations
between the U(1)$_X$ charges of Higgs and matter fields.
[In the following, we use
$Q_1=Q_{H_d}$ and $Q_2=Q_{H_u}$ for notational convenience.] The effective
$\mu$ parameter is generated by the vacuum expectation value $\langle S\rangle$
of the scalar $S$-field. \s

Compared with the MSSM, the USSM Higgs sector is extended by a single scalar
state. The neutralino sector includes an additional pair of higgsino and
gaugino states, while the chargino sector remains unaltered.
The complexity of phenomena increases dramatically by this extension but the
structure remains transparent if the original and the new degrees of freedom are
coupled weakly as naturally demanded [see below]. \s

The supersymmetric particle spectrum of the USSM has received limited
attention so far in the
literature \cite{u1ssm0,u1ssm3,u1ssm1,u1ssm2,Suematsu:1997tv}.
In this report we attempt a systematic analytical analysis of the
neutralino system, based on the well-motivated assumption of weak
coupling between the original MSSM and the new additional
gaugino/higgsino subsystem. In contrast to the MSSM where exact
solutions of the mass spectrum and mixing parameters can be
constructed mathematically in closed form (see
e.g. Ref.~\cite{ckmz}), this is not possible anymore for the
supersymmetric U(1)$_X$ model in which the eigenvalue equation for the
masses is a 6th order polynomial equation.  However, analogously to
the NMSSM \cite{NMSSM_CMZ}, if the mass scales of the supersymmetric
particles are set by higgsino and gaugino parameters of the order
the supersymmetry (SUSY)-breaking scale,
$M_{\rm SUSY}\sim\mathcal{O}(10^3$~GeV), while the interaction between the
new singlet and the MSSM fields is of the order of the electroweak
scale, $v\sim\mathcal{O}(10^2$~GeV), then the perturbative expansion of the
solution in $v/M_{\rm SUSY}$ provides an excellent approximation to
the mass spectrum and yields a good understanding of
the main features of the mixing matrix.\s

Once the masses and mixings are determined, the couplings of the neutralinos to
the electroweak gauge bosons and to the scalar/fermionic matter particles are
fixed. Decay widths and production rates can subsequently be predicted for
squark cascades at the LHC \cite{LHC}
and pair production in $e^+e^-$ collisions
at linear colliders \cite{ILC}. Of particular interest
are the radiative transitions between neutralinos in cross-over zones,
where the masses of two neutralinos are nearly degenerate. \s

The report is organized as follows. In Sect.$\,$\ref{sec:sec2} we
first describe the general basis of the neutralino sector in the USSM.
Subsequently, for the naturally expected weak coupling between the MSSM and
the new subsystem, the properties of the new higgsino and gaugino
are derived in Sect.$\,$\ref{sec:sec3}. It is shown to what extent the
properties of the standard neutralinos are modified. The spectrum and the
mixings are determined analytically in a weak-coupling perturbative
expansion.  The neutralino masses are determined to second-order,
whereas the mixing matrix elements are determined to first-order
in the weak coupling. The accuracy of the perturbative results will
become apparent by comparing the analytic approximations with the numerical
solutions, thereby demonstrating that a satisfactory understanding
of the system can be achieved. As an illustration we will study
the limit in which the gaugino mass parameters are significantly
larger than the higgsino mass parameters, where both sets
of parameters are assumed to
be much larger than the electroweak scale. A general
description of the neutralino couplings and decay widths is given
in Sect.$\,$\ref{sec:sec4}, including photon transitions. We also discuss
production cross sections in $e^+e^-$ collisions and cascade
decay chains of squarks at the LHC that involve neutralinos. Section
\ref{sec:conclusion} summarizes and concludes this report.
Technical details of the analytical diagonalization procedures for the
$6\times 6$ neutralino mass matrix for non-degenerate and degenerate
levels are given in three appendices.\s

\section{The USSM Neutralino Sector}
\label{sec:sec2}
\setcounter{equation}{0}

\subsection{Supersymmetric kinetic mixing}

In a theory with two U(1) gauge symmetries, the two sectors can mix,
consistently with all gauge symmetries, through the coupling of the
kinetic parts of the two gauge bosons \cite{kinetic_mixing_loop}. In the
basis in which the couplings between matter and gauge fields have
the canonical minimal-interaction form, the pure gauge part of the
Lagrangian for the U(1)$_Y$$\times$U(1)$_X$ theory can be written
\begin{eqnarray}
{\cal L}_{\rm gauge}
 =
 -\frac{1}{4} Y^{\mu\nu} Y_{\mu\nu}
 -\frac{1}{4} X^{\mu\nu} X_{\mu\nu}
 -\frac{\sin\chi}{2} Y^{\mu\nu} X_{\mu\nu}\,,
\end{eqnarray}
where the parameter $\sin\chi$ is introduced to characterize the gauge kinetic
mixing \cite{kinetic_mixing}. This Lagrangian generalizes to
\begin{eqnarray} \label{lgauge}
{\cal L}_{\rm gauge}
 = \frac{1}{32}\int d^2\theta
   \left\{\hat{W}_Y \hat{W}_Y + \hat{W}_X \hat{W}_X
        +2\sin\chi\, \hat{W}_Y \hat{W}_X\right\}\,,
\end{eqnarray}
in a supersymmetric theory, where $\hat{W}_Y$ and $\hat{W}_X$ are the chiral
superfields associated with the two gauge symmetries.\footnote{The normalization
of the superfield $\hat{W} =\overline  D^2D\hat{V}$
follows the conventions of Ref.~\cite{bailinlove}, where $\hat{V}$ is the
corresponding vector superfield.} \s

The gauge/gaugino part of the Lagrangian can be converted back to
the canonical form by the following GL(2,$\mathbb{R}$) transformation of the
superfields \cite{kinetic_mixing_loop,kinetic_mixing,kinetic_mixing_string}:
\begin{eqnarray} \label{kinmixmatrix}
\left(\begin{array}{c}
      \hat{W}_Y \\
      \hat{W}_X
      \end{array}\right)
 =
\left(\begin{array}{cc}
         1   & -\tan\chi \\
         0   & 1/\cos\chi
      \end{array}\right)
\left(\begin{array}{l}
        \hat{W}_B \\
        \hat{W}_{B'}
       \end{array}\right)\,,
\end{eqnarray}
which acts on the gauge boson and gaugino components of the chiral
superfields in the same form. The transformation alters the
U(1)$_Y\times$U(1)$_X$ part of the covariant derivative to
\begin{align}
D_\mu =&\;\partial_\mu + i g_Y Y B_\mu
   +i \left(- g_Y Y \tan\chi +\frac{g_X}{\cos\chi} Q_X \right) B'_\mu  \\
    =&\;\partial_\mu + i g_Y Y B_\mu +i g_X Q'_X B'_\mu\,.
\end{align}
The choice of the kinetic mixing matrix in the form given
by Eq.$\,$(\ref{kinmixmatrix}) is motivated by the fact that
the hypercharge sector of the Standard Model is left unaltered by
this transformation, and the new effects are separated in the $X$ sector (see,
e.g., Ref.~\cite{Abel:2006qt} for an alternative choice).
Consequently, the effective U(1)$_X$ charge is shifted from its original
value $Q_X$ to
\begin{eqnarray}
Q'_X=\frac{Q_X}{\cos\chi} - \frac{g_Y}{g_X} Y \tan\chi\,.
\end{eqnarray}
Specifically, the U(1)$_X$ charge of any field is shifted by
an amount proportional to their hypercharge $Y$ and the mixing parameter
$\sin\chi$. Thus, as a result of the kinetic
mixing, new interactions among the gauge
bosons and matter fields are generated even for matter fields with
zero U(1)$_X$ charge originally. \s

In grand unification theories the two U(1) groups are orthogonal at
the unification scale but small mixing \cite{kinetic_mixing_loop} can be
induced through loop effects when the theory evolves down to the electroweak
scale.  In string theories, kinetic mixing can be induced at
the tree level \cite{kinetic_mixing_string};
however, such mixing effects must
remain small in order to guarantee the general agreement between SM analyses
and precision data in a natural way \cite{Zprime_bound}.\s

\subsection{The USSM neutralino mass matrix}

The Lagrangian of the neutralino system follows from the
superpotential in Eq.$\,$(\ref{eq:superpotential}), complemented by the
gaugino SU(2)$_L$, U(1)$_Y$ and U(1)$_X$ mass terms
of the soft-supersymmetry breaking electroweak Lagrangian:
\begin{eqnarray} \label{gauginomass}
{\cal L}^{\rm gaugino}_{\rm mass}
 &=& -\frac{1}{2} M_2 \widetilde W^a\widetilde W^a
-\frac{1}{2} M_Y \tilde{Y} \tilde{Y}
   -\frac{1}{2} M_X \tilde{X} \tilde{X}
   - M_{YX} \tilde{Y}\tilde{X} + {\rm h.c.} \nonumber\\
 &=&  -\frac{1}{2} M_2 \widetilde W^a\widetilde W^a
-\frac{1}{2} M_1  \tilde{B}  \tilde{B}
     -\frac{1}{2} M'_1 \tilde{B}' \tilde{B}'
     -M_K \tilde{B}\tilde{B}' + {\rm h.c.}\,,
\end{eqnarray}
where the $\widetilde W^a$ ($a=1,2,3$), $\tilde{Y}$ and $\tilde{X}$  are
the (two-component) SU(2)$_L$, U(1)$_Y$ and U(1)$_X$ gaugino fields, and
\begin{eqnarray} \label{Mfactors}
M_1\equiv M_Y\,, \qquad
M'_1 \equiv \frac{M_X}{\cos^2\chi}-\frac{2\sin\chi}{\cos^2\chi} M_{YX} +
M_Y \tan^2\chi\,,\qquad
M_K\equiv \frac{M_{YX}}{\cos\chi}-M_Y \tan\chi\,.
\end{eqnarray}
In parallel to the gauge kinetic mixing discussed in Sect.$\,$2.1,
the Abelian gaugino mixing mass parameter $M_{YX}$ is assumed small compared
with the mass scales of the gaugino and higgsino fields. \s

After breaking the electroweak and U(1)$_X$
symmetries spontaneously due to non-zero vacuum expectation
values of the iso-doublet and the iso-singlet Higgs fields,
\begin{eqnarray}
\langle H_u \rangle
  = \frac{\sin\beta}{\sqrt{2}}
    \left(\begin{array}{c}
              0 \\
              v
          \end{array}\right),\qquad
\langle H_d \rangle
  = \frac{\cos\beta}{\sqrt{2}}
    \left(\begin{array}{c}
              v \\
              0
          \end{array}\right),\qquad
\langle S \rangle = \frac{1}{\sqrt{2}} v_s\,,
\end{eqnarray}
the doublet higgsino mass and the doublet higgsino-singlet higgsino
mixing parameters,
\begin{equation}
\mu\equiv \lambda \frac{v_s}{\sqrt{2}} \;\;\;\;\; {\rm{and}}
\;\;\; \;\; \mu_\lambda\equiv\lambda \frac{v}{\sqrt{2}}\,,
\end{equation}
are generated.
The USSM neutral gaugino-higgsino mass matrix can be written in the following
block matrix form\footnote{Although our initial exploratory analysis is carried
out at tree-level, loop corrections can easily be included following the
procedures of Ref.~\cite{SPA}.},
\begin{eqnarray}
{\cal M}_6 =\left(\begin{array}{cc}
      {\cal M}_4  & X  \\
      X^T       & {\cal M}_2
                 \end{array}\right)\,,
\label{eq:general_mass_matrix}
\end{eqnarray}
where ${\cal M}_4$ is the neutral gaugino-higgsino mass matrix of the MSSM,
${\cal M}_2$ corresponds to the new sector containing the singlet higgsino
(singlino) and the new
U(1)-gaugino $\tilde B'$ that is orthogonal to the bino~$\tilde B$, and
$X$ describes the coupling of the two sectors via the neutralino mass matrix.
More explicitly,
in a basis of two-component spinor fields $\xi\equiv(\tilde{B}, \tilde{W}^3,
\tilde{H}^0_d, \tilde{H}^0_u, \tilde{S}, \tilde{B}')^T$, the full
neutralino mass matrix is given by \cite{u1ssm2}:
\begin{eqnarray}
{\cal M}_6 =\left(\begin{array}{cccc|cc}
  M_1  &   0   & -m_Z\, c_\beta\, s_W  &  m_Z\, s_\beta\, s_W  &  0 & M_K \\
   0   &  M_2  &  m_Z\, c_\beta\, c_W  & -m_Z\, s_\beta\, c_W  &  0 & 0  \\
-m_Z\, c_\beta\, s_W &  m_Z\, c_\beta\, c_W  &   0
                     & -\mu & -\mu_\lambda\,s_\beta & Q'_1 m_v\, c_\beta \\
 m_Z\, s_\beta\, s_W & -m_Z\, s_\beta\, c_W  & -\mu
                     &   0  & -\mu_\lambda\,c_\beta & Q'_2 m_v\, s_\beta
  \\[2mm]
 \cline{1-6}
  0    &   0   & -\mu_\lambda\,s_\beta & -\mu_\lambda\, c_\beta
       &  0 &  Q'_S m_s  \\
 M_K  &   0   &  Q'_1 m_v\, c_\beta & Q'_2 m_v\, s_\beta & Q'_S m_s
       &  M_1'
                   \end{array}\right)\,,
\label{eq:mass_matrix}
\end{eqnarray}
where the various gaugino mass parameters $M_1$, $M_2$, $M_1^\prime$
and $M_K$ have been defined in Eqs.$\,$(\ref{gauginomass})
and (\ref{Mfactors}).
Notice the absence of a diagonal mass parameter of the new
singlino in contrast to the NMSSM where the cubic
self-interaction generates this singlet mass term \cite{NMSSM_CMZ}.
Two additional mass mixing parameters,
\begin{equation}
m_v\equiv g_X v\;\;\; \;\;{\rm and}\;\;\;\;\; m_s \equiv  g_X v_s\,,
\end{equation}
are generated after gauge symmetry breaking and the effective
charges $Q'_{1}$, $Q'_{2}$ and $Q'_S$ are defined by
\begin{eqnarray}
Q'_1 \equiv\frac{Q_1}{\cos\chi}+\frac{1}{2}\frac{g_Y}{g_X} \tan\chi,\quad
Q'_2 \equiv\frac{Q_2}{\cos\chi}-\frac{1}{2}\frac{g_Y}{g_X} \tan\chi,\quad
Q'_S \equiv\frac{Q_S}{\cos\chi}\,,
\end{eqnarray}
in terms of the $Q_i$ defined below Eq.$\,$(\ref{eqi}).
As usual, $\tan\beta\equiv v_2/v_1$ is the
ratio of the vacuum expectation values of the two neutral SU(2)
Higgs doublet fields, $s_\beta\equiv\sin\beta$, $c_\beta\equiv\cos\beta$, and
$s_W, c_W$ are the sine and cosine of the electroweak mixing angle
$\theta_W$.\s

In general, the neutralino mass matrix ${\cal M}_6$ is
a complex symmetric matrix.  To diagonalize this matrix,
we introduce a unitary matrix $N^6$ such that
\begin{eqnarray} \label{ndef}
\tilde{\chi}^0_k
    = N^6_{k\ell}\, ( \tilde{B}, \tilde{W}^3, \tilde{H}_d,
                      \tilde{H}_u, \tilde{S}, \tilde{B}')_\ell\,,
\end{eqnarray}
where the physical neutralino states are ordered by some convention.
A typical choice, motivated by experimental analyses, is the ordering of
$\tilde{\chi}^0_k$ $[k =1,..,6]$ according to ascending
mass values.  As an intermediate step, we shall often refer to an auxiliary
convention, in which the ordering of states $\tilde{\chi}^0_{k'}$,
denoted by primed
subscripts, follows the order of the
original $(\tilde{B}, \tilde{W}^3, \tilde{H}^0_d, \tilde{H}^0_u,
\tilde{S}, \tilde{B}')$ basis.\s

Given the neutralino mass matrix ${\cal M}_6$,
the physical neutralino masses $m^{ph}_k$, which are real non-negative
numbers, and the neutralino mixing matrix elements $N^6_{k\ell}$ can be
calculated. The mass term in the Lagrangian is given by:
\begin{eqnarray} \label{lmass}
-{\cal L}_{\rm mass}&=&\half\, \xi^T\!  {\cal M}_6\, \xi + {\rm h.c.}
= \half\, \sum_{k=1}^6
m^{ph}_k\, \tilde\chi_k^0\widetilde\chi_k^0+ {\rm h.c.} \,,
\end{eqnarray}
The transformation of the two-component fields
generates the diagonalized mass matrix for the physical neutralino states,
\begin{eqnarray} \label{takagifact}
(N^6)^{*}\, {\cal M}_6\, (N^6)^{\dagger} = {\rm
  diag}(m^{ph}_1\,,\,m^{ph}_2\,,\,\ldots\,,\, m^{ph}_6)\,,\qquad m^{ph}_k\geq 0\,.
\end{eqnarray}
Mathematically, this transformation is the Takagi
diagonalization \cite{takagi,horn,takcompute,takcompute2,hahn}
of a general complex symmetric matrix; see Appendix~A for further details.
Physically,  the unitary matrix $N^6$ determines the couplings
of the mass-eigenstates $\tilde{\chi}^0_k$ to other particles. \s

If ${\cal M}_6$ is complex, then CP is violated in the
neutralino sector of the theory
if no diagonal matrix of phases $P$ exists
such that $P^T{\cal M}_6 P$ is real.  If $P$ exists, then the neutralino
interaction-eigenstates can be rephased to produce a real neutralino mass
matrix, and the neutralino sector is CP-conserving.\footnote{In
this context, the
neutralino sector refers to the neutralino kinetic energy and mass
terms, plus terms that couple the neutralinos to the gauge bosons.  In
this restricted sector, the neutralinos would be states of definite CP
quantum number.  Of course, it is possible to introduce CP-violating
interactions through the neutralino couplings to other particles, e.g.
matter particles of the USSM.  In this case, radiative corrections could
transmit these effects into the neutralino mass matrix.}
If ${\cal M}_6$ is real, then the Takagi
diagonalization of Eq.$\,$(\ref{takagifact}) still applies but can easily
be carried out in two steps. First the real symmetric matrix
${\cal M}_6$ can be diagonalized by an orthogonal matrix~$V^6$:
\begin{eqnarray} \label{vmv}
V^6\, {\cal M}_6\, (V^6)^T = {\rm diag}
(m_1\,,\, m_2\,,\,\ldots\,,\, m_6)\,,
\end{eqnarray}
where the eigenvalues
$m_k$ are real but not necessarily positive.
The Takagi diagonalization of
${\cal M}_6$, which yields real non-negative diagonal mass elements,
can then be achieved in a second step by taking $m^{ph}_k=|m_k|$ and
defining the unitary matrix $N^6$ in Eq.$\,$(\ref{takagifact})
by $N^6= (P^6 V^6)^*$, where $P^6$ is a
diagonal phase matrix with elements $P^6_{k\ell}=\varepsilon_k^{1/2}
\, \delta_{k\ell}$.  Here,
$\varepsilon_k\equiv m_k/m^{ph}_k=\pm 1$ is the sign of $m_k$, which
is also proportional to the CP-quantum number \cite{carruthers}
of the neutralino $\tilde\chi_k^0$.  More precisely, the \textit{relative}
CP-quantum numbers of $\tilde\chi_k^0$ and $\tilde\chi^0_\ell$,
which is the physical quantity of interest,
is given by $\varepsilon_k\varepsilon_\ell$.\s

Although the ordering of states $\tilde{\chi}^0_{k}$ in ascending mass
values is convenient, it is often useful to adopt an intermediate
auxiliary convention.  Note that the neutralino mass matrix is easily
diagonalized in the limit of $M_K=v=0$ (i.e., before the coupling of
the MSSM with the new gaugino/singlino block is introduced).  In this
limit, ${\cal M}_6$ is real after rephasing the neutralino
interaction-eigenstates (if necessary).  That is, without loss of
generality, we can choose $M_1$, $M_1^\prime$, $M_2$ and $\mu$ to be
real in this limit, in which case Eq.$\,$(\ref{vmv}) yields the
following mass eigenvalues: $m_{k'}=\{ M_1, M_2, \mu, -\mu, m_{5'},
m_{6'}\}$, where $m_{5',6'}=\half
M_1'\left[1\mp\sqrt{1+(2Q'_Sm_s/M'_1)^2\,}\right]$ (with
$m_{5'}<m_{6'}$).  Away from this limit, the mass-eigenstates
$\tilde{\chi}^0_{k'}$ will be defined such that their masses are
continuously connected to the masses of the corresponding states in
the $M_K=v=0$ limit.  This defines an alternative ordering of the
states $\tilde{\chi}^0_{k'}$ which will be indicated with primed
subscripts.\s

We shall present a set of techniques for computing analytic approximations
of the physical neutralino masses, $m^{ph}_{k'}$ and the corresponding
neutralino mixing matrix elements $N^6_{k'\ell'}$.  As previously indicated,
the primed subscripts denote that these quantities refer
to the physical neutralino states $\tilde{\chi}^0_{k'}$, whose ordering is
specified above.  Of course, at the end of the computation, one can
convert to an ascending mass ordering convention by an appropriate
relabeling of the states, masses and mixing matrix elements.\s

\section{Small Mixing Scenarios}
\label{sec:sec3}
\setcounter{equation}{0}
\subsection{General analysis}
\label{genanalysis}

It is well known that the MSSM neutralino mass matrix ${\cal M}_4$
can be diagonalized analytically (see, e.g., Ref.~\cite{NMSSM_CMZ}).
In contrast, the
diagonalization of the new USSM $6\times 6$ neutralino mass matrix
${\cal M}_6$ cannot be performed analytically in closed form.
However, the case of physical interest is one in which
both the couplings of the MSSM higgsino doublets
to the singlet higgsino and to the U(1)$_X$ gaugino, and the coupling of the
U(1)$_Y$ and U(1)$_X$ gaugino singlets are weak, i.e. the
elements of the $4\times 2$
submatrix $X$ in Eq.$\,$(\ref{eq:general_mass_matrix}) are small.  Then,
an approximate analytical solution can be found following the procedure
given in Appendix~B.\s

As an initial step, the $4\times 4$ MSSM
submatrix ${\cal M}_4$ and the new $2\times 2$ singlino-U(1)$_X$
gaugino submatrix ${\cal M}_2$ are separately diagonalized:
\begin{eqnarray}
\overline{\cal M}^{D}_4 &=& N^{4\,*} {\cal M}_4 N^{4\,\dagger}
= {\rm diag} (\overline{m}_{1'}, \overline{m}_{2'}, \overline{m}_{3'},
\overline{m}_{4'})\,, \\
\overline{\cal M}^{D}_2 &=& N^{2\,*} {\cal M}_2 N^{2\,\dagger}
= {\rm diag} (\overline{m}_{5'}, \overline{m}_{6'})\,,
\end{eqnarray}
where the $\overline m_{k'}$ are real and non-negative.  Here we use
primed subscripts to indicate that the neutralino states
are continuously connected to the corresponding states in the
$M_K=v=0$ limit, as discussed at the end of Sect.$\,$\ref{sec:sec2}.
The above procedure
results in a partial Takagi diagonalization of the full neutralino
mass matrix, ${\cal M}_6$:
\begin{eqnarray}
\overline{\cal M}_6\equiv \left(\begin{array}{cc}
  N^{4\,*} & \mathds{O} \\
 \mathds{O}^T &  N^{2\,*}
 \end{array}\right)\,
\left(\begin{array}{cc}
      {\cal M}_4  & X  \\
      X^T       & {\cal M}_2
                 \end{array}\right)\,
\left(\begin{array}{cc} N^{4\,\dagger} & \mathds{O} \\
     \mathds{O}^T &
  N^{2\,\dagger}\end{array}\right)
= \left(\begin{array}{cc}
\overline{\cal M}^{D}_4 & N^{4\,*}XN^{2\,\dagger}
\\ N^{2\,*}X^T N^{4\,\dagger} &
\overline{\cal M}^{D}_2 \end{array}\right)\,.
\end{eqnarray}
where $\mathds{O}$ is a $4\times 2$ matrix of zeros.
The upper left and lower right blocks of $\overline{\cal M}_6$ are
diagonal with real non-negative entries, but the upper right and lower
left off-diagonal blocks are non-zero.\s

Performing a block-diagonalization of $\overline{\cal M}_6$ will
remove the non-zero off-diagonal blocks while leaving the diagonal
blocks approximately diagonal up to second order, due to the weak coupling
of the two subsystems.  That is,
\begin{eqnarray}
{\cal M}^D_6 = \overline{N}^{6*}_B\, \overline{\cal M}_6\,
  \overline{N}^{6\dagger}_B
= {\rm diag} ({m}^{ph}_{1'}, {m}^{ph}_{2'}, {m}^{ph}_{3'}, {m}^{ph}_{4'},
              {m}^{ph}_{5'},m^{ph}_{6'})\,,
\end{eqnarray}
where
\begin{eqnarray} \label{calndef}
\overline{N}^6_B  \simeq
\left(\begin{array}{cc}
    \mathds{1}_{4\times 4}
    - \frac{1}{2} \Omega \Omega^\dagger
  &  \Omega \\[1mm]
    -\Omega^\dagger & \mathds{1}_{2\times 2}
   -\frac{1}{2} \Omega^\dagger \Omega
      \end{array} \right)\times
      {\rm diag}(e^{-i\phi_{1'}}\,,\,
\ldots\,,\,e^{-i\phi_{6'}})\,.
\end{eqnarray}
A detailed derivation will be presented in Appendix B.
The elements of the $4\times 2$ mixing matrix $\Omega$ are given by:
\begin{eqnarray}
\Re\Omega_{i'j'}\equiv \frac{\Re(N^{4\,*}XN^{2\,\dagger})_{i'j'}}
{\overline m_{i'}-\overline m_{j'}}\,,\label{romega}
\qquad\qquad
\Im\Omega_{i'j'}\equiv \frac{\Im(N^{4\,*}XN^{2\,\dagger})_{i'j'}}
{\overline m_{i'}+\overline m_{j'}}\,,
\end{eqnarray}
with $i'=1',\ldots,4'$ and $j'=5',6'$.  After the block-diagonalization,
the upper left $4\times 4$ and the lower right $2\times 2$ blocks need
not be re-diagonalized up to second order in the small mixing $X$
between the blocks, but the corresponding eigenvalues are shifted
 $\overline m_{k'}\to m^{ph}_{k'}$  to second order in the small mixing.
The physical neutralino masses $m^{ph}_{k'}$ are given by:
\begin{eqnarray}
m^{ph}_{i'} & \simeq &\overline m_{i'}+\sum_{j'=5}^6\left\{
 \frac{[\Re(N^{4\,*}XN^{2\,\dagger})_{i'j'}]^2}{\overline m_{i'}-\overline m_{j'}}
+ \frac{[\Im(N^{4\,*}XN^{2\,\dagger})_{i'j'}]^2}{\overline m_{i'}+\overline m_{j'}}
\right\}\,,
\qquad [i'=1',\ldots,4']\,, \label{realmi}\\[5pt]
m^{ph}_{j'} & \simeq &\overline m_{j'}-\sum_{i'=1}^4\left\{
 \frac{[\Re(N^{4\,*}XN^{2\,\dagger})_{i'j'}]^2}{\overline m_{i'}-\overline m_{j'}}
- \frac{[\Im(N^{4\,*}XN^{2\,\dagger})_{i'j'}]^2}{\overline m_{i'}+\overline m_{j'}}
\right\}\,,
\qquad [j'=5'\,,\,6']\,,\label{realmj}
\end{eqnarray}
The diagonal matrix of phases is chosen such that the $m^{ph}_{k'}$
are real and non-negative, with the phases $\phi_{k'}$ given by:
\begin{eqnarray}
\phi_{i'} &\simeq & -\sum_{j'=5}^6\,\frac{\overline m_{j'}}{\overline
 m_{i'}(\overline m_{i'}^2 -\overline m_{j'}^2)}
 \Re(N^{4\,*}XN^{2\,\dagger})_{i'j'}\,\Im(N^{4\,*}XN^{2\,\dagger})_{i'j'}\,,
 \qquad [i'=1',\ldots,4']\,,\\[5pt] \phi_{j'} &\simeq
 &\phm\sum_{i'=1}^4\,\frac{\overline m_{i'}}{\overline m_{j'}(\overline
 m_{i'}^2 -\overline m_{j'}^2)}
 \Re(N^{4\,*}XN^{2\,\dagger})_{i'j'}\,\Im(N^{4\,*}XN^{2\,\dagger})_{i'j'}\,,
 \qquad [j'=5',6']\,.
\end{eqnarray}
\s\vskip -0.4cm

The (perturbative) Takagi diagonalization of the neutralino mass matrix ${\cal M}_6$
has now been achieved, with the (real and non-negative) neutralino masses given
by Eqs.$\,$(\ref{realmi}) and (\ref{realmj}), and the neutralino mixing matrix
given by:
\begin{eqnarray}
N^6=\overline{N}^6_B \left(\begin{array}{cc}N^4 & \mathds{O}\\
    \mathds{O}^{\bf T} & N^2
\end{array}\right)\,.
\end{eqnarray}
\s\vskip -0.4cm

The validity of the perturbative expansion relies on the assumption
that\footnote{Since the $\overline m_{k'}$ are non-negative, and by
definition of order $M_{\rm SUSY}$, the
conditions $|\Im(N^{4\,*}X N^{2\,\dagger})_{i'j'}
/(\overline m_{i'}+\overline m_{j'})|\ll 1$ are automatically satisfied.}
\begin{eqnarray} \label{validity}
\left|\frac{\Re(N^{4\,*}X N^{2\,\dagger})_{i'j'}}
{\overline m_{i'}-\overline m_{j'}}\right|\ll 1\,,
\end{eqnarray}
for all choices of $i'=1',\ldots,4'$ and $j'=5',6'$.  That is, only
degeneracies between the $4\times 4$ block $\overline{\cal M}_4^D$ and the
$2\times 2$ block $\overline{\cal M}_2^D$ are potentially problematic.
In particular, in the so-called
cross-over zones in which the masses $\overline{m}_{i'}
\simeq \overline{m}_{j'}$ exhibit a near degeneracy and the
corresponding residue $\Re(N^{4\,*}XN^{2\,\dagger})_{i'j'}\neq 0$,
mixing effects are enhanced and the analytical formalism
in Appendix C must be applied.
\s

\subsection{The case of a real neutralino mass matrix}

We shall present numerical case studies under the assumption that
the parameters of the neutralino mass matrix are real.  The general
analysis then simplifies, since a real symmetric mass matrix can always
be diagonalized by a similarity transformation, $V{\cal M}V^T$,
where $V$ is real and orthogonal.
Since some of the mass eigenvalues of a real symmetric matrix may be negative,
we complete the Takagi diagonalization, $N^*{\cal M}N^\dagger$,
by introducing a suitable
diagonal matrix of phases $P$ and identifying the unitary neutralino mixing
matrix by $N=(PV)^*$, as indicated below Eq.$\,$(\ref{vmv}).
In this case, the (perturbative)
neutralino mass matrix diagonalization can be performed using the
three-step procedure of Ref.~\cite{NMSSM_CMZ}: \s
\vspace{0.1in}

\noindent {\bf{[1]}} \underline{\it Diagonalization of the submatrices
  ${\cal M}_4$ and ${\cal M}_2$}\\
In the first step, we diagonalize the (real symmetric)
MSSM matrix ${\cal M}_4$:
\begin{eqnarray}
\widetilde{\cal M}^{D}_4 = V^4 {\cal M}_4 (V^{4})^T
  = {\rm diag} (\tilde{m}_{1'}, \tilde{m}_{2'}, \tilde{m}_{3'}, \tilde{m}_{4'})\,.
\end{eqnarray}
The mass eigenvalues, which are real but need not be non-negative,
are denoted by $\tilde{m}_{i'}$ for $i' = 1',..,4'$.
The orthogonal diagonalization matrix $V^4$ is given explicitly in
Ref.~\cite{ckmz} for the most general choice of gaugino and
higgsino mass parameters. Simple analytic forms for the neutralino
mass and mixing matrix elements can be found
in limits where either the gaugino parameters are much larger
than the higgsino parameter or vice versa \cite{gunhab}. \s

The exact analytic diagonalization of the new $2 \times 2$ submatrix ${\cal M}_2$
singlet higgsino-U(1)$_X$ gaugino submatrix ${\cal M}_2$ is
straightforward.  The matrix:
\begin{eqnarray}
{\cal M}_2 = \left(\begin{array}{cc}
            0       &  Q'_S m_s \\
            Q'_S m_s &  M'_1
                  \end{array}\right)
\end{eqnarray}
is diagonalized by an orthogonal rotation $V^2$ as
\begin{eqnarray}
\widetilde{\cal M}^D_2 = V^2 {\cal M} (V^{2})^T
                   = {\rm diag}\, (\tilde{m}_{5'},\, \tilde{m}_{6'})\,.
\end{eqnarray}
The eigenvalues $\tilde{m}_{5',6'}$ are given by
\begin{eqnarray}
\tilde{m}_{5',6'} =
       \frac{M'_1}{2}\left(1\mp \sqrt{1+ (2Q'_S m_s/M'_1)^2} \right)\,.
\label{eq:exact_new_masses}
\end{eqnarray}
The orthogonal diagonalization matrix $V^2$ is given by:
\begin{eqnarray}
 V^2 =\left(\begin{array}{cc}
      \cos\theta_s  & -\sin\theta_s \\
      \sin\theta_s  &  \phm\cos\theta_s
     \end{array}\right)\,,
\label{eq:exact_new_rotation}
\end{eqnarray}
where the angle $\theta_s$ satisfies the relations:
\begin{eqnarray}
\cos\theta_s = \frac{\left(\sqrt{1+x^2}+1\right)^{1/2}}{
                     \sqrt{2}\left(1+x^2\right)^{1/4}}\qquad
                     \mbox{\rm and}\qquad
\sin\theta_s = {\rm sign}(x)\frac{\left(\sqrt{1+x^2}-1\right)^{1/2}}{
                     \sqrt{2}\left(1+x^2\right)^{1/4}}\,,
\end{eqnarray}
with $x\equiv 2 Q'_S m_s/M'_1$.\s

Two limits are of particular interest:\\
\noindent {\bf (i)} If $m_s \gg |M'_1|$, then the masses
and the mixing parameters are approximately given by
\begin{eqnarray}
\tilde{m}_{5'} \simeq - |Q'_S| m_s\,,\qquad
\tilde{m}_{6'} \simeq  |Q'_S| m_s\,,\quad\mbox{and}\quad
\sin\theta_s \simeq {\rm sign}(x)/\sqrt{2}\,,
\end{eqnarray}
corresponding to maximal mixing due to the large off-diagonal
entries in the mass matrix ${\cal M}_2$.

\noindent {\bf (ii)} In the opposite limit,
$|M'_1|\gg m_s$, and the mass eigenvalues and mixing angle
are approximately given by
\begin{eqnarray}
\tilde{m}_{5'} \simeq -Q'^2_S m^2_s/M'_1\,,\quad
\tilde{m}_{6'} \simeq M'_1+Q'^2_S m^2_s/M'_1\,,\quad \mbox{and}\quad
\sin\theta_s \simeq Q'_S m_s/M'_1\,.
\end{eqnarray}
This is a typical see-saw type mixing phenomenon. The heavy 6th
state is a U(1)$_X$ gaugino-dominated state, whereas
the 5th neutralino state is a singlet-higgsino dominated
state. \s

\noindent {\bf{[2]}} \underline{\it Block-diagonalization of ${\cal M}_6$}\\
We can now perform a block-diagonalization of ${\cal M}_6$:
\begin{eqnarray}
V^6{\cal M}^6 (V^6)^T = \widetilde{V}^6_B
\left(\begin{array}{cc} {\cal M}^{\prime\,D}_4 & V^{4}XV^{2\,T}
\\ V^{2}X^T V^{4\,T} & {\cal M}^{\prime\,D}_2 \end{array}\right)
\widetilde{V}^{6T}_B=
{\rm diag}(m_{1'}\,,\,\ldots\,,\,m_{4'}\,,\, m_{5'}\,,\, m_{6'})\,,
\end{eqnarray}
where
\begin{eqnarray} \label{calvdef}
\widetilde{V}^6_B & \simeq &
\left(\begin{array}{cc}
    \mathds{1}_{4\times 4}
    - \frac{1}{2} \Omega \Omega^T
  &  \Omega \\[1mm]
    -\Omega^T & \mathds{1}_{2\times 2}
   -\frac{1}{2} \Omega^T \Omega
      \end{array} \right)\,,
\end{eqnarray}
and the elements of the real matrix $\Omega$ are given by
[cf. Eq.$\,$(\ref{romega})]:
\begin{eqnarray} \label{omegareal}
\Omega_{i'j'}\equiv \frac{(V^{4}XV^{2\,T})_{i'j'}}
{\tilde m_{i'}-\tilde m_{j'}}\,,
\end{eqnarray}
with $i'=1',..,4'$ and $j'=5',6'$.
That is, the orthogonal matrix $V^6$ is
conveniently split into the matrices $V^4$ and $V^2$ that diagonalize the
$4\times 4$ and $2\times 2$ submatrices ${\cal M}_4$ and ${\cal M}_2$
respectively, and into the matrix $\widetilde{V}^6_B$ that performs the
subsequent block-diagonalization \cite{NMSSM_CMZ}:
\begin{eqnarray}
V^6\, \simeq\,
 \widetilde{V}^6_B\,\left(\begin{array}{cc}
          V^4  &  \mathds{O} \\[1mm]
          \mathds{O}^T  &  V^2
        \end{array}\right)\,.
\label{eq:mixing_matrix}
\end{eqnarray}
After the block-diagonalization,
the mass eigenvalues are shifted to second order in the perturbation $X$.
The shifts are given by [cf. Eq.$\,$(\ref{realmi}) and (\ref{realmj})]:
\begin{eqnarray} \label{massshift}
&& m_{i'} = \tilde{m}_{i'}+ \sum^{6'}_{j'=5'}\frac{[(V^4 X V^{2T})_{i'j'}]^2}
                                       {\tilde{m}_{i'}-\tilde{m}_{j'}}\,,
         \qquad \,[\,i'=1',..,4']\,,\\
&& m_{j'} = \tilde{m}_{j'}-\sum^{4'}_{i'=1'}\frac{[(V^4 X V^{2T})_{i'j'}]^2}
                                      {\tilde{m}_{i'}-\tilde{m}_{j'}}\,,
         \qquad [\,j'=5',\,6']\,.
\end{eqnarray}
As expected, the eigenvalues fulfill the trace formula
\begin{eqnarray} \label{sumrule}
\sum^{6'}_{k'=1'} m_{k'} = M_1+M_2+M'_1\,,
\end{eqnarray}
which is independent of the higgsino mass and the mixing
parameters.\s

The perturbative results obtained above are valid if
$|(V^4 XV^{2\,T})_{i'j'}/(\tilde m_{i'}-\tilde m_{j'})|\ll 1$ for all
possible choices of $i'$ and $j'$.
In the regime of near-degeneracy, $\tilde m_{i'}\simeq \tilde m_{j'}$,
the perturbation theory breaks down, and the analytic approach of
Appendix C must be employed.  Note that $\tilde
m_{i'}=-\tilde m_{j'}$ is \textit{not} a case of
mass-eigenvalue degeneracy, so that the
perturbative results obtained above should be reliable.  This may
seem to be in conflict with results of the previous subsection,
since the latter corresponds to the \textit{degenerate}
case of $\overline m_{i'}=\overline m_{j'}$, where we identify the
\textit{positive} masses $\overline m_{k'}=|\tilde m_{k'}|$ in
the notation of Sect.~\ref{genanalysis}.  However, a more careful
analysis reveals that the condition given by
Eq.$\,$(\ref{validity}) does \textit{not} apply,
since in the case of opposite sign mass
eigenvalues, the residue $\Re(N^{4\,*}XN^{2\,\dagger})_{i'j'}=
\Re(iV^4 XV^{2\,T})_{i'j'}=0$.\footnote{As in step {\bf{[3]}} below,
we identify $N^M=(P^M V^M)^*$ for $M=2$ and 4, respectively, and
$(P^4)^{-1}_{i'i'}(P^2)^{-1}_{j'j'}=-i$ for opposite sign mass
eigenvalues $\tilde m_{i'}$ and $\tilde m_{j'}$.}
\s

The higgsino doublet-singlet and the higgsino doublet-U(1)$_X$
gaugino mixings generate additional singlino and U(1)$_X$ gaugino
components in the wave functions of the original MSSM neutralinos
$\tilde{\chi}^0_{i'}$ $[\,i'=1',.. ,4'\, ]$ of the size
\begin{eqnarray}
V^6_{i'j'} \, \approx\, \sum_{k'=5'}^{6'}\, \Omega_{i'k'} V^2_{k'j'}\quad
[i'=1',.., 4'; j'=5',6']
\end{eqnarray}
which is linear in the mixing parameter to first approximation as expected
for off-diagonal elements. Reciprocally, the
MSSM gaugino/higgsino
components and the singlino and U(1)$_X$ gaugino components in the
wave functions of $\tilde{\chi}^0_{5'}$ and $\tilde{\chi}^0_{6'}$ are
reduced to
\begin{eqnarray}
&& V^6_{j'i'}\, \approx\, -\sum_{l'=1'}^{4'} \Omega_{l'j'}
   V^4_{l'i'}\qquad\qquad\ \ \
              \;\;[i'=1',..,4';\, j'=5',6']\nonumber \\[-1mm]
&& V^6_{j'k'}\, \approx\, V^2_{j'k'}-\frac{1}{2}\,\left(
  \Omega^T \Omega V^2\right)_{j'k'} \qquad [j',k'=5',6']
\end{eqnarray}
with $V^6_{j'k'}$ differing from $V^2_{j'k'}$ only to second order in
the mixing, as expected for diagonal elements.\s

\noindent {\bf{[3]}} \underline{\it Ensuring that the physical neutralino masses
are non-negative}\\
The diagonalization of a real symmetric matrix by an
orthogonal similarity transformation
produces a diagonal matrix with real but not necessarily non-negative
elements.  Hence, some of the eigenvalues $m_{k'}$ will typically be negative.
Defining the unitary matrix $N^6=(P^6 V^6)^*$, where $P^6$ is a diagonal
matrix whose $k'k'$ element is $1$ [$i$] if $m_{k'}$ is non-negative [negative],
the Takagi diagonalization of the neutralino mass matrix is achieved with
non-negative neutralino masses. In particular, the unitary neutralino mixing
matrix $N^{6\,*}$ appears (instead of the real orthogonal matrix $V^6$) in
the corresponding Feynman rules involving the neutralino mass-eigenstates.\s

\subsection{Large gaugino mass parameters}

To illustrate the previous general discussion we shall first give a
detailed parametric analysis in the limit in which all gaugino
masses are much larger than the higgsino masses, and both sets much
larger than the electroweak and the kinetic mixing scales,
i.e. $M_1, M_2, M'_1 \gg \mu, v_s \gg v, M_K$.
All neutralino mass matrix parameters will be taken real.\s

\noindent
{\bf{[1]}} Starting again with the diagonalization of the MSSM
submatrix ${\cal M}_4$, the diagonalization matrix $V^4$ defined in
Eq.$\,$(\ref{eq:mixing_matrix}) can be parameterized up to second order
according to standard MSSM procedure (see, e.g., Ref.~\cite{ckmz}), as
\begin{eqnarray}
V^4\simeq \left(\begin{array}{cc}
            V_G    &  \mathds{O}  \\[1mm]
             \mathds{O}^T    &  V_H
      \end{array}\right)
\left(\begin{array}{cc}
             \mathds{1}_{2\times 2}   & V_x       \\[1mm]
              -V^T_x &   \mathds{1}_{2\times 2}
      \end{array}\right)
\left(\begin{array}{cc}
            \mathds{1}_{2\times 2}    & \mathds{O} \\[1mm]
            \mathds{O}^T    &  R_{\pi/4}
      \end{array}\right)\,.
\label{eq:V_mixing}
\end{eqnarray}
The effect of the
$2\times 2$  rotation $R_{\pi/4}\equiv(\mathds{1}-i\tau_2)/\sqrt{2}$
[where $\boldsymbol{\vec\tau}\equiv(\tau_1\,,\,\tau_2\,,\,\tau_3)$
are the $2\times 2$ Pauli matrices]
is to shift the $\{34\}$ off-diagonal elements $[-\mu,-\mu]$ onto the
diagonal axis $[\mu, -\mu]$. The matrix, $V_x$,
\begin{eqnarray} \label{cpm}
V_x\, = \,
 \left(\begin{array}{cc}
        -c_+\, s_W\, m_Z/M_1  & \ \  -c_-\, s_W\, m_Z/M_1\\[3mm]
         c_+\, c_W\, m_Z/M_2  & \ \  c_-\, c_W\, m_Z/M_2
        \end{array}\right)\,,
\end{eqnarray}
with the abbreviations $c_\pm \equiv (c_\beta\pm s_\beta)/\sqrt{2}$,
removes the mixing between the blocks of the two gaugino and the two
higgsino states. $V_G$ and $V_H$ rescale the
gaugino and higgsino blocks themselves:
\begin{eqnarray}
&& V_G \approx \mathds{1}_{2\times 2} -\frac{1}{2}
        \left(\begin{array}{cc}
             s^2_W \, m^2_Z/M^2_1 &  0 \\[2mm]
               0    &   c^2_W\,  m^2_Z/M^2_2
            \end{array}\right)\,,\nonumber\\[2mm]
&& V_H \approx \mathds{1}_{2\times 2}-\frac{1}{2}
        \left(\begin{array}{cc}
            (1+s_{2\beta})\, M_{12}^{\prime\prime 2} m^2_Z/2 M^2_1 M^2_2
         &  0 \\[2mm]
            0
         &  (1-s_{2\beta})\, M_{12}^{\prime\prime 2} m^2_Z/2 M^2_1 M^2_2
            \end{array}\right)\,,
\end{eqnarray}
with $M_{12}^{\prime\prime\,2} \equiv M^2_1 c^2_W+ M^2_2 s^2_W$.
$V_G$ and $V_H$
relate to a diagonal form of the gaugino-higgsino mass matrix for large
$M_{1,2}$ and $\mu$. Their off-diagonal matrix elements are of second order
and can be omitted consistently as they would only affect the eigenvalues
at fourth order. \s

The $2\times 2$ diagonalization matrix defined in Eq.$\,$(\ref{eq:mixing_matrix})
can be parameterized up to second order as
\begin{eqnarray}
V^2 \approx \left(\begin{array}{cc}
     1 - Q'^2_S m^2_s/2M^{\prime 2}_1  & - Q'_S m_s/M'_1 \\
      Q'_S m_s/M'_1 & 1 - Q'^2_S m^2_s/2M^{\prime 2}_1
                 \end{array}\right)\,.
\label{eq:2nd_rotation1}
\end{eqnarray}
The $2\times 2$ matrix $V^2$ generates a diagonal form of the
singlino-U(1)$_X$ gaugino mass matrix for $M'_1\gg m_s$.\s

After these steps are performed, the $4\times 4$ and $2\times 2$
mass submatrices are diagonal and the complete symmetric mass
matrix ${\cal M}_6$ takes the intermediate form
\begin{eqnarray}
\hspace{-0.3in}
\left(\begin{array}{cc} V^{4} &  \mathds{O} \\  \mathds{O}^T &  V^{2}
\end{array}\right)\,{\cal M}_6\,
\left(\begin{array}{cc} V^{4\,T} &  \mathds{O} \\  \mathds{O}^T &
  V^{2\,T}\end{array}\right)\simeq
\left(\begin{array}{cccc|cc}
  \tilde{m}_{1'} &  &  &  & 0 & M_K  \\[1mm]
   & { }\hskip 1mm \tilde{m}_{2'}&  &  & 0 & 0  \\[1mm]
   &  & \tilde{m}_{3'} &  & + \mu_\lambda c_-  & Q'_- m_v \\[4mm]
   &  &  & \tilde{m}_{4'} & -\mu_\lambda c_+ &  Q'_+ m_v \\[2mm]
   \cline{1-6}
   0 & 0 & +\mu_\lambda c_- & -\mu_\lambda c_+  & \tilde{m}_{5'} &  \\[2mm]
   M_K & 0 & Q'_- m_v & Q'_+ m_v &  & \tilde{m}_{6'}
      \end{array}\right),
\label{eq:1st_rotated_matrix}
\end{eqnarray}
where, in obvious notation, zero elements of the diagonal blocks
are suppressed for easier
reading, and $Q'_\pm \equiv (Q'_1 c_\beta \pm Q'_2
s_\beta)/\sqrt{2}$. The diagonal elements $\tilde m_{k'}$ are given by
\begin{align}
\tilde{m}_{1'} &= M_{1} + \frac{m^2_Z}{M_{1}} s^2_W\,,  &
\tilde{m}_{3'} &=\phm\mu - \frac{m^2_Z M_{12}}{M_1 M_2}c_+^2\,,
                 & \tilde{m}_{5'} &= \mu_\kappa \,,      \nonumber\\
\tilde{m}_{2'} &= M_{2} + \frac{m^2_Z}{M_{2}} c^2_W\,, &
\tilde{m}_{4'} &=-\mu - \frac{m^2_Z M_{12}}{M_1 M_2}c_-^2\,,
                 & \tilde{m}_{6'} &= M'_1 -\mu_\kappa\,,
\label{eq:1st_mass_spectrum}
\end{align}
where $c_\pm$ is defined below Eq.$\,$(\ref{cpm}) and
\begin{eqnarray}
M_{12}\equiv M_1 c^2_W + M_2 s^2_W\,,\qquad\quad
\mu_\kappa \equiv -Q_S'^2 m_s^2/M_1'\,.
\end{eqnarray}
The parameter $\mu_\kappa$ can be identified with the
NMSSM-type singlino mass parameter \cite{NMSSM_CMZ}.
Note that $\tilde{m}_{5'}=\mu_\kappa$ is small compared to
all the other neutralino masses in the limit of large gaugino mass parameters
considered in this subsection.\s

\noindent
{\bf [2]} The block-diagonalization of the
6-dimensional intermediate matrix [Eq.$\,$(\ref{eq:1st_rotated_matrix})]
can be performed by choosing the proper form of $\Omega$ in $V^6$.
In the limit of large gaugino mass parameters
and small singlino mass $\mu_\kappa \ll \mu\ll M_1, M_2, M'_1$, the $4\times 2$
mixing matrix $\Omega$ is reduced to the simple expression
\begin{eqnarray}
\Omega \approx \left(\begin{array}{cc}
   0  \quad &  {M_K}/{(M_1-M'_1)} \\
   0  \quad &  0 \\
   {\mu_\lambda c_-}/{\mu} \quad
   & -{Q'_- m_v}/{M'_1} \\[2mm]
     {\mu_\lambda c_+}/{\mu} \quad
   & -{Q'_+ m_v}/{M'_1}
                    \end{array}\right)\,.
\label{eq:large_gaugino_omega}
\end{eqnarray}
As a result of the block diagonalization of
Eq.$\,$(\ref{eq:1st_rotated_matrix}), the mass eigenvalues
are shifted according to Eq.$\,$(\ref{massshift}).
The resulting mass eigenvalues to the desired order are given by:
\begin{align}
m_{1'} &\approx M_1 + \frac{m^2_Z}{M_1} s^2_W
+\frac{M_K^2}{M_1-M^{\prime}_1}\,,&
m_{4'} &\approx -\mu - \frac{m^2_ZM_{12}}{M_1 M_2} c^2_-
             - \frac{\mu^2_\lambda c^2_+}{\mu}
             + \frac{Q^{\prime\,2}_+ m^2_v}{M'_1}\,,
     \nonumber\\
m_{2'} &\approx M_2 + \frac{m^2_Z}{M_2} c^2_W\,,                     &
 m_{5'} &\approx \mu_\kappa + \frac{\mu^2_\lambda}{\mu} s_{2\beta}
                   \,,
     \nonumber\\
 m_{3'} &\approx \mu - \frac{m^2_ZM_{12}}{M_1 M_2} c^2_+
             + \frac{\mu^2_\lambda c^2_-}{\mu}
             +\frac{Q^{\prime\,2}_+ m^2_v}{M'_1}\,,        &
 m_{6'} &\approx M'_1-\mu_\kappa
             + \frac{m^2_v(Q'^2_+ + Q'^2_-)}{M'_1}-\frac{M^2_K}{M_1-M'_1} \,.
\label{eq:mass_spectrum1}
\end{align}
Note that the sum rule given by Eq.$\,$(\ref{sumrule}) is satisfied.
\s

As expected, while the
large SU(2) gaugino mass $m_{2'}$ is not affected by the singlino and
the U(1)$_X$ gaugino, the MSSM U(1) mass $m_{1'}$ is affected by
the U(1) kinetic mixing.
All the higgsino states are modified by the interactions between the
MSSM and the new subsystem. The value of $m_{5'}$ is raised
by the interaction with the MSSM higgsinos, but remains small
nevertheless.  \s

The mixing in the wave-functions is described by the components of
$\Omega$ alone since the $4\times 4$ matrix $V^4$ and the $2\times
2$ matrix $V^2$ deviate from unity only to second order in the small
parameters of the order of the SUSY scales [$i = 1',..,4'$]:
\begin{align}
&V^6_{i'5'} \approx \frac{\mu_\lambda}{\mu}
         \left(0,\, 0,\, c_{-},\, c_{+}\right)_{i'}\,,&
  &V^6_{5'5'} \approx 1-\frac{Q'^2_S m^2_s}{2M'^2_1}
                       -\frac{\mu^2_\lambda}{2\mu^2}\,,
                       \nonumber\\
&V^6_{5'i'} \approx -\frac{\mu_\lambda}{\mu}
         \left(0,\, 0,\, c_\beta,\, s_\beta\right)_{i'}\,,&
  &V^6_{5'6'} \approx -\frac{Q'_S m_s}{M'_1}\,,
                       \nonumber\\
  &V^6_{i'6'} \approx \left(\frac{M_K}{M_1-M'_1},\, 0,\,
         -\frac{Q'_-m_v}{M'_1},\, -\frac{Q'_+ m_v}{M'_1}\right)_{i'}\,,&
  &V^6_{6'5'} \approx \frac{Q'_S m_s}{M'_1}\,,
    \nonumber\\
&V^6_{6'i'} \approx \left(\frac{-M_K}{M_1-M'_1},\, 0,\,
         \frac{Q'_1 m_v c_\beta}{M'_1},\,
         \frac{Q'_2 m_v s_\beta}{M'_1}\, \right)_{i'}\,,  &
  &V^6_{6'6'} \approx 1-\frac{Q'^2_S m^2_s}{2M'^2_1}
         -\frac{m^2_v(Q_+^{\prime\,2}+Q_-^{\prime\,2})}{2M'^2_1}
         -\frac{M^2_K}{2(M_1-M_1^{\prime})^2}\,.
\label{eq:v5}
\end{align}
The non-trivial mixing between two U(1) gaugino states, elements
$\{1'6'\}$ and $\{6'1'\}$, is generated by the non-zero Abelian gauge
kinetic and mass mixing with non-zero $M_K$.
The analysis above fails when $M_1\approx M'_1$; this
region of near degeneracy can be handled analytically using the results of
Appendix C.\s

In Eqs.$\,$(\ref{eq:mass_spectrum1}) and (\ref{eq:v5}), perturbative corrections
up to second order have been included for the masses and diagonal mixing matrix
elements, whereas only the first order corrections have been given for the
off-diagonal mixing matrix elements.  This follows the usual procedure of
stationary perturbation theory in quantum mechanics, which associates
second-order corrections to the eigenvalues with the first-order corrections
to the wave function.  Consequently, the zeros that appear in some of the matrix
elements of $V^6_{k'\ell'}$, should be interpreted as approximate.  For example,
$V^6_{2'6'}$ and $V^6_{6'2'}$ are expected to receive higher
order perturbative corrections and hence be shifted away from zero.  Nevertheless,
the fact that the magnitude of these matrix elements are so suppressed will have
some dramatic consequences for the behavior of the $\tilde\chi_{2'}^0$ and
$\tilde\chi_{6'}^0$ masses in regions of near-degeneracy.\s

\noindent
{\bf [3]} The final step is to identify $N^6=(P^6 V^6)^*$, where $P^6$ is a diagonal
matrix whose $k'k'$ element is $1$ ($i$) if $m_{k'}$ is non-negative (negative).
The physical masses $m^{ph}_k$ are given by the absolute values of the $m_{k}$ given
above. The neutralino states can then be reordered in ascending (non-negative) mass
if desired.\s

The results of this subsection are easily generalized for the case of
$M_1$, $M_2$, $M'_1$, $\mu$, $v_s\gg v$, $M_K$.  As long
as the MSSM gaugino and higgsino parameters,
$M_{1,2}$ and $\mu$ remain significantly larger than the electroweak scale $v$,
the couplings between the MSSM and the new fields, generated by $X$,
remain weak and the diagonalization
of the mass matrix can still be performed analytically.  However, instead of
the approximate values $\tilde{m}_{5',6'}$ in
Eqs.$\,$(\ref{eq:1st_rotated_matrix})
and (\ref{eq:1st_mass_spectrum}) the exact solutions
(\ref{eq:exact_new_masses})
must be used, and for $V^2$ the general rotation matrix
(\ref{eq:exact_new_rotation}) must be inserted. The approximation ceases to
be valid at isolated points where $X/(\tilde{m}_{i'}-\tilde{m}_{j'})$ is no
longer a small perturbation, due to the degeneracy of mass eigenvalues
$\tilde{m}_{i'}\approx\tilde{m}_{j'}$. In these cross-over zones the
analysis described in Appendix C must be applied.  \s

\subsection{An illustrative example}
\label{sec:sec3.3}

To illustrate the properties of the two new neutralinos and the
impact of the coupling of the two subsystems on the original MSSM
neutralinos, we study, numerically and analytically, the evolution of the
neutralino mass spectrum and representative examples for the mixing of the
particles from a very light new U(1)$_X$ gaugino across typical MSSM mass
scales up to very high scales.
Gauge kinetic mixing has only a small impact on the spectrum and it will
therefore be neglected in the illustrative example.
Throughout the evolution, including all intermediate regions, the
coupling between the new states and the MSSM states remains weak, apart
from regions of mass degeneracy. The evolution affects primarily the
spectrum of the two new neutralino states. In the initial limit, $M'_1$ small,
two medium-heavy degenerate states, $m_{5',6'} \sim O(v_s)$, are realized
in the spectrum. At the end of the chain, $M'_1$ large,
the spectrum is of a see-saw type, including one heavy and one nearly
zero-mass state. \s

As an illustrative
example, we take $M_2=1.5$ TeV, $m_s=1.2$ TeV, $\mu=0.3$ TeV and $M_K=0$, and
we assume the gaugino unification relation $M_1=(5/3)\tan^2\theta_W M_2
\approx 0.5 M_2$. Also, for the numerical analysis in this paper, we set
$\tan\beta=5$. We adopt the $N$-model charge assignments \cite{u1ssm3},
\begin{eqnarray}
Q_1 = -\frac{3}{2\sqrt{10}}\,,\qquad
Q_2 = -\frac{2}{2\sqrt{10}}\,,\qquad
Q_S = \frac{5}{2\sqrt{10}}\,.
\end{eqnarray}
For definiteness we fix the gauge coupling at $g_X \simeq 0.46$, evolved
from its E$_{6}$ unification value of
$\sqrt{5/3}\, g_Y$ down to the electroweak scale;
however the results are not very sensitive to this assumption.
We could also choose to fix $M_X$ at its gaugino unification value
under the assumption that all gaugino masses unify at the
grand unification scale.
This would correspond to a value of $M_1'\approx M_X=M_1=750$~GeV (neglecting
kinetic mixing effects).  However, to illustrate the structure of the
system in various scenarios, we shall be slightly more general
by allowing $M_1'$ to vary over a large range of values
($0\leq M_1'\leq 5$~TeV). \s

To be specific, we choose the evolution with $M_1'$
\begin{eqnarray}
\begin{array}{lccccccccc}
{\rm from:} &  M_1' & \ll & v & \ll & \mu & \ll & M_1,\, M_2,\, v_s & &  \nonumber \\
{\rm to:}   &       &  & v & \ll & \mu & \ll & M_1,\, M_2,\, v_s & \ll & M_1' \,.
\nonumber
\end{array}
\end{eqnarray}
\begin{figure}[t!]
\begin{center}
\includegraphics[height=10.cm,width=11.cm,angle=0]{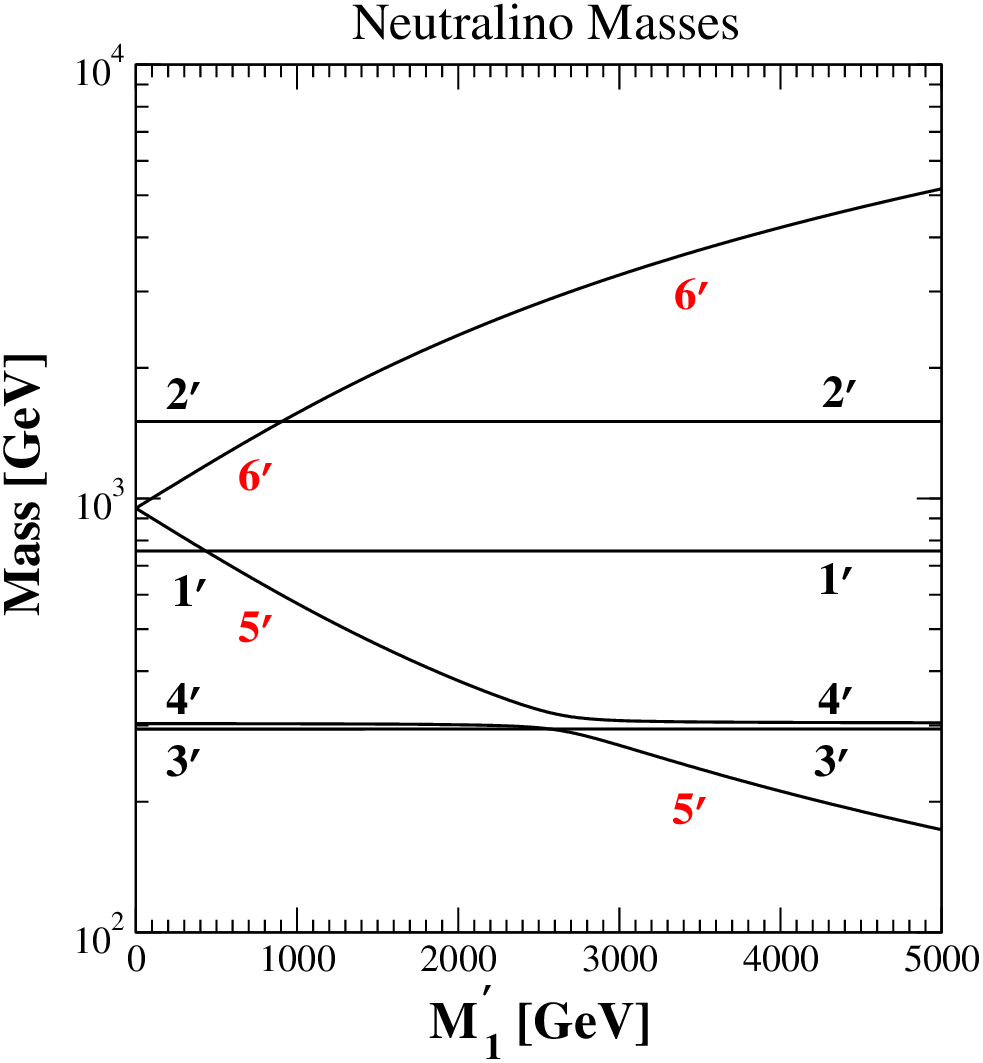}\hskip 1.cm
\end{center}
\vskip -0.5cm \caption{\it The evolution of the six  neutralino masses
             when varying the U(1)$_X$ gaugino mass parameter $M'_1$.
             The values used for the parameters are given in the text.
             The numbers with primes characterize the nature of
             the neutralino states connected with
             the ordering of the states when evolving from $M'_1=0$.
             Note that the $2'$ and $6'$ curves and the $4'$ and $5'$ curves,
            respectively, do not actually touch.  This can be seen more
            clearly in Fig.$\,$\ref{fig:fig7}, where these near intersection regions
            are expanded. The $1'$ and $5'$ curves, corresponding to opposite-sign
            mass eigenvalues, intersect for small $M'_1$ but affect each other
            only weakly.}
\label{fig:fig1}
\end{figure}
The evolution of the six (positive) neutralino masses\footnote{The
eigenvalues $4'$ and $5'$ of the mass matrix
[Eq.$\,$(\ref{eq:1st_rotated_matrix})] are negative, while
all the other eigenvalues are positive. Level crossing will therefore occur
only between $2'$-$6'$ and $4'$-$5'$ when
$M_1'$ is increased.  The physical neutralino masses are given by
the absolute values of the corresponding mass eigenvalues.} and the
values of two typical $V^6$ mixing elements, $\{5'4'\}$ and $\{5'6'\}$,
are shown in Figs.$\,$\ref{fig:fig1} and \ref{fig:fig2}.
The neutralino state mixings are exemplified by the $V^6$ matrix elements
$\{5'4'\}$ and $\{5'6'\}$ as representative for gaugino and
higgsino mixings of the MSSM and the new states, as well
as the mixing among the new gaugino and singlino states themselves.\s

\begin{figure}[ht!]
\begin{center}
\includegraphics[height=10.cm,width=11.cm,angle=0]{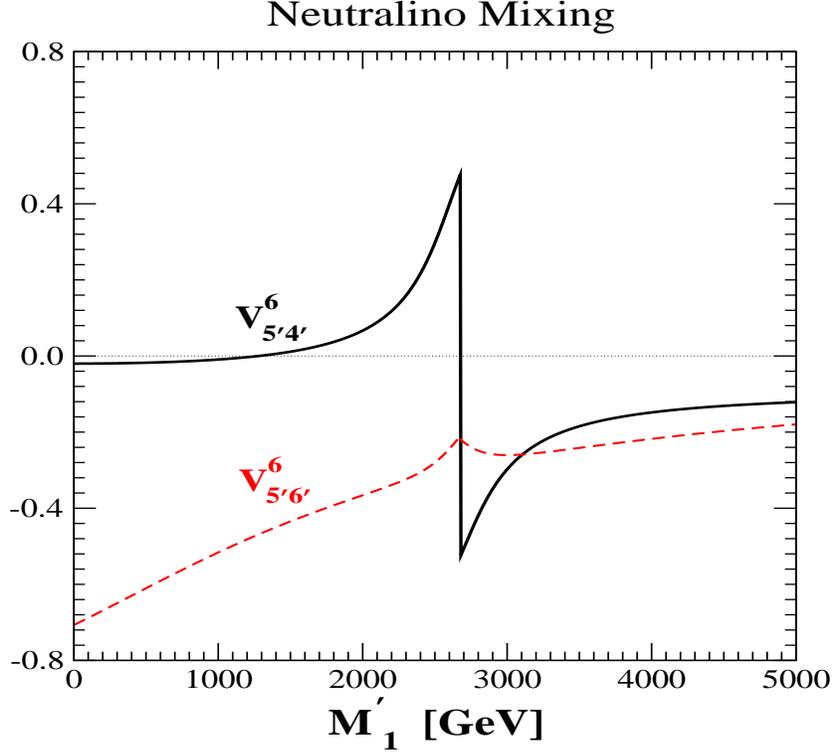}\hskip 1.2cm
\end{center}
\vskip -0.5cm \caption{\it The evolution of two representative mixing
             matrix elements when the U(1)$_X$ gaugino mass parameter
             $M'_1$ is varied from small to large values. Large variations
             of the $5'4'$ parameter occur
             in the cross-over zone near $M'_1=2.6$
             TeV.}
\label{fig:fig2}
\end{figure}

When the new U(1)$_X$ gaugino mass parameter $M'_1$ is varied from small
to very large values, the pattern of neutralino masses evolves in an
interesting way, as shown in Fig.$\,$\ref{fig:fig1}. For small $M'_1$ the
set of parameters chosen in the previous paragraph, leads to a heavy
SU(2) MSSM gaugino $\tilde{\chi}^0_{2'}$. It is followed by the two new
states, mixed maximally in the U(1)$_X$ gaugino and singlino sector,
$\tilde{\chi}^0_{5'}$ and $\tilde{\chi}^0_{6'}$. The fourth heaviest state is
the U(1) MSSM gaugino $\tilde{\chi}^0_{1'}$. The lightest states are the two
MSSM higgsinos $\tilde{\chi}^0_{4'}$ and $\tilde{\chi}^0_{3'}$.
If $M'_1$ is shifted to higher values, the mass eigenvalues in the new sector
move apart, generating strong cross-over patterns whenever a mass
from the new block comes close to one of the MSSM masses. This is realized
at small $M'_1 \approx \tilde{m}_{2'} - Q^{\prime 2}_Sm^2_s/
\tilde{m}_{2'} \approx 0.91$ TeV for the neutralino $\tilde{\chi}^0_{6'}$ in
the new block and the SU(2) MSSM neutralino $\tilde{\chi}^0_{2'}$; later
between the new-block state $\tilde{\chi}^0_{5'}$ and the MSSM higgsino
$\tilde{\chi}^0_{4'}$ for
$M'_1 \approx \tilde{m}_{4'}-Q^{\prime 2}_S m^2_s/\tilde{m}_{4'} \approx 2.68$
TeV. For very large $M'_1$, $\tilde{\chi}^0_{5'}$ approaches the singlino
state with a small mass value $|\tilde{m}_{5'}|\sim Q'^2_S m^2_s/M'_1$,
and $\tilde{\chi}^0_{6'}$ the pure U(1)$_X$ gaugino state with a very large
mass $\tilde{m}_{6'}\sim M'_1$.\s

Outside the cross-over regions the approximate analytical mass spectra
nearly coincide with the exact (numerically computed) solutions for the
eigenvalues as demonstrated in Table~\ref{tab:tab1} for three $M_1'$ values.
\s
\begin{table}[ht!]
\caption{\label{tab:tab1} {\it Comparison between the exact and approximate
    neutralino masses $m_{\tilde{\chi}^0_i}$ [in GeV] for three values of $M_1'$.
    The values of the other parameters are defined in the text.}}
\begin{center}
\begin{tabular}{|c|r|r|r|r|r|r|r|r|r|}
\hline
\!\!$\tilde{\chi}^0_i$\!\!   & \multicolumn{3}{c|} {$M'_1 = 400$ GeV }
       & \multicolumn{3}{|c|} {$M'_1 = 2000$ GeV }
       & \multicolumn{3}{|c|} {$M'_1 = 4000$ GeV}\\
\cline{2-10}
\!\!$m$ [GeV]\!\!  & Exact & Appr. & ${ \Delta m/m}$
       & Exact & Appr. & {$\Delta m/m$}
       & Exact & Appr. & {$\Delta m/m$} \\
\hline\hline
  1    & 294.0  & 295.8   &  0.6\%
       & 294.1  & 295.9   &  0.6\%
       & 211.6  & 211.4   & -0.1\% \\
  2    & 302.7  & 303.2   &  0.1\%
       & 301.0  & 301.4   &  0.1\%
       & 294.2  & 296.0   &  0.6\% \\
  3    & 756.5  & 755.6   & -0.1\%
       & 380.3  & 380.3   &  0.0\%
       & 304.7  & 305.3   &  0.2\% \\
  4    & 770.1  & 770.1   &  0.0\%
       & 756.5  & 755.6   & -0.1\%
       & 756.5  & 755.6   & -0.1\% \\
  5    & 1170.6 & 1170.6  &  0.0\%
       & 1504.8 & 1504.3  &  0.0\%
       & 1504.8 & 1504.3  &  0.0\% \\
  6    & 1504.8 & 1504.3  &  0.0\%
       & 2379.0 & 2379.0  &  0.0\%
       & 4213.9 & 4213.9  &  0.0\% \\
\hline
\end{tabular}
\end{center}
\end{table}

The mixing pattern is more directly reflected in the elements of the
rotation matrix $V^6$, as shown in Fig.$\,$\ref{fig:fig2}. For zero kinetic
mixing, $\tilde\chi^0_{5'}$ and $\tilde\chi^0_{6'}$ do not overlap with
the U(1) MSSM gaugino, since $V^6_{1'5'}$, $V^6_{1'6'}\approx 0$.
Their overlap with the MSSM higgsinos, $V^6_{5'4'}$,
is small except in the cross-over zone. The mixing $V^6_{5'6'}$ between
the new U(1)$_X$ gaugino and singlino states is reduced
from maximal mixing $-1/\sqrt{2}$ for vanishing U(1)$_X$ gaugino mass
parameter $M'_1$ to nearly zero mixing at asymptotically large $M'_1$.
The moderate change in the $5'$-$4'$ cross-over zone is a reflection
of the $5'4'$ variations by unitarity of the neutralino mixing matrix.\s

\section{Neutralino Production and Decays}
\label{sec:sec4}
\setcounter{equation}{0}

Neutralino production rates in various channels and decay properties in various
modes are affected by the mixing of the neutralino states and
by the mass and kinetic mixings of the gauge bosons associated with the broken
U(1)$_X$ and SU(2)$\times$U(1)$_Y$ gauge symmetries.  \s

The $Z$ and $Z'$ bosons can mix through kinetic coupling, as
analyzed before, and mass mixing induced by the exchange of
the Higgs fields, for example, charged under both U(1)$'$s. The
resulting $Z$ and $Z'$ mixing is described by the mass-squared
matrix
\begin{eqnarray}
M^2_{ZZ'} =\left(\begin{array}{cc}
       m^2_Z    & \Delta^2_Z \\
       \Delta^2_Z &  m^2_{Z'}
                 \end{array}\right)\,,
\end{eqnarray}
where the matrix elements are given by
\begin{eqnarray}
&& m^2_Z    = \frac{1}{4} g^2_Z v^2\,, \nonumber\\
&& m^2_{Z'} =  g^2_X v^2 \left( Q'^2_1 c^2_\beta + Q'^2_2
s^2_\beta\right)
             + g^2_X v^2_S Q'^2_S\,,\nonumber\\
&& \Delta^2_Z = \frac{1}{2} g_Z g_X  v^2
              \left(Q'_1 c^2_\beta -Q'_2 s^2_\beta\right)\,,
\end{eqnarray}
and where $g^2_Z\equiv g^2_2+g^2_Y$. The eigenvalues of
$M_{ZZ'}^2$ and the $Z$ and $Z'$ mixing angle follow from
\begin{eqnarray}
&& m^2_{Z_1, Z_2}=\frac{1}{2}\left( m^2_Z+m^2_{Z'}
           \mp \sqrt{(m^2_Z-m^2_{Z'})^2+4 \Delta^4_Z}\right)\,,\nonumber\\
&& \tan2\theta_{ZZ'} =-2\Delta^2_Z/(m^2_{Z'}-m^2_Z)\,.
\end{eqnarray}
The phenomenological constraints typically require this mixing angle to
be less than a few times $10^{-3}$~\cite{Zprime_bound}, although values as
much as ten times larger may be possible in some models with a light $Z'$ and
reduced couplings \cite{Leike}.\s

For the neutralino production processes in $e^+e^-$ annihilation it
is sufficient to consider the neutralino-neutralino-$Z_{1,2}$
vertices
\begin{eqnarray}
&& \langle \tilde{\chi}^0_{iL}|Z_1|\tilde{\chi}^0_{jL}\rangle
 = -g_Z{\cal Z}_{ij} \cos\theta_{ZZ'}
   -g_X {\cal Z}^{\prime}_{ij} \sin\theta_{ZZ'}\,,
   \nonumber\\
&& \langle \tilde{\chi}^0_{iL}|Z_2|\tilde{\chi}^0_{jL}\rangle
 = +g_Z{\cal Z}_{ij} \sin\theta_{ZZ'}
   -g_X {\cal Z}^{\prime }_{ij} \cos\theta_{ZZ'}\,,
\label{eq:coupling_set1}
\end{eqnarray}
with $i,j=1,..,6$ and $g_Z=g_2/c_W$; $L \to R$ can be switched by substituting
${\cal Z}_{ij} \to -{\cal Z}^*_{ij}$ and ${\cal Z}^\prime_{ij}
\to -{\cal Z}^{\prime\,*}_{ij}$.  Explicitly, the couplings
${\cal Z}_{ij}$ and ${\cal Z}'_{ij}$ are given in terms of the USSM
neutralino mixing matrix $N$ by\footnote{For simplicity of notation,
the USSM neutralino mixing matrix $N^6$ will be denoted by $N$ in this
section.}
\begin{eqnarray}
&& {\cal Z}_{ij}=\frac{1}{2}\left(N_{i3} N^{*}_{j3}
                                 -N_{i4} N^{*}_{j4}\right)\,,\nonumber\\
&& {\cal Z}'_{ij} = Q'_1 N_{i3} N^{*}_{j3}
                   +Q'_2 N_{i4} N^{*}_{j4}
                   +Q'_S N_{i5} N^{*}_{j5}\,.
\end{eqnarray}
Sfermion $t/u$-channel exchanges require the fermion-sfermion-neutralino
vertices (with the fermion masses neglected):
\begin{eqnarray}
&& \langle \tilde{\chi}^0_{iR}|\tilde{f}_L|f_L\rangle
 = -\sqrt{2}
    \left[g_2(I^f_{3} N^{*}_{i2} +(e_f-I^f_{3}) N^{*}_{i1} t_W)
          +g_X Q'_{f_L} N^{*}_{i6}\right]\,,
    \nonumber\\
&& \langle \tilde{\chi}^0_{iL}|\tilde{f}_R|f_R\rangle
 = + \sqrt{2}\left[ g_2 \, e_f\, t_W N_{i1} + g_X\, Q'_{f_R} N_{i6}\right]\,.
\label{eq:coupling_set2}
\end{eqnarray}
In Eq.$\,$(\ref{eq:coupling_set2}) the coupling to the higgsino component,
which is proportional to the fermion mass, has been neglected.
These would have to be included if one were to study, e.g., the neutralino interaction
with the top quark and squark.\s

For completeness, we also provide the fermion-fermion-$Z_{1,2}$ vertices:
\begin{eqnarray} \label{eq:coupling_set3}
&& \langle f_L|Z_1|f_L\rangle = -g_Z (I^f_3-e_f s^2_W) \cos\theta_{ZZ'}
                                -g_X Q'_{f_L} \sin\theta_{ZZ'}\,,\nonumber\\
&& \langle f_L|Z_2|f_L\rangle = +g_Z (I^f_3-e_f s^2_W) \sin\theta_{ZZ'}
                                -g_X Q'_{f_L} \cos\theta_{ZZ'}\,.
\end{eqnarray}
When switching from $L\to R$ in Eq.$\,$(\ref{eq:coupling_set3}),
the corresponding SU(2)$\times$U(1)
and U(1)$_X$ charges must be changed accordingly.
$I^f_{3}\equiv I^f_{3L}$
is the SU(2) isospin component (note that $I^f_{3R}=0$), $e_f$ is the
electric charge of the fermion $f$ and $Q'_{f_{L,R}}$ are the
effective U(1)$_X$ charges of the left/right-handed fermions.\s

The neutralino production and decay properties in the USSM model
with the additional gaugino and singlino states depend crucially on
their masses with respect to the MSSM neutralino masses. If they are
much heavier than the other states, they
will rarely be produced and so are practically unobservable. In
contrast, if the singlino is lighter than the other states, a
singlino-dominated state will be the lightest supersymmetric particle
(LSP) into which the other
neutralino states will decay, possibly through cascades.\s

In the following subsections, we present a brief description of the general
formalism of neutralino production and the subsequent cascade decays of
the neutralinos. Once charges and mixing matrices are generalized to
the present U(1)$_X$ theory, the phenomenological infrastructure
for cross sections and decay widths can be copied from the MSSM.\s

\subsection{Singlino Production in $e^+e^-$ Annihilation}
\label{sec:sec4.2}

The production processes of a neutralino pair in $e^+e^-$
annihilation,\footnote{Recall that the numbering $\tilde{\chi}^0_i$
[$i=1,\ldots,6$] of the neutralinos [without primed subscripts]
refers to ascending mass ordering.}
\begin{eqnarray}
e^+e^-\rightarrow \tilde{\chi}^0_i\tilde{\chi}^0_j\quad
[i,j=1\mbox{--}6]\,,
\end{eqnarray}
are generated by $s$-channel $Z_1$ and $Z_2$ exchanges, and $t$-
and $u$-channel $\tilde{e}_{L,R}$ exchanges. The transition
amplitudes,
\begin{eqnarray}
T\left(e^+e^-\rightarrow\tilde{\chi}^0_i\tilde{\chi}^0_j\right)
 = \frac{e^2}{s}\, Q_{\alpha\beta}
   \left[\bar{v}(e^+)  \gamma_\mu P_\alpha  u(e^-)\right]
   \left[\bar{u}(\tilde{\chi}^0_i) \gamma^\mu P_\beta
               v(\tilde{\chi}^0_j)\right]\,,
\label{eq:neutralino production amplitude}
\end{eqnarray}
are built up by products of chiral neutralino currents
and chiral fermion currents, coupled by bilinear ``charges'' $Q_{LL}$,
$Q_{LR}$ {\it etc}. The four generalized bilinear charges correspond
to independent helicity amplitudes, describing the neutralino
production processes for polarized electrons/positrons \cite{ckmz}.
They can be parameterized by the fermion and neutralino currents and
the propagators of the exchanged (s)particles as follows: 
\begin{eqnarray}
&& Q_{LL}=+\frac{D_{Z_1}}{s_W^2c_W^2}\, F_{1L} {\cal Z}_{1ij}
          +\frac{D_{Z_2}}{s_W^2c_W^2}\, F_{2L} {\cal Z}_{2ij}
          -\frac{D_{uL}}{s_W^2}\, L_i L^*_j\,,\nonumber\\
&& Q_{LR}=-\frac{D_{Z_1}}{s_W^2c_W^2}\, F_{1L} {\cal Z}^*_{1ij}
          -\frac{D_{Z_2}}{s_W^2c_W^2}\, F_{2L} {\cal Z}^*_{2ij}
          +\frac{D_{tL}}{s_W^2}\, L^*_i L_j\,,\nonumber\\
&& Q_{RL}=+\frac{D_{Z_1}}{s_W^2c_W^2}\, F_{1R} {\cal Z}_{1ij}
          +\frac{D_{Z_2}}{s_W^2c_W^2}\, F_{2R} {\cal Z}_{2ij}
          +\frac{D_{tR}}{s_W^2}\, R_i R^*_j\,,\nonumber\\
&& Q_{RR}=-\frac{D_{Z_1}}{s_W^2c_W^2}\, F_{1R} {\cal Z}^*_{1ij}
          -\frac{D_{Z_2}}{s_W^2c_W^2}\, F_{2R} {\cal Z}^*_{2ij}
          -\frac{D_{uR}}{s_W^2}\, R^*_i R_j\,.
\end{eqnarray}

The first two terms in each
bilinear charge are generated by $Z_1$ and $Z_2$ exchanges and the
third term by selectron exchange; $D_{Z_{1,2}}$, $D_{tL,R}$ and
$D_{uL,R}$ denote the scaled $s$-channel $Z_{1,2}$ propagators and the
$t$- and $u$-channel left/right-type selectron propagators
\begin{eqnarray}
 D_{Z_{1,2}}=\frac{s}{s-m^2_{Z_{1,2}}+im_{Z_{1,2}}\Gamma_{Z_{1,2}}}
  \qquad {\rm and}\qquad
 D_{(t,u)L,R}=\frac{s}{(t,u)-m^2_{\tilde{f}_{L,R}}}\,,
\end{eqnarray}
with $s=(p_{e^-}+p_{e^+})^2$, $t=(p_{e^-}-p_{\tilde{\chi}^0_i})^2$
and $u=(p_{e^-}-p_{\tilde{\chi}^0_j})^2$ denoting the Mandelstam
variables for neutralino pair production in $e^+e^-$ collisions. The
couplings $F_{iL,R}$ of the gauge bosons $Z_i$ ($i=1,2$) to a
fermion pair are given by
\begin{eqnarray}
&& F_{1L} = +\left(I^f_3 - e_f s^2_W \right) c_{ZZ'}
            \!+\!\frac{g_X}{g_Z} Q'_{f_L} s_{ZZ'};\ \qquad
   F_{1R} = - e_f s^2_W\, c_{ZZ'} + \frac{g_X}{g_Z} Q'_{f_R} s_{ZZ'}\,,
            \nonumber\\
&& F_{2L} = -\left(I^f_3 - e_f s^2_W \right) s_{ZZ'}
            \!+\!\frac{g_X}{g_Z} Q'_{f_L} c_{ZZ'};\ \qquad
   F_{2R} = + e_f s^2_W s_{ZZ'} + \frac{g_X}{g_Z} Q'_{f_R} c_{ZZ'}\,,
\end{eqnarray}
where $s_{ZZ'}\equiv\sin\theta_{ZZ'}$,
$c_{ZZ'}\equiv\cos\theta_{ZZ'}$, $I^f_3=-1/2$ and $e_f=-1$ for the electron
charges. Finally, the matrices ${\cal Z}_{1,2ij}$
and the vectors $L_i$ and $R_i$ are defined by ($t_W=\tan\theta_W$)
\begin{eqnarray}
&& {\cal Z}_{1ij}
  =+\frac{1}{2} \left(N_{i3}N^{*}_{j3}-N_{i4}N^{*}_{j4}\right) c_{ZZ'}
  +\frac{g_X}{g_Z} \left(Q'_1 N_{i3} N^{*}_{j3}
                        +Q'_2 N_{i4} N^{*}_{j4}
                        +Q'_S N_{i5} N^{*}_{j5}\right) s_{ZZ'}\,, \nonumber\\
&& {\cal Z}_{2ij}
  =-\frac{1}{2} \left(N_{i3}N^{*}_{j3}-N_{i4}N^{*}_{j4}\right) s_{ZZ'}
  +\frac{g_X}{g_Z} \left(Q'_1 N_{i3} N^{*}_{j3}
                        +Q'_2 N_{i4} N^{*}_{j4}
                        +Q'_S N_{i5} N^{*}_{j5}\right) c_{ZZ'}\,, \nonumber\\
&& L_i = +I^f_3 N_{i2} + (e_f-I^f_3) t_W N_{i1}
         +\frac{g_X}{g_2} Q'_{f_L} N_{i6}\,, \nonumber\\
&& R_i = -e_f t_W N_{i1} + \frac{g_X}{g_2} Q'_{f_R} N_{i6}\,.
\label{eq:Zgg}
\end{eqnarray}
\s\vskip -0.4cm

The $e^+e^-$ annihilation cross sections follow from the squares of
the relevant couplings,
\begin{eqnarray}
\sigma\left[e^+e^-\rightarrow \tilde{\chi}^0_i
\tilde{\chi}^0_j\right]
 &=&{\cal S}_{ij}\, \frac{\pi \alpha^2}{2 s}\lambda_{\rm PS}^{1/2}
  \int^1_{-1}
      \bigg\{\left[ 1-(\mu^2_i-\mu^2_j)^2
           +\lambda_{\rm PS} \cos^2\Theta\right] {\cal Q}_1
     \nonumber\\
  && \hskip 3.cm   + 4\mu_i\mu_j {\cal Q}_2
           +2\lambda_{\rm PS}^{1/2} {\cal Q}_3 \cos\Theta\bigg\}
    \, d\cos\Theta\,,
\end{eqnarray}
where ${\cal S}_{ij}=(1+\delta_{ij})^{-1}$ is a statistical factor
which is equal to 1 for $i\neq j$ and
$1/2$ for $i=j$; $\mu_i= m_{\tilde{\chi}^0_i}/\sqrt{s}$, $\Theta$ is
the polar angle of the produced neutralinos; and $\lambda_{\rm PS}\equiv
\lambda_{\rm PS}(1,\mu^2_i,\mu^2_j)$ denotes the familiar $2$-body
phase space function. The quartic charges ${\cal Q}_i$ ($i=1,2,3$)
are given by products of bilinear charges:
\begin{eqnarray}
&& {\cal Q}_1 = \frac{1}{4}\left[ |Q_{RR}|^2+|Q_{LL}|^2
                           +|Q_{RL}|^2+|Q_{LR}|^2\right]\,,\nonumber\\
&& {\cal Q}_2 =\frac{1}{2} \Re \left[ Q_{RR} Q^*_{RL} + Q_{LL} Q^*_{LR}
                          \right]\,,\nonumber\\
&& {\cal Q}_3 = \frac{1}{4}\left[ |Q_{RR}|^2+|Q_{LL}|^2
                           -|Q_{RL}|^2-|Q_{LR}|^2\right]\,.
\label{eq:quartic_charges}
\end{eqnarray}
The integration over the polar angle $\Theta$ can easily be performed
analytically.\s

\begin{figure}[ht!]
\begin{center}
\includegraphics[height=10.cm,width=11.cm,angle=0]{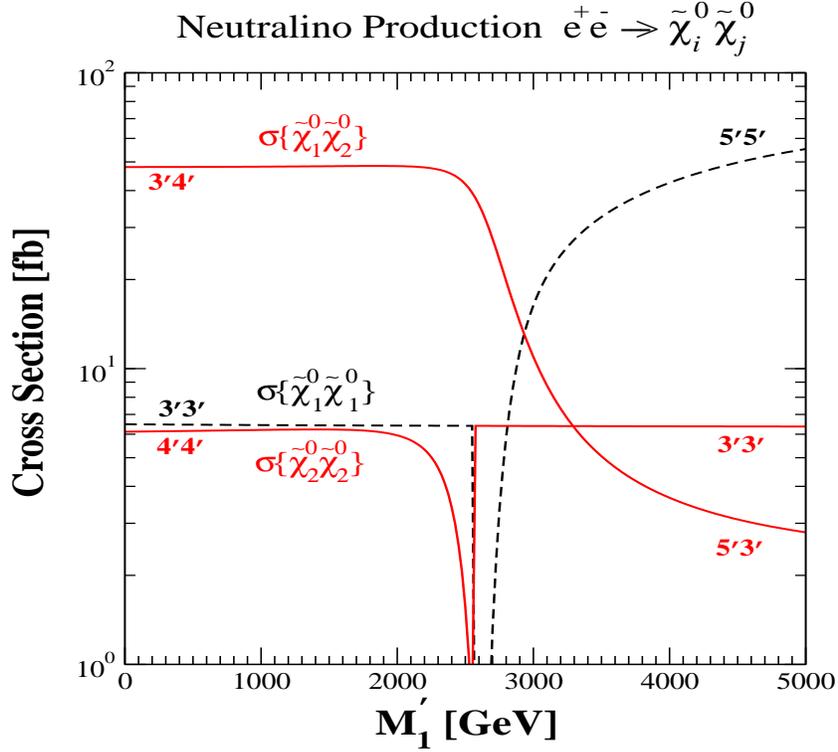}
\end{center}
\vskip -0.5cm \caption{\it The production cross sections
             for $\tilde{\chi}^0_1\tilde{\chi}^0_1$,
             $\tilde{\chi}^0_1\tilde{\chi}^0_2$ and
             $\tilde{\chi}^0_2\tilde{\chi}^0_2$
             neutralino pairs in $e^+e^-$ collisions when varying the U(1)$_X$
             gaugino mass parameter $M'_1$. The R and L selectron masses are
             chosen as
             $m_{\tilde{e}_{R,L}}=701$ GeV in this example.}
\label{fig:fig3}
\end{figure}

The production cross sections for
the three pairings of the two lightest neutralinos, $\{11\}, \{12\}$ and
$\{22\}$, are illustrated in Fig.$\,$\ref{fig:fig3} as functions of $M_1'$.
For the parameter set defined in Sect.$\,$\ref{sec:sec3.3},
the corresponding $Z'$ mass is $M_{Z_2}=949$ GeV and the $ZZ'$ mixing angle is
$\theta_{ZZ'}=3.3\times 10^{-3}$;
these parameters are compatible with existing limits \cite{Zprime_bound,Leike}.
The center-of-mass energy of the $e^+e^-$ collider is set to 800 GeV. Of course,
if $Z_2$ is in the reach of the collider, running on the $Z_2$ resonance would
be the most natural way to explore all the facets of the new particle sector
in an optimal way.\s

For small values of $M'_1$ the cross section
$\sigma\{\tilde{\chi}^0_1\tilde{\chi}^0_2\}$
is of similar size as the MSSM prediction
for the mixed higgsino pairs, $\tilde{\chi}^0_{3'}\tilde{\chi}^0_{4'}$
(cf. Fig.$\,$\ref{fig:fig1}), modified
only by the $Z'$ contribution. However, at and beyond the cross-over points
with the new singlino type neutralino $\tilde{\chi}^0_{5'}$, dramatic
changes set in for pairs involving the lightest neutralino. Since the couplings
of the mixed pair, $\tilde{\chi}^0_{5'}\tilde{\chi}^0_{3'}$, are suppressed to
both the $Z$ and $Z'$ vector bosons, the cross section
$\sigma\{\tilde{\chi}^0_1
\tilde{\chi}^0_2\}$ drops significantly. In contrast,
the rising $\tilde{\chi}^0_{5'}\tilde{\chi}^0_{5'}$ coupling to $Z'$ increases
the cross section of the diagonal pair
$\sigma\{\tilde{\chi}^0_1\tilde{\chi}^0_1\}$
with rising $M'_1$. [The  cross section for the diagonal pair
$\sigma\{\tilde{\chi}^0_2\tilde{\chi}^0_2\}$ does not change as the MSSM
higgsino character is modified only transiently in the $5'$-$4'$
cross-over zone.]\s

\begin{table}[b!]
\caption{\it \label{tab:tab2} Comparison of production cross sections
         between the MSSM and the USSM. The value for $M'_1$
         is set to zero in the USSM.  For other values of $M'_1$ see
         Fig.$\,$\ref{fig:fig3}.}
\begin{center}
\begin{tabular}{|c|c|c|c|}
\hline
 Cross Section [fb]  & $\sigma\{\tilde{\chi}^0_1\tilde{\chi}^0_1\}$
                     & $\sigma\{\tilde{\chi}^0_1\tilde{\chi}^0_2\}$
                     & $\sigma\{\tilde{\chi}^0_2\tilde{\chi}^0_2\}$ \\
\hline\hline
USSM  & $6.5$ & $48.0$  & $6.1$  \\
\hline
MSSM  & $1.7\times 10^{-3}$  & $67.1$  & $8.5\times 10^{-3}$   \\
\hline
\end{tabular}
\end{center}
\end{table}

The presence of the extra gauge boson $Z_2$ with a mass of $\sim1$ TeV
alters the neutralino-pair production cross sections
$\sigma\{\tilde{\chi}^0_1\tilde{\chi}^0_1\}$ and
$\sigma\{\tilde{\chi}^0_2\tilde{\chi}^0_2\}$ in the USSM
significantly compared with the MSSM, as demonstrated in
Table\,\ref{tab:tab2}. The production of light neutralino pairs,
diagonal pairs in particular, are greatly enhanced although the light
neutralino masses are nearly identical in the two models.\s

\subsection{Neutralino cascade decays and sfermion decays}
\label{sec:sec4.3}

If kinematically allowed, the two-body decays of neutralinos
into a neutralino and
an electroweak gauge bosons $Z_{1,2}$ are among the dominant channels.
The widths of the decays, $\tilde{\chi}^0_i \rightarrow \tilde{\chi}^0_j
Z_k$ ($k=1,2$), are given by
\begin{eqnarray}
\Gamma[\tilde{\chi}^0_i\!\rightarrow\!\tilde{\chi}^0_j Z_k]=
\frac{g_Z^2 \lambda_{\rm PS}^{1/2}}{16\pi m_{\tilde{\chi}^0_i}}\!\!
   \left\{\! |{\cal Z}^2_{kij}|\!\!
   \left[\!\frac{(m^2_{\tilde{\chi}^0_i}\!-\!m^2_{\tilde{\chi}^0_j})^2}{
       m^2_{Z_k}}\!+\!m^2_{\tilde{\chi}^0_i}\!+\!m^2_{\tilde{\chi}^0_j}
                    \!-\!2 m^2_{Z_k}\!\right]
         + 6 m_{\tilde{\chi}^0_i} m_{\tilde{\chi}^0_j}
           \real({\cal Z}^2_{kij})\!\!\right\}\,,
\end{eqnarray}
where $\lambda_{\rm PS} \equiv\lambda_{\rm PS}(1,m^2_{\tilde{\chi}^0_j}/
m^2_{\tilde{\chi}^0_i}, m^2_{Z_k}/m^2_{\tilde{\chi}^0_i})$, with ${\cal
Z}_{1ij}$ and ${\cal Z}_{2ij}$ defined in Eq.$\,$(\ref{eq:Zgg}).\s

\begin{figure}[b!]
\begin{center}
\includegraphics[height=10.cm,width=11.cm,angle=0]{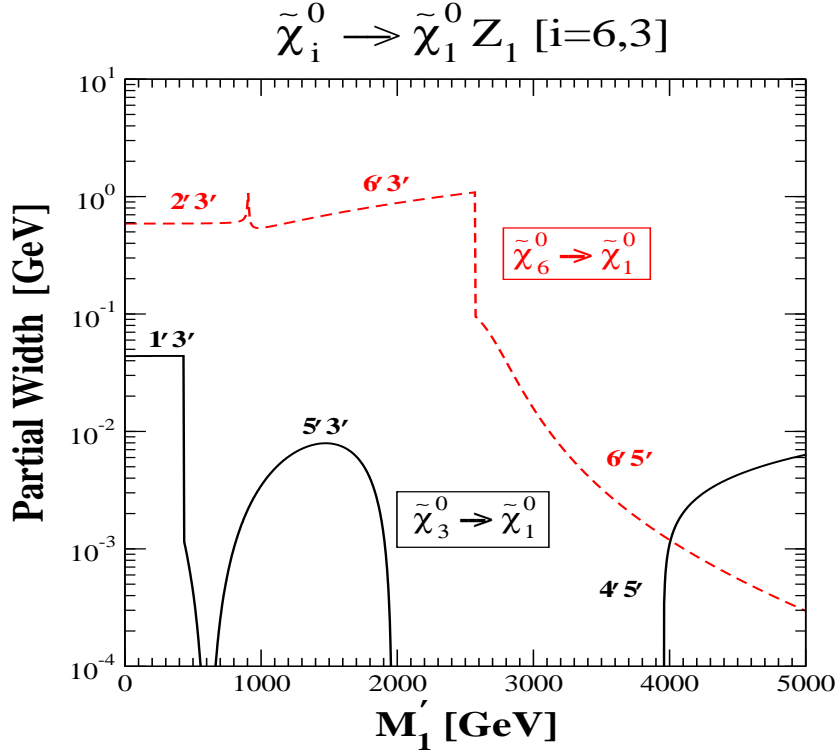}
\end{center}
\caption{\it The evolution of the neutralino decays $\tilde{\chi}^0_i\to
             \tilde{\chi}^0_1 Z_1$ [$i=6,3$] when varying the U(1)$_X$
             gaugino mass parameter $M'_1$. The notation follows the previous
             figures.}
\label{fig:fig4}
\end{figure}

Two examples, $\tilde{\chi}^0_i\to\tilde{\chi}^0_1 Z_1$ for $i=6,3$,
illustrate the evolution of the widths with $M'_1$ in Fig.$\,$\ref{fig:fig4}.
The neutralinos
$\tilde{\chi}^0_6$ and $\tilde{\chi}^0_1$ are identified with the MSSM SU(2)
gaugino and the lighter of the MSSM higgsinos for small $M'_1$, and
with the U(1)$_X$ gaugino and the singlino for large $M'_1$, respectively
[cf. Fig.$\,$\ref{fig:fig1}].
Even after $\tilde{\chi}^0_6$ crosses to the U(1)$_X$ gaugino at the
$2'$-$6'$ cross-over zone, the width increases due to
an increasing phase space factor (due to the increasing mass difference)
and the fact that
$\tilde{\chi}^0_6$ has a significant singlino component. However, once the
state $\tilde{\chi}^0_1$ becomes singlino-dominated above the $5'$-$4'$
cross-over zone, the
width of the decay $\tilde{\chi}^0_6\to\tilde{\chi}^0_1 Z_1$ drops dramatically
as the mixing between the U(1)$_X$ gaugino and the singlino state is strongly
suppressed for large $M'_1$.
The state $\tilde{\chi}^0_3$ is the MSSM U(1) gaugino
for small $M'_1$, the singlino-dominated state for moderate $M'_1$ and the
heavier
of the MSSM higgsinos for large $M'_1$. As the $\tilde{\chi}^0_3$ mass drops,
even only slightly, the two-body
decay $\tilde{\chi}^0_3\to\tilde{\chi}^0_1 Z_1$ is kinematically forbidden
for moderate $M'_1$. However, the mode is kinematically allowed again and its
magnitude increases when the mass of the singlino-dominated $\tilde{\chi}^0_1$
decreases sufficiently. \s

Similarly, the two-body decays of the charginos into
a neutralino and the $W^\pm$ gauge boson
are expected to be among the dominant channels if kinematically allowed.  The
widths of the decays, $\tilde{\chi}^\pm_i\rightarrow \tilde{\chi}^0_j\, W^\pm$,
are given by
\begin{eqnarray}
&& \hskip -0.8cm\Gamma[\tilde{\chi}^\pm_i\rightarrow
\tilde{\chi}^0_j\, W^\pm] = \frac{g^2_2 \lambda_{\rm PS}^{1/2}}{16\pi
m_{\tilde{\chi}^\pm_i}}
   \left\{\frac{|{\cal W}_{Lij}|^2+|{\cal W}_{Rij}|^2}{2}
         \left[\frac{(m^2_{\tilde{\chi}^\pm_i}-m^2_{\tilde{\chi}^0_j})}{
                    m^2_W}+m^2_{\tilde{\chi}^\pm_i}+m^2_{\tilde{\chi}^0_j}
                    -2 m^2_W\right]\right.\nonumber\\
       &&\left.\hskip 5.cm  -6\, m_{\tilde{\chi}^\pm_i} m_{\tilde{\chi}^0_j}
           \,\real({\cal W}_{Lij}{\cal W}^*_{Rij})\right\}\,,
\end{eqnarray}
where $\lambda_{\rm PS}\equiv\lambda_{\rm PS}(1, m^2_{\tilde{\chi}^0_j}/
m^2_{\tilde{\chi}^\pm_i}, m^2_W/m^2_{\tilde{\chi}^\pm_i})$ and the
${\cal W}_{L,R}$ are defined as
\begin{eqnarray}
&& {\cal W}_{Lij}= U^*_{Li1} N_{j2}
                 +\frac{1}{\sqrt{2}} U^*_{Li2} N_{j3},\qquad
   {\cal W}_{Rij}= U^*_{Ri1} N^{*}_{j2}
                 -\frac{1}{\sqrt{2}} U^*_{Ri2} N^{*}_{j4}\,.
\end{eqnarray}
The unitary matrices $U_L$ and $U_R$ diagonalize the chargino mass
matrix via the singular value decomposition \cite{horn}
\mbox{$U_R {\cal M}_C U^\dagger_L ={\rm diag}
\bigl\{m_{\tilde{\chi}^\pm_1}, m_{\tilde{\chi}^\pm_2}\bigr\}$}.
Explicit formulae for the chargino masses and mixing matrices can be
found in Refs.~\cite{haberkane,ref:chargino}.\s

At the LHC, sfermion decays, $\tilde{f}\rightarrow f \tilde{\chi}^0_i$
can produce complex cascades, as heavier neutralinos are often produced
in the initial decay and subsequently decay through a number of steps before
the lightest neutralino (which is presumably the LSP) is
produced to end the chain.  Thus, cascade decays are of
great experimental interest at the LHC. The width of the sfermion 2-body
decay into a fermion and a neutralino follows from
\begin{eqnarray}
\Gamma[\tilde{f}\rightarrow f \tilde{\chi}^0_i]= \frac{g^2_2
\lambda_{\rm PS}^{1/2}}{16\pi m_{\tilde{f}}} \, |g_{\tilde{f}i}|^2
\left(m^2_{\tilde{f}}-m^2_{\tilde{\chi}^0_i}-m^2_f\right)\,,
\end{eqnarray}
where $\lambda_{\rm PS}\equiv\lambda_{\rm PS}(1,m^2_f/m^2_{\tilde{f}},
m^2_{\tilde{\chi}^0_i}/m^2_{\tilde{f}})$, the couplings
$g_{\tilde{f}_L i} =L_i$ and $g_{\tilde{f}_R
i} =R_i$ are defined in terms of the neutralino
mixing matrix $N$ and the appropriate fermion charges
in Eq.$\,$(\ref{eq:Zgg}).\s

The rates for the reverse decays, neutralino decays to sfermions plus fermions,
$\tilde{\chi}^0_i\rightarrow\tilde{f}\bar f$, $\bar{\tilde f} f$ are given by the
corresponding partial widths\footnote{As the decay rates into $\tilde{f}\bar f$ and
$\bar{\tilde f} f$ are the same, we shall henceforth denote \textit{either} of the
final states by $\tilde ff$.}
\begin{eqnarray}
\Gamma[\tilde{\chi}^0_i\rightarrow\tilde{f}f] =\frac{g^2_2
\lambda_{\rm PS}^{1/2} N^f_C}{32\pi\, m_{\tilde{\chi}^0_i}} \,
 |g_{\tilde{f}i}|^2
  \left(m^2_{\tilde{\chi}^0_i}+m^2_f-m^2_{\tilde{f}}\right)\,,
\end{eqnarray}
with the same couplings as before, $\lambda_{\rm PS}\equiv
\lambda_{\rm PS}(1, m^2_{\tilde{f}}/m^2_{\tilde{\chi}^0_i},
m^2_f/m^2_{\tilde{\chi}^0_i})$ and the color factor $N^f_C=1,3$
for leptons and quarks, respectively. [Analogous expressions hold for
chargino decays.]\s

Supersymmetric particles will be analyzed at the LHC primarily in cascade decays
of some initially produced squark or gluino. In the U(1)$_X$ extended model,
the cascade chains may be extended compared with the MSSM by an additional step
due to the presence of two new neutralino states, for example,
\begin{eqnarray*}
\tilde{u}_R\, \to\, u \tilde{\chi}^0_6 \,  \to\, u [Z_1 \tilde{\chi}^0_5]
\, \to\,  u Z_1 [\ell \tilde{\ell}_R]\,  \to\,  u Z_1 \ell [\ell \tilde{\chi}^0_1]\,.
\end{eqnarray*}
At each step in the decay chain, we have placed the decay products from the previous
step within brackets.
The end result of the cascade above is the final state
$uZ_1\ell\ell\tilde\chi^0_1$ with visible particles/jet $u$, $Z_1\simeq Z$, and
two $\ell$'s.\s

\begin{figure}[t!]
\begin{center}
\includegraphics[height=10.cm,width=16.cm,angle=0]{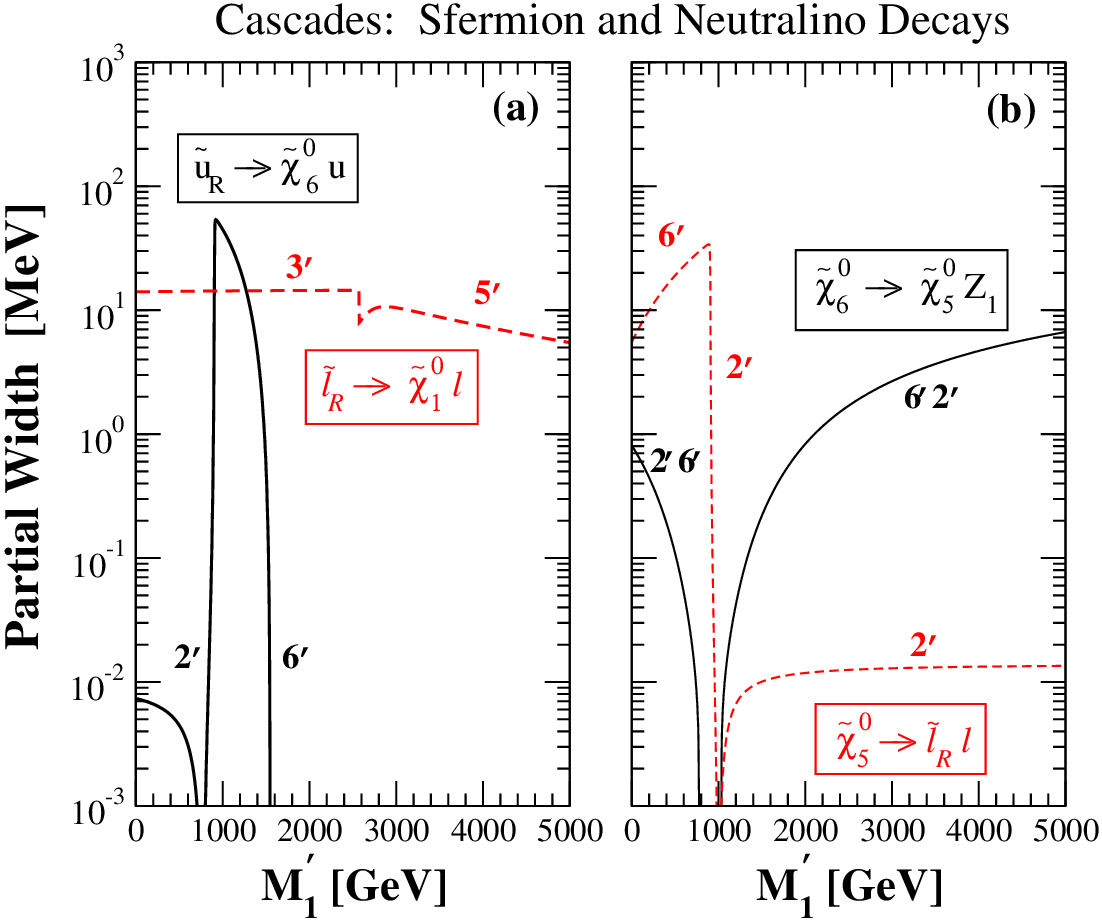}
\end{center}
\caption{\it The evolution of (a) the sfermion decays, $\tilde{u}_R\to
            \tilde{\chi}^0_6u$ and $\tilde{\ell}_R\to
            \tilde{\chi}^0_1 \ell$, and (b) the neutralino decays,
            $\tilde{\chi}^0_6\to\tilde{\chi}^0_5 Z$ and
            $\tilde{\chi}^0_5\to\tilde{\ell}_R\ell$, when varying the
            U(1)$_X$ gaugino mass parameter $M'_1$. The R-type slepton and
            R-type $u$-squark masses are $m_{\tilde{\ell}_R}=701$ GeV and
            $m_{\tilde{u}_R}=2000$ GeV, respectively. }
\label{fig:fig5}
\end{figure}

For the parameter set introduced earlier, the partial widths involved in the cascade
steps are shown for evolving $M'_1$ in Fig.$\,$\ref{fig:fig5}. The sfermion decays
$\tilde{u}_R\to\tilde{\chi}^0_6 u$ and $\tilde{\ell}_R\to\tilde{\chi}^0_1\ell$
are shown in the left panel and the neutralino decays $\tilde{\chi}^0_6\to
\tilde{\chi}^0_5 Z_1$ and $\tilde{\chi}^0_5\to\tilde{\ell}_R\ell$ are shown in
the right panel. The first step $\tilde{u}_R \to \tilde{\chi}^0_6 u$
corresponds to the decay of the R-type
$u$-squark to $ \tilde{\chi}^0_6$, which coincides with
the MSSM SU(2) gaugino for small $M'_1$ and, after the $2'$-$6'$ cross-over
zone, with the U(1)$_X$ gaugino.
The width increases
dramatically before the decay is forbidden kinematically for $M'_1$
larger than $1.5$ TeV. The second step $\tilde{\chi}^0_6 \to
\tilde{\chi}^0_5 Z_1 $ in this cascade chain corresponds to the decay of the
MSSM SU(2) gaugino  for small $M'_1$, changing to the U(1)$_X$ gaugino
decay thereafter. The dependence of this two-body decay mode on $M'_1$ is
mainly of kinematic nature; the decay is not allowed for $M'_1$ between
$\sim 0.8$ TeV and $\sim 1.0$ TeV. Just beyond the $2'$-$6'$ cross-over zone,
it increases very sharply and keeps increasing moderately with $M'_1$
thereafter. The pattern for the
third decay $\tilde{\chi}^0_5\to\tilde{\ell}_R\ell$ is mainly determined by
the size of the U(1)$_Y$ and U(1)$_X$ gauge components of the state
$\tilde{\chi}^0_5$. Before the $2'$-$6'$ cross-over zone the state
is a U(1)$_X$ gaugino so that the width is large. But the width is strongly
suppressed for moderate and large $M'_1$ for which $\tilde{\chi}^0_5$ is a
MSSM SU(2) gaugino with very small mixing with the two MSSM U(1) and U(1)$_X$
gauginos. The width for the final decay $\tilde{\ell}_R\to\tilde{\chi}^0_1
\ell$ remains moderate as the U(1) components of the $\tilde{\chi}^0_1$ state
are small. Beyond the cross-over zone, the width decreases with the
suppressed U(1)$_X$ component.\s

\begin{table}[t!]
\caption{\it \label{tab:tab3} The comparison of decay widths
         between the USSM and the MSSM.
         The state $\tilde{\chi}^0_i$ in the table denotes the
         second heaviest neutralino, i.e. $\tilde{\chi}^0_5$ in the USSM and
         $\tilde{\chi}^0_3$ in the MSSM.  The value of $M'_1$ is set to zero
         in the USSM. For other values of $M'_1$ see
         Fig.$\,$\ref{fig:fig5}. }
\begin{center}
\begin{tabular}{|c|c|c|c|}
\hline
 Decay Width [MeV]  & $\Gamma[\tilde{u}_R\to \tilde{\chi}^0_i u]$
                    & $\Gamma[\tilde{\chi}^0_i\to\tilde{\ell}_R\ell]$
                    & $\Gamma[\tilde{\ell}_R\to\tilde{\chi}^0_1\ell]$ \\
\hline\hline
USSM  & $130.0$ & $5.5$  & $14.1$  \\
\hline
MSSM  & $3294.6$ & $18.9$ & $15.0$   \\
\hline
\end{tabular}
\end{center}
\end{table}

Conventional chains like $\tilde{q}\to q\tilde{\chi}^0_i\to q[\ell\tilde{\ell}]
\to q\ell[\ell\tilde{\chi}^0_1]$ may also be observed in the U(1)$_X$
extended model. However, the partial widths in the USSM can be very different
from the MSSM. As an example, we consider the cascade chains,
in which the intermediate neutralino state $\tilde{\chi}^0_i$
is the second heaviest neutralino, i.e. $\tilde{\chi}^0_5$ in the USSM and
$\tilde{\chi}^0_3$ in the MSSM. As demonstrated in Table\,\ref{tab:tab3}, the
width for the decay of $\tilde{u}_R$ to the second heaviest neutralino in the
USSM is much smaller than in the MSSM.\s

These cascade chains should only be taken as representative theoretical
examples. A systematic phenomenological survey needs significantly more
detailed analyses.\s

\subsection{Decays to Higgs bosons}

The USSM Higgs sector includes two Higgs doublets $H_u$ and $H_d$ as
well as the SM singlet field $S$ \cite{u1ssm3,Barger:2006dh,Demir:2003ke,
Barger:2006rd}.
Their interactions are determined by the gauge interactions and  the
superpotential in Eq.$\,$(\ref{eq:superpotential}). Including soft SUSY
breaking terms and radiative corrections, the resulting effective Higgs
potential
consists of four parts:
\begin{eqnarray} \label{vh}
V_H = V_F + V_D + V_{\rm soft} + \Delta V\,,
\end{eqnarray}
where the $F$, $D$ and soft-breaking terms $V_F, V_D$ and $V_{\rm soft}$
are given by
\begin{eqnarray}
V_F&=& |\lambda|^2|H_u\cdot H_d|^2 + \lambda^2 |S|^2(|H_u|^2+|H_d|^2)\,,
     \nonumber\\
V_D&=& \frac{g_Z^2}{8}\left(|H_d|^2-|H_u|^2\right)^2
      +\frac{g^2_2}{2}\left(|H_u|^2|H_d|^2-|H_u\cdot H_d|^2\right)
      +\frac{g^2_X}{2}\left(Q'_1 |H_d|^2+Q'_2|H_u|^2+Q'_S|S|^2\right)^2\,,
      \nonumber\\
V_{\rm soft} &=& m^2_1 |H_d|^2+m^2_2 |H_u|^2 + m^2_S |S|^2
      +\left( \lambda A_\lambda S\, H_u\cdot H_d + {\rm h.c}\right)\,,
\end{eqnarray}
with $H_u\cdot H_d\equiv H^+_u H^-_d - H^0_u H^0_d$.
The structure of the $F$ term $V_F$ is the same as in the NMSSM
without the self-interaction of the singlet field. However
the $D$ term $V_D$ contains a new ingredient: the terms proportional to $g^2_X$
are  $D$-term contributions due to the extra U(1)$_X$ which are
not present in the MSSM or NMSSM. The soft SUSY breaking terms are
collected in $V_{\rm soft}$.  The tree-level Higgs potential is
CP-conserving \cite{Demir:2003ke}. That is, one can rephase the Higgs fields
to absorb the phases of the potentially complex coefficient $\lambda A_\lambda$.
Thus, without loss of generality, we will assume that these parameters are real.\s

The term $\Delta V$ in Eq.$\,$(\ref{vh}) represents the
radiative corrections to the Higgs effective potential \cite{radcors}.
The dominant contributions at one-loop are generated by top quark and scalar
top quark (stop) loops due
to the large Yukawa couplings; these terms are the same as in the MSSM.
All the other model-dependent contributions do not contribute
significantly at one-loop order \cite{Barger:2006dh}. Therefore, we
will ignore these subdominant model-dependent radiative corrections in the
following analysis.\s

The set of soft SUSY breaking parameters in the tree-level Higgs
potential includes the soft masses $m^2_1, m^2_2$ and $m^2_S$ and the
trilinear coupling $A_\lambda$. Radiative corrections are affected
by many other soft SUSY breaking parameters that generate masses of
scalar tops and their mixings: the SU(2) and U(1)
soft SUSY breaking scalar top masses $m_Q, m_U$, the stop trilinear parameter
$A_t$, the supersymmetric mass scale $M_{\rm SUSY}$ and, spuriously, the
renormalization scale $Q$. To simplify the analysis of the Higgs spectrum it
is useful to express the soft masses $m^2_1, m^2_2, m^2_S$ in terms
of $v_s, v, \tan\beta$ and the other parameters. The tree-level Higgs
masses and couplings depend on four variables only: $\lambda, v_s, \tan\beta$
and $A_\lambda$. In the numerical analysis, we take 1 TeV for the
new parameters, $m_Q, m_U, A_t, Q, M_{\rm SUSY}$ and $A_\lambda$. \s

Decays involving Higgs bosons can be quite different for different
Higgs boson mass spectra. We first decompose the neutral Higgs states into
real and imaginary parts as follows:
\begin{eqnarray}
H_d^0 &=& \frac{1}{\sqrt{2}} \left( v \cos\beta + h \cos\beta
          - H \sin\beta + i A\, \sin\beta \sin\varphi \right)\,,\nonumber\\
H_u^0 &=& \frac{1}{\sqrt{2}} \left( v \sin\beta + h \sin\beta
          + H \cos\beta + i A\, \cos\beta \sin\varphi \right)\,,\nonumber\\
S &=& \frac{1}{\sqrt{2}}\left( v_s + N + i A\, \cos\varphi \right)\,,
\end{eqnarray}
where the CP-odd mixing angle $\varphi$ is determined by
$\tan\varphi=2v_s/v\sin2\beta$ and all the Goldstone states are removed by
adopting the unitary gauge. Subsequently the CP-even states $(h, H, N)$
are rotated onto the mass eigenstates $H_i$ ($i=1,2,3$), labeled in order of
ascending
mass, by applying the orthogonal rotation matrix $O^H$:
\begin{eqnarray}
(H_1, H_2, H_3)_k\, =\, (h,H,N)_a \, O^H_{ak}\,,
\end{eqnarray}
The resulting Higgs mass spectrum consists of three CP-even scalars, one
CP-odd scalar, and two charged Higgs bosons.\s

Generally, the width of a 2-body neutralino or chargino $\tilde
\chi_i$ decay to a neutralino or chargino $\tilde \chi_j$ and a
Higgs boson $\phi_k$ ($H_{1,2,3}$ or $A$) is given by
\begin{eqnarray}
\mbox{ }\hskip -1.0cm \Gamma[{\tilde \chi}_i \to {\tilde \chi}_j \phi_k]
  = \frac{g^2_2\lambda_{\rm PS}^{1/2}}{32 \pi m_{{\tilde \chi}_i}}
    \left\{\left( m_{{\tilde \chi}_i}^2 + m_{{\tilde \chi}_j}^2
           -m_{\phi_k}^2\right)
         \left( |C^L_{ijk}|^2+|C^R_{ijk}|^2 \right)
          +4 m_{{\tilde \chi}_i} m_{{\tilde\chi}_j}
          \real\left( C^L_{ijk} C^{R \, *}_{ijk}\right) \right\}\,,
\label{eq:genwidth}
\end{eqnarray}
where $\lambda_{\rm PS} \equiv \lambda_{\rm PS}(1,m_{{\tilde
\chi}_j}^2/m_{{\tilde \chi}_i}^2, m_{\phi_k}^2/m_{{\tilde
\chi}_i}^2)$ and the left/right couplings $C^{L/R}_{ijk}$ must be
specified in each individual case.\\ \vspace*{-2.5mm}

{\bf (i)} For the decay of a neutralino $\tilde \chi^0_i$ to a
neutralino $\tilde \chi^0_j$ and a scalar Higgs boson $H_k$,
\mbox{${\tilde \chi^0}_i \to {\tilde \chi^0}_j H_k$,} the couplings
are given by,
\begin{eqnarray}
C^R_{ijk}({\tilde \chi^0}_i \to {\tilde \chi^0}_j H_k )
 &=&-\frac{1}{2} \left[ (N_{i2}\!-\!N_{i1} t_W)
   (N_{j3} c_{\beta}\! -\!N_{j4} s_{\beta})
  \!-\!\sqrt{2} \frac{\lambda}{g_2} \left( N_{i3} s_{\beta}
    \! +\! N_{i4} c_{\beta} \right)
    N_{j5} \right. \nonumber\\
   &&\left.\hskip 0.5cm + 2 \frac{g_X}{g_2} N_{i6}\left(Q'_1 N_{j3} c_\beta
     + Q'_2 s_\beta N_{j4}\right)\right]\, O^H_{1k} \nonumber \\
  \!&&+\!\frac{1}{2} \left[(N_{i2}\!-\!N_{i1} t_W)
   (N_{j3} s_{\beta}\! +\! N_{j4} c_{\beta})
  \!+\!\sqrt{2} \frac{\lambda}{g_2}
        \left( N_{i3} c_{\beta}\! -\! N_{i4} s_{\beta} \right)
         N_{j5}\right. \nonumber\\
   &&\left.\hskip 0.5cm + 2 \frac{g_X}{g_2} N_{i6}\left(Q'_1 N_{j3} s_\beta
      - Q'_2 c_\beta N_{j4}\right)\right]\, O^H_{2k}
   \label{eq:sij} \nonumber \\
 \!&&+\!\frac{1}{2} \left[\sqrt{2} \frac{\lambda}{g_2} N_{i3} N_{j4}
     -2 \frac{g_X}{g_2} Q'_S N_{i6} N_{j6} \right]\, O^H_{3k}
     \quad + \quad (i \leftrightarrow j)\,,\\[8pt]
C^L_{ijk}( {\tilde \chi^0}_i \to {\tilde \chi^0}_j H_k )
   &=&C^{R \, *}_{ijk}( {\tilde \chi^0}_i \to {\tilde \chi^0}_j H_k )\,.
\end{eqnarray}
While the first term in each of the two square brackets of
Eq.$\,$(\ref{eq:sij}) are reminiscent of the MSSM couplings
$\tilde{\chi}_i^0 {\tilde \chi}_j^0 h$ and $\tilde{\chi}_i^0 {\tilde
\chi}_j^0 H$ respectively, the other terms are genuinely new in
origin, arising from the extra interaction terms in the USSM
superpotential and the extra U(1)$_X$ gauge interactions. \s

\begin{figure}[t]
\vskip 0.3cm
\begin{center}
\includegraphics[height=10.cm,width=16.cm]{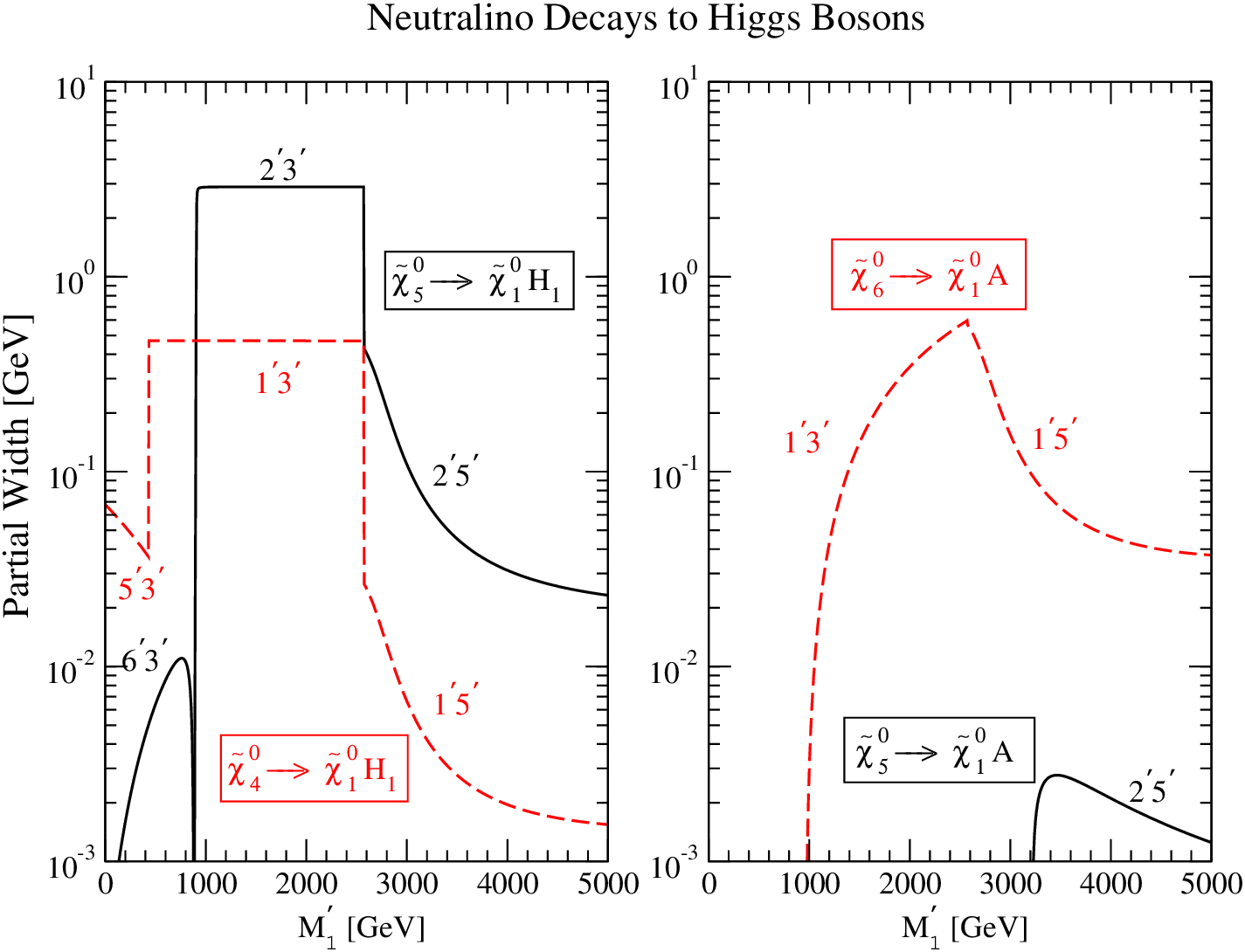} \hspace{0.3cm}
\end{center}
\vskip -0.3cm \caption{\it The decay widths for $\tilde \chi_{5,4}^0 \to
       \tilde\chi_1^0 H_1$ (left) and $\tilde \chi_{5,6}^0 \to
             \tilde \chi_1^0 A_1$ (right) for the parameter set
             given in the text. For the purposes of example, the Higgs mass
             parameter $M_A$ is set to $1.25$ TeV. }
\label{fig:chitochiH}
\end{figure}

The partial widths for the kinematically allowed decays $\tilde \chi_{4,5}^0
\to \tilde \chi_1^0 H_1$ are shown in the left panel of
Fig.$\,$\ref{fig:chitochiH}
as a function of $M'_1$. In the areas in which $\tilde{\chi}^0_{4,5}$
and $\tilde{\chi}^0_1$ nearly coincide with the MSSM neutralinos, the partial
widths do not depend on $M'_1$.\\
\vspace*{-2.5mm}

{\bf (ii)} Similarly, a 2-body neutralino decay to a neutralino and
a CP-odd Higgs boson, ${\tilde \chi}_i^0 \to {\tilde \chi}_j^0
A$, follows Eq.$\,$(\ref{eq:genwidth}) with the left/right couplings
given by
\begin{eqnarray}
C^R_{ij} ( {\tilde \chi^0}_i \to {\tilde \chi^0}_j A )
 \!&=&\! -\frac{i}{2} \left[  (N_{i2}\!-\!N_{i1} t_W)
     (N_{j3} s_{\beta}\! -\! N_{j4} c_{\beta})
    \!+\!\sqrt{2} \frac{\lambda}{g_2}
       \left(N_{i3} c_{\beta}\!+\! N_{i4} s_{\beta} \right)
    N_{j5}\right.\nonumber\\
 &&\left.\hskip 0.5cm +2 \frac{g_X}{g_2} N_{i6} \left(Q'_1 N_{j3} s_\beta
   +Q'_2 N_{j4} c_\beta\right) \right]\, \sin\varphi \nonumber \\
 \!&&\!-\frac{i}{2}
      \left[\sqrt{2} \frac{\lambda}{g_2} N_{i3} N_{j4}
      +2 \frac{g_X}{g_2} Q'_S N_{i6} N_{j5}\right]\, \cos\varphi
     \quad + \quad (i \leftrightarrow j)\,, \label{eq:pij} \\
C^L_{ij}( {\tilde \chi^0}_i \to {\tilde \chi^0}_j A )
 &=& C^{R \, *}_{ij}( {\tilde \chi^0}_i \to {\tilde \chi^0}_j A )\,.
\end{eqnarray}
Again, only the first term in the square brackets is similar to the
MSSM ${\tilde \chi}_i^0 {\tilde \chi}_j^0 A$ coupling. \s

The widths for the kinematically allowed decays
$\tilde{\chi}^0_{5,6}\rightarrow \tilde{\chi}^0_1 A$ are shown in
the right panel of Fig.$\,$\ref{fig:chitochiH} as a function of $M'_1$ in the
benchmark scenario.  In contrast to the scalar case,
only a few decays are kinematically allowed since the CP-odd scalar $A$ is
heavy. \\ \vspace*{-2.5mm}

{\bf (iii)} For completeness, we describe the decays of charginos to
a neutralino and charged Higgs boson ${\tilde \chi}_i^- \to
{\tilde \chi}_j^0 H^-$ ($i=1,2; j=1,2,\ldots,6$). These follow a similar
pattern, but with the last index of the coupling removed:
\begin{eqnarray}
\mbox{ }\hskip -1.cm C^L_{ij}( {\tilde \chi}_i^- \to {\tilde \chi}_j^0 H^-)
 \!&=&\! -  s_\beta\!\! \left[ N^*_{j3} U^*_{Li1}
     \!-\! \frac{1}{\sqrt{2}} \left(N^*_{j2}
     \!+\! N^*_{j1} t_W \right)U^*_{Li2} \right]
     \!-\!\sqrt{2} \frac{g_X}{g_2} Q'_1 N^*_{j6} U^*_{Li2}
     \!-\! \frac{\lambda}{g_2} c_\beta N^*_{j5} U^*_{Li2}\,, \\
\mbox{ }\hskip -1.cm C^R_{ij}( {\tilde \chi}_i^- \to {\tilde \chi}_j^0 H^-)
 \!&=&\! - c_\beta\!\! \left[ N_{j4} U^*_{Ri1}
     \!+\! \frac{1}{\sqrt{2}} \left(N_{j2}
     \!+\! N_{j1} t_W\right) U^*_{Ri2} \right]
     \!-\!\sqrt{2} \frac{g_X}{g_2} Q'_2 N_{j6}U^*_{Ri2}
     \!-\!\frac{\lambda}{g_2} s_\beta N_{i5} U^*_{Ri2}\,,
\label{eq:cij}
\end{eqnarray}
The same left/right couplings determine
the decays of neutralinos to charginos and charged Higgs boson
${\tilde \chi}_j^0 \to {\tilde \chi}_i^+ H^-$ ($i=1,2; j=1,2,\ldots,6$).
For the parameters chosen here, the large mass of the charged Higgs
boson allows kinematically only decays of the
heavier chargino $\tilde{\chi}^\pm_2$ to the lightest neutralino
$\tilde{\chi}^0_1$ and $H^\pm$. \\ \vspace*{-2.5mm}

\begin{figure}[b!]
\vskip 0.3cm
\begin{center}
\includegraphics[height=10.cm,width=16.cm]{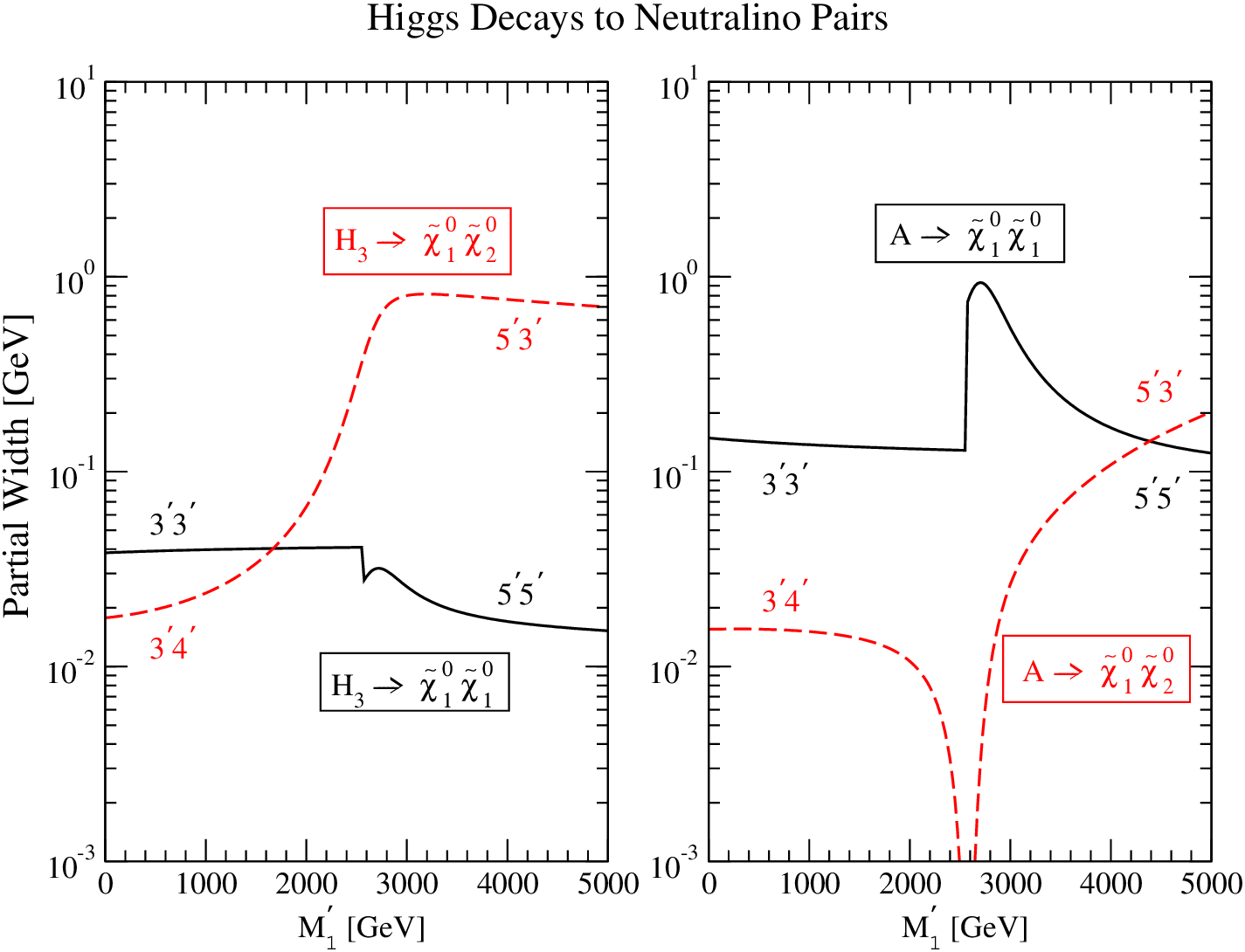} \hspace{0.3cm}
\end{center}
\vskip -0.3cm \caption{\it The decay widths for $H_3\to \tilde \chi_1^0
              \tilde\chi_{1,2}^0$ (left) and $A \to
             \tilde{\chi}^0_1\tilde \chi_{1,2}^0$ (right) for the same
             parameters as in Fig.$\,$\ref{fig:chitochiH}.}
\label{fig:Htochichi}
\end{figure}

{\bf (iv)} It is also possible for Higgs bosons to decay
into neutralino/chargino states, for example the decays $H_i \to {\tilde
\chi}^0_1 {\tilde \chi}^0_j$, $A \to {\tilde \chi}^0_1 {\tilde
\chi}^0_j$ and $H^{\pm} \to {\tilde \chi}^0_1 {\tilde
\chi}^{\pm}_i$. Clearly this is kinematically
possible only for the heavier Higgs states. The general form of the
width for these decays $\phi_i \to {\tilde \chi}_j {\tilde \chi}_k$
($\phi_i = H_i,\,A,\,H^{\pm}$), is given by the crossing of
Eq.$\,$(\ref{eq:genwidth}):
\begin{eqnarray}
\mbox{ }\hskip -1.1cm \Gamma[\phi_i \to {\tilde \chi}_j {\tilde \chi}_k]
  \!=\! {\cal{S}}_{jk} \frac{g^2_2\lambda_{\rm PS}^{1/2}}{16 \pi m_{\phi_i}}
      \left\{\left( m_{\phi_i}^2 - m_{{\tilde \chi}_j}^2
      - m_{{\tilde \chi}_k}^2 \right)
      \left( |C^L_{ijk}|^2+|C^R_{ijk}|^2 \right)
      -4 m_{{\tilde \chi}_j} m_{{\tilde \chi}_k}
     \real\!\!\left( C^L_{ijk} C^{R \, *}_{ijk}\right)
     \right\},
\label{eq:genwidth2}
\end{eqnarray}
where $\lambda_{\rm PS} \equiv \lambda_{\rm PS}(1,m_{{\tilde
\chi}_j}^2/m_{\phi_i}^2, m_{{\tilde \chi}_j}^2/m_{\phi_i}^2)$ and
${\cal{S}}_{jk}=(1+\delta_{jk})^{-1}$ is the usual statistical factor.
The couplings $C^{L/R}_{ijk}$ are related to their neutralino/chargino decay
counterparts in the obvious way:
\begin{eqnarray}
    C^{L/R}_{ijk} ( H_i \to {\tilde \chi^0}_j {\tilde \chi^0}_k)
&=& C^{L/R}_{kji} ( {\tilde \chi^0}_k \to {\tilde \chi^0}_j H_i )\,, \\
    C^{L/R}_{ij} ( A \to {\tilde \chi^0}_i {\tilde \chi^0}_j)
&=& C^{L/R}_{ji} ( {\tilde \chi^0}_j \to {\tilde \chi^0}_i A )\,, \\
    C^{L/R}_{ij} ( H^+ \to {\tilde \chi^0}_i {\tilde \chi^+}_j)
&=& C^{L/R}_{ji} ( {\tilde \chi^-}_j \to {\tilde \chi^0}_i H^- )\,.
\end{eqnarray}
As an example, the Higgs boson decays $H_3, A\to \tilde \chi_1^0 \tilde\chi_{1,2}^0$
are displayed in Fig.$\,$\ref{fig:Htochichi}.
For small $M'_1$ the $\tilde{\chi}_{3'}^0 \tilde{\chi}_{4'}^0 H_3 /A$
couplings in the decays $H_3,A\to\tilde{\chi}^0_1\tilde{\chi}^0_2$
are suppressed while for large $M'_1$
the $\tilde{\chi}_{5'}^0 \tilde{\chi}_{3'}^0 H_3 /A$
couplings are no longer suppressed. The rapid
changes in the $5'$-$4'$ cross-over zone are generated by
interference effects between the Yukawa and the gauge interaction
terms. Similar interference effects, though less significant, occur for the
decays $H_3,A\to \tilde \chi_1^0 \tilde\chi_1^0$ near the cross-over zone. \s

\subsection{Neutralino radiative decays}

In the cross-over zones of the neutralino mass eigenvalues, the gaps between
the neutralino masses become very small. As a result, standard decay
channels are almost shut and photon transitions between neutralino states
\cite{HaWy} become enhanced. These photon transitions are
particularly important in the cross-over zone $4'$-$5'$ at
$M'_1 \simeq 2.6$ TeV [cf. Fig.$\,$\ref{fig:fig1}].
The proximity of the two heavier states to the lightest neutralino
dramatically reduces the rates of all other decay
modes so that the radiative decays
\begin{equation}
\tilde{\chi}^0_2\, ,\; \tilde{\chi}^0_3 \to \tilde{\chi}^0_1 + \gamma
\qquad \mbox{\rm and} \qquad
\tilde{\chi}^0_3  \to \tilde{\chi}^0_2 + \gamma\,,
\label{eq:nrd}
\end{equation}
become non-negligible modes. Of course, also the $\gamma$
transitions are phase-space suppressed in cross-over zones but less strongly
than the competing standard channels due to the vanishing photon mass, even
for 3-particle decays into a lighter neutralino and lepton- or
light-quark pair. \s

The effective couplings $g_{\tilde{\chi}^0_i \tilde{\chi}^0_j \gamma}$ in the
partial decay widths
\begin{equation}
\Gamma[\tilde{\chi}^0_i \to \tilde{\chi}^0_j \gamma] \, =\,
       \frac{g^2_{\tilde{\chi}^0_i \tilde{\chi}^0_j \gamma}}{8\pi}\,
       \frac{(m^2_{\tilde{\chi}^0_i} - m^2_{\tilde{\chi}^0_j})^3}{
              m^5_{\tilde{\chi}^0_i}}\,,
\end{equation}
are of magnetic or electric dipole type depending on the relative CP
quantum numbers of $\tilde\chi^0_i$ and $\tilde\chi^0_j$. The couplings are
generated by triangle graphs of sfermion/fermion, chargino/$W$-boson
and chargino/charged Higgs-boson lines.  The sum of all two-point graphs
associated with the photon line and attached to the neutralino legs by
a $Z$-boson line vanish in the non-linear $R$-gauge
\cite{HaWy}. The $\gamma$ transition amplitudes are finally complex
combinations of mixing matrix elements with reduced triangle functions.
\s

\begin{figure}[t!]
\begin{center}
\vskip 0.2cm
\includegraphics[height=8.5cm,width=16.5cm,angle=0]{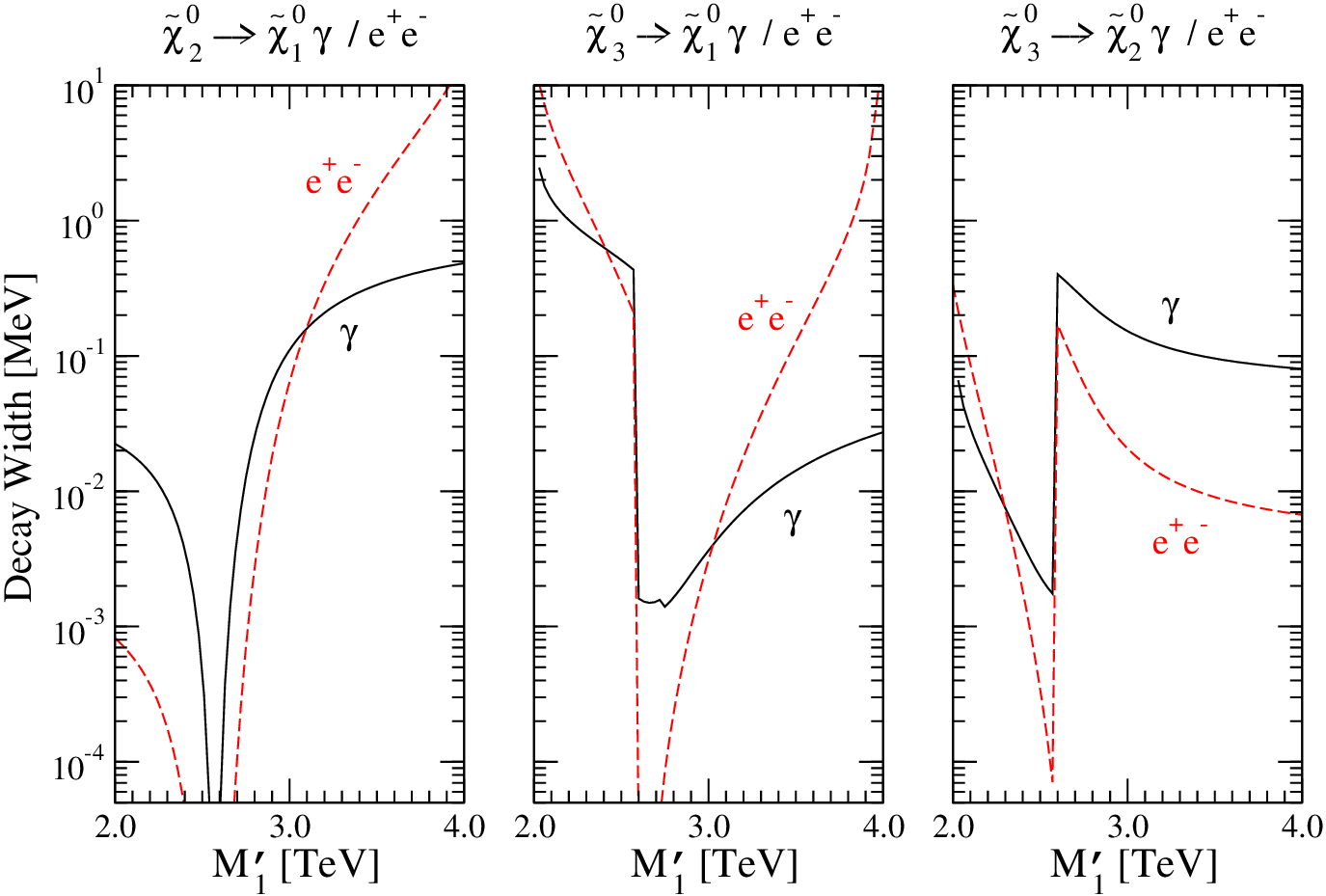}\hskip 1.cm
\end{center}
\vskip -0.5cm
\caption{\it Comparison between photon transitions [full lines]
  and ordinary lepton-pair decays [dashed lines] of neutralinos in the
  $3'$-$4'$-$5'$ cross-over zone of the reference point. The U(1)$_X$
  gaugino mass parameter $M'_1$ is varied between 2 and 4 TeV for comparison
  with the side-band wings; the charged Higgs mass, affecting the photon loop
  couplings is chosen 1.25 TeV and the trilinear parameters are set to 1~TeV. }
\label{fig:fig6}
\end{figure}

For the $\gamma$ transitions of Eq.$\,$(\ref{eq:nrd}), the
partial widths are displayed in Fig.$\,$\ref{fig:fig6} for the set of
parameters
chosen earlier. In this example the three lightest neutralino states are
predominantly of higgsino type so that decays to lepton pairs are allowed
through
couplings mediated by virtual $Z$-bosons which in fact are the dominant modes.
Therefore, for illustrative comparison, the partial widths for standard
electron-pair decays $\tilde{\chi}^0_i \to \tilde{\chi}^0_j \, e^+e^-$ are
also shown. Evidently the branching fractions for radiative decays in general
drastically change in the complex $4'$-$5'$ cross-over zone near
$M'_1 \simeq 2.6$ TeV compared with the side-band wings.\s

\section{Summary and Conclusions}
\label{sec:conclusion}
\setcounter{equation}{0}

In this report we have investigated the neutralino sector in the
U(1)$_X$ extension of the minimal Supersymmetric Standard Model, as
suggested by many GUT and superstring models. The extended model has
attractive features which solve several problems of the
MSSM. It provides a natural solution of the $\mu$-problem without
creating cosmological problems. The upper limit of the light
Higgs mass is somewhat
increased and the growing fine tuning in this sector is reduced. \s

While the MSSM neutralino sector is already quite complex,
the complexity increases dramatically in the extended model due to
two additional degrees of freedom. However, the small coupling
between the original four MSSM states and the two new states offers
an elegant analytical solution of this problem within a perturbative
expansion. We have worked out this solution in detail for the mass
spectrum and the mixing of the states.\s

The expansion in the parameter  $v/M_{\rm SUSY}$, the
ratio of the electroweak scale $v$ over the generic
supersymmetry-breaking scale $M_{\rm SUSY}$,
leads to an excellent approximation of the exact solutions. Even in
the cross-over zones in which two mass eigenvalues are nearly
degenerate, proper adaption of the analytical formalism provides an
accurate description
of the system. Thus, in the limit in which $M_{\rm SUSY}$ is
sufficiently above the electroweak scale, the neutralino
system of the
U(1)$_X$-extended supersymmetric standard model is under good analytical
control and its features are theoretically well understood.
\s

A few examples of mass spectra, widths for cascade decays at LHC, decays
to Higgs bosons, photon
transitions and production cross sections in $e^+e^-$ collisions illustrate
the characteristic features of the model.
\s

\subsection*{Acknowledgments}

We acknowledge useful discussions with D. Jarecka and H.S. Lee.
The work was supported in part by the Korea Research Foundation
Grant (KRF-2006-013-C00097), by KOSEF through CHEP at Kyungpook
National University, in part by U.S. Department of Energy grant number
DE-FG02-04ER41268, by the Deutsche Forschungsgemeinschaft, and by
the Polish Ministry of Science and Higher Education Grant No~1~P03B~108~30,
the EC Grant No~Contract MTKD-CT-2005-029466, and the NATO Grant
PST.\,CLG.\,980066.
S.Y.C. is grateful for the hospitality during a visit to DESY, and P.M.Z.
to the Santa Cruz Institute for Particle Physics (SCIPP)
at the University of California, Santa Cruz.  \s


\setcounter{equation}{0}
\setcounter{section}{1}
\def\thesection{\Alph{section}}
\renewcommand{\theequation}{{\rm \thesection.\arabic{equation}}}

\section*{Appendix A: Takagi diagonalization of a complex symmetric matrix}

In quantum field theory, the most general neutral fermion mass matrix,
$M$, is complex and symmetric.  To identify the physical eigenstates,
$M$ must be diagonalized.\footnote{An alternative method---the standard
diagonalization of the hermitian matrix $M^\dagger M$, which
is commonly advocated in the literature, fails to identify the
physical states in the case of mass-degenerate fermions,
as noted below Eq.~(\ref{app:diagmm}).}
However, the equation that governs
the identification of the physical fermion states
is \textit{not} the standard unitary similarity transformation.  Instead
it is a different diagonalization equation that was discovered by
Takagi \cite{takagi}, and rediscovered many times since
\cite{horn}.\footnote{Subsequently, it was recognized in Ref.~\cite{horn2}
that the Takagi diagonalization was first established for nonsingular
complex symmetric matrices by Autonne \cite{autonne}.}
Despite this illustrious history, the mathematics of the Takagi
diagonalization is relatively unknown among physicists.  Thus, in
this appendix we present a self-contained introduction to the Takagi
diagonalization of a complex symmetric matrix.  After presenting some
background material and a constructive proof of Takagi's result,
we provide, as a pedagogical example, the explicit Takagi
diagonalization of an arbitrary $2\times 2$ matrix.  The latter will be
particularly useful for considering cases in which there is a near-degeneracy
in mass between two of the neutral fermions.

\subsection*{A.1\quad General analysis}

Consider a system of $n$ two component fermion fields $\xi\equiv
(\xi_1\,,\,\xi_2\,,\,\ldots\,,\,\xi_n)^T$, whose physical masses
are governed by the Lagrangian
\begin{eqnarray}
{\cal L}_{\rm mass} = \half\, \xi^T M\, \xi + {\rm h.c.}
\end{eqnarray}
In general, the mass matrix
$M$ is an $n\times n$  complex symmetric matrix.  In order
to identify the physical masses $m_i$ and the corresponding physical
fermion fields $\chi_i$, one introduces a unitary matrix $U$ such that
$\xi=U\chi$ and demands that $\xi^T M\, \xi=\sum_i
m_i\chi_i\chi_i$.  This corresponds to the Takagi diagonalization of
a complex symmetric matrix,\footnote{If $U=N^\dagger$, we
obtain the form of the Takagi diagonalization used in
Eqs.$\,$(\ref{takagifact}) and (\ref{takaginm}).} which is governed by
the following theorem \cite{takagi,horn}:

{\bf Theorem:}
For any complex symmetric $n\times n$ matrix $M$,
there exists a unitary matrix $U$ such that:\footnote{In Ref.~\cite{horn},
Eq.$\,$(\ref{app:takagi}) is
called the Takagi factorization of a complex symmetric matrix.
We choose to refer to this as Takagi diagonalization to emphasize
and contrast this with the more standard diagonalization of
normal matrices by a unitary similarity transformation.  In
particular, not all \textit{complex} symmetric matrices are
diagonalizable by a similarity transformation, whereas complex
symmetric matrices
are \textit{always} Takagi-diagonalizable.}
\begin{eqnarray}
\label{app:takagi}
U^{T} M\, U = M_D = {\rm diag}(m_1,m_2,\ldots,m_n)\,,
\end{eqnarray}
where the $m_k$ are real and non-negative.  \s

The $m_k$ are \textit{not} the eigenvalues of $M$.  Rather, the $m_k$
are the so-called \textit{singular values} of the symmetric matrix
$M$, which are defined to be the non-negative square roots of the
eigenvalues of $M^\dagger M$.  To compute the singular values, note
that:
\begin{eqnarray} \label{app:diagmm}
U^\dagger  M^\dagger M U= M_D^2={\rm diag}(m^2_1,m^2_2,
     \ldots,m^2_n)\,.
\end{eqnarray}
Since $M^\dagger M$ is hermitian, it can be diagonalized by a unitary
similarity transformation.
Although $U$ can be determined from Eq.$\,$(\ref{app:diagmm}) in cases of
non-degenerate singular values, the case of degenerate singular
values is less straightforward.
For example, if $M=
\bigl(\begin{smallmatrix}0\,\, & 1 \\ 1\,\, & 0\end{smallmatrix}\bigr)$,
the singular value 1 is doubly-degenerate, but Eq.$\,$(\ref{app:diagmm})
yields $U^\dagger U= \mathds{1}_{2\times 2}$, which does not specify $U$.
Below, we shall present a constructive method for determining $U$
that is applicable in both the non-degenerate and the degenerate cases.\s

Eq.$\,$(\ref{app:takagi}) can be rewritten as $MU=U^*M_D$, where the
columns of $U$ are orthonormal.  If we denote the $k$th column of
$U$ by $v_k$, then,
\begin{eqnarray} \label{mvks}
Mv_k=m_k v_k^*\,,
\end{eqnarray}
where the $m_k$ are the singular values and the vectors $v_k$
are normalized to have unit norm.  Following Ref.~\cite{takcompute},
the $v_k$ are called the {\it Takagi vectors} of the symmetric complex
$n\times n$ matrix $M$.
The Takagi vectors corresponding to non-degenerate
non-zero [zero] singular
values are unique up to an overall sign [phase].
Any orthogonal [unitary] linear
combination of Takagi vectors corresponding to a set of degenerate
non-zero [zero] singular values is also a Takagi vector
corresponding to the same singular value.  Using these results,
one can determine the degree of non-uniqueness of the matrix $U$.
For definiteness,
we fix an ordering of the diagonal elements of $M_D$.
If the singular values of $M$ are distinct, then the matrix
$U$ is uniquely determined up to multiplication by a diagonal matrix whose
entries are either $\pm 1$.  If there are
degeneracies corresponding to non-zero singular values,
then within the degenerate subspace, $U$
is unique up to multiplication on the right by an arbitrary orthogonal
matrix.  Finally, in the subspace corresponding to zero singular
values, $U$
is unique up to multiplication on the right by an arbitrary unitary
matrix.\s

We shall establish the Takagi diagonalization of a complex symmetric
matrix by formulating an algorithm for constructing $U$.  A
method will be provided for determining the orthonormal Takagi vectors $v_k$
that make up the columns of $U$.
This is achieved by rewriting the $n\times n$ complex
matrix equation Eq.$\,$(\ref{mvks}) [with $m$ real and non-negative]
as a $2n\times 2n$ real matrix equation \cite{dreeschoi}:
\begin{eqnarray} \label{eigprob}
M_S\, \left(\begin{array}{c} \Re v \\ \Im v\end{array}\right)\equiv
\left(\begin{array}{cc} \phm\Re M & \quad -\Im M \\ -\Im M & \quad -\Re M
\end{array}\right)\,\left(\begin{array}{c} \Re v \\ \Im v\end{array}\right) =
m \left(\begin{array}{c} \Re v \\ \Im v\end{array}\right)
\,,\ \ {\rm where}~m\geq 0\,.
\end{eqnarray}
Since $M=M^T$, the $2n\times 2n$ matrix $M\ls{S}$ defined by
Eq.$\,$(\ref{eigprob})
is a real symmetric matrix.  In particular, $M_S$ is diagonalizable by a real
orthogonal similarity transformation, and its eigenvalues are real.
Moreover, if $m$ is an eigenvalue of $M\ls{S}$
with eigenvector $(\Re v\,,\,\Im v)$, then $-m$ is an eigenvalue of
$M\ls{S}$ with (orthogonal) eigenvector $(-\Im v\,,\,\Re v)$.
This observation proves that $M_S$ has an equal number of positive and
negative eigenvalues and an even number of zero
eigenvalues.\footnote{Note that
$(-\Im v\,,\,\Re v)$ corresponds to replacing $v_k$ in Eq.$\,$(\ref{mvks})
by $i v_k$.  However, for $m<0$ these solutions are not relevant for
Takagi diagonalization (where the $m_k$ are by definition non-negative).
The case of $m=0$ is considered in footnote~\ref{fn0}. \label{fn1}}
Thus, Eq.$\,$(\ref{mvks}) has been converted into an ordinary
eigenvalue problem for a real symmetric matrix.  Since $m\geq 0$, we
solve the eigenvalue problem $M\ls{S} u = mu$ for the
eigenvectors corresponding to the non-negative
eigenvalues.\footnote{For $m=0$, the
corresponding vectors $(\Re v\,,\,\Im v)$ and
$(-\Im v\,,\,\Re v)$ are two
linearly independent eigenvectors of $M_S$; but these yield only one
independent Takagi vector $v$ (since $v$ and $iv$ are
linearly dependent).  See footnote \ref{fn1}. \label{fn0}}
It is
straightforward to prove that the total number of linearly independent
Takagi vectors is equal to $n$.  Simply note that the orthogonality of
$(\Re v_1\,,\,\Im v_1)$ and $(-\Im v_1\,,\,\Re v_1)$ with
$(\Re v_2\,,\,\Im v_2)$ implies that $v_1^\dagger v_2=0$.\s

Thus, we have derived a constructive method for obtaining the
Takagi vectors $v_k$.  If there are degeneracies, one can always
choose the $v_k$ in the degenerate subspace to be orthonormal.
The Takagi vectors then make up the
columns of the matrix $U$ in Eq.$\,$(\ref{app:takagi}).  A numerical
package for performing the Takagi diagonalization of a complex symmetric
matrix has recently been presented in Ref.~\cite{hahn}
(see also Refs.~\cite{takcompute,takcompute2} for
previous numerical approaches to Takagi diagonalization).\s

\subsection*{A.2\quad Example: Takagi diagonalization of a $\boldsymbol{2\times 2}$
complex symmetric matrix}

The Takagi diagonalization of a $2\times 2$ complex symmetric matrix can
be performed analytically.\footnote{The main results of this subsection
have been obtained, e.g., in Ref.~\cite{hahn}.  Nevertheless, we
provide some of the details here, which include minor improvements
over the results previously obtained.}  The result is somewhat more
complicated than the standard diagonalization of a $2\times 2$
hermitian matrix by a unitary similarity transformation.
Nevertheless, the corresponding analytic formulae for the Takagi
diagonalization will prove useful in Appendix C in the treatment of
nearly degenerate states.  Consider the complex symmetric matrix:
\begin{eqnarray}
M=\left(\begin{array}{cc} a & c \\ c & b\end{array}\right)\,,
\end{eqnarray}
where $c\neq 0$ and, without loss of generality, $|a|\leq|b|$.
We parameterize the $2\times 2$ unitary matrix $U$
in Eq.$\,$(\ref{app:takagi}) by \cite{murnaghan}:
\begin{eqnarray} \label{vp}
U=VP=
\left(\begin{array}{cc} \cos\theta & e^{i\phi}\sin\theta \\
-e^{-i\phi}\sin\theta & \cos\theta\end{array}\right)\,
\left(\begin{array}{cc} e^{-i\alpha} & 0 \\ 0 &
e^{-i\beta}\end{array}\right)\,,
\end{eqnarray}
where $0\leq\theta\leq\pi/2$ and
$0\leq \alpha\,,\,\beta\,,\,\phi<2\pi$.  However, we may restrict
the angular parameter space further.  Since the normalized Takagi vectors are
unique up to an overall sign if the corresponding singular values are
non-degenerate and non-zero,\footnote{In the case of
a zero singular value or a pair of degenerate of
singular values, there is more freedom in defining the Takagi
vectors as discussed below Eq.$\,$(\ref{mvks}).
These cases will be treated separately at the
end of this subsection. \label{fn}}
one may restrict $\alpha$ and $\beta$ to the range $0\leq\alpha\,,\,\beta<\pi$
without loss of generality.  Finally, we may restrict $\theta$ to the
range $0\leq\theta\leq\pi/4$.  This range corresponds to one of two
possible orderings of the singular values in the diagonal matrix $M_D$.\s

Using the transformation (\ref{vp}), we can rewrite
the Takagi equation (\ref{app:takagi}) as follows:
\begin{eqnarray} \label{vstar}
\left(\begin{array}{cc} a & c \\ c & b\end{array}\right)V=
V^*\left(\begin{array}{cc} \sigma_1& 0 \\ 0 & \sigma_2\end{array}\right)\,,
\end{eqnarray}
where
\begin{eqnarray} \label{msigma}
\sigma_1\equiv m_1\, e^{2i\alpha}\,,\qquad {\rm and}
\qquad  \sigma_2\equiv m_2\, e^{2i\beta}\,,
\end{eqnarray}
with real and non-negative $m_k$.
Multiplying out the matrices in Eq.$\,$(\ref{vstar}) yields:
\begin{eqnarray}
\sigma_1 &=& a- c\, e^{-i\phi}t_\theta
= b\, e^{-2i\phi}-c\, e^{-i\phi}t^{-1}_\theta\,,\label{sig1}\\
\sigma_2 &=& b+ c\, e^{i\phi}t_\theta \,\,\,\,=
a\, e^{2i\phi}+c\, e^{i\phi}t^{-1}_\theta\,,\label{sig2}
\end{eqnarray}
where $t_\theta\equiv \tan\theta$.
Using either Eq.$\,$(\ref{sig1}) or (\ref{sig2}),
one immediately obtains a simple equation for
$\tan 2\theta=2(t_\theta^{-1}-t_\theta)^{-1}$:
\begin{eqnarray} \label{t2th}
\tan 2\theta=\frac{2c}{b\, e^{-i\phi}-a\,e^{i\phi}}\,.
\end{eqnarray}
Since $\tan 2\theta$ is real, it follows that
$bc^*\, e^{-i\phi}-ac^*\, e^{i\phi}$ is real and must
be equal to its complex conjugate.  The resulting equation can be solved
for $e^{2i\phi}$:
\begin{eqnarray}
e^{2i\phi}=\frac{bc^*+a^*c}{b^*c+ac^*}\,,
\end{eqnarray}
or equivalently
\begin{eqnarray}\label{eiphi}
e^{i\phi}=\frac{bc^*+a^*c}{|bc^*+a^*c|}\,.
\end{eqnarray}
The (positive) choice of sign in Eq.$\,$(\ref{eiphi}) follows from
the fact that $\tan 2\theta\geq 0$
(since by assumption, $0\leq\theta\leq\pi/4$), which implies
$0\leq c^*(b\,e^{-i\phi}-a\,e^{i\phi})=|c|^2(|b|^2-|a|^2)$
after inserting the results of Eq.$\,$(\ref{eiphi}).  Since $|b|\geq|a|$
by assumption, the asserted inequality holds as required.\s

Inserting the result for $e^{i\phi}$ back into Eq.$\,$(\ref{t2th}) yields:
\begin{eqnarray}
\tan 2\theta=\frac{2|bc^*+a^*c|}{|b|^2-|a|^2}\,.
\end{eqnarray}
One can compute $\tan\theta$ in terms of $\tan 2\theta$ for
$0\leq\theta\leq\pi/4$:
\begin{eqnarray} \label{tantheta}
\tan\theta&=&\frac{1}{\tan2\theta}\left[\sqrt{1+\tan^2 2\theta}-1
\right]\nonumber \\[6pt]
&=&\frac{|a|^2-|b|^2+\sqrt{(|b|^2-|a|^2)^2+
4|bc^*+a^*c|^2}}{2|bc^*+a^*c|}\,,\\[6pt]
&=& \frac{2|bc^*+a^*c|}{|b|^2-|a|^2+\sqrt{(|b|^2-|a|^2)^2+
4|bc^*+a^*c|^2}}\,.\label{tanth3}
\end{eqnarray}
Starting from Eqs.$\,$(\ref{sig1}) and (\ref{sig2}), it
is now straightforward, using Eqs.$\,$(\ref{eiphi}) and
(\ref{tantheta}), to compute the squared magnitudes of $\sigma_k$:
\begin{eqnarray} \label{mk2}
m_k^2=|\sigma_k|^2=\frac{1}{2}\left[|a|^2+|b|^2+2|c|^2\mp
\sqrt{(|b|^2-|a|^2)^2+4|bc^*+a^*c|^2}\right]\,,
\end{eqnarray}
with $|\sigma_1|\leq |\sigma_2|$.  This ordering of the $|\sigma_k|$
is governed by the convention that $0\leq\theta\leq\pi/4$ (the opposite
ordering would occur for $\pi/4\leq\theta\leq\pi/2$).  Indeed, one can
check explicitly that the $|\sigma_k|^2$ are the eigenvalues of
$M^\dagger M$, which provides the more direct way of computing the
singular values.\s

The final step of the computation is the determination of the angles
$\alpha$ and $\beta$ from Eq.$\,$(\ref{msigma}).
Inserting Eqs.$\,$(\ref{eiphi}) and (\ref{tanth3}) into
Eqs.$\,$(\ref{sig1}) and (\ref{sig2}), we end up with:
\begin{eqnarray}
\alpha &=& \nicefrac{1}{2}\arg\bigl\{a(|b|^2-|\sigma_1|^2)-b^*c^2\bigr\}
\label{alphadef}\,,\\
\beta &=& \nicefrac{1}{2}\arg\bigl\{b(|\sigma_2|^2-|a|^2)+a^*c^2\bigr\}\,.
\label{betadef}
\end{eqnarray}

If ${\rm det}~M=ab-c^2=0$ (with $M\neq {\bf 0}$) , then there is one singular
value which is equal to zero.  In this case, it is easy to verify that
$\sigma_1=0$ and $|\sigma_2|^2={\rm Tr}~(M^\dagger
M)=|a|^2+|b|^2+2|c|^2$.  All the results obtained above remain valid,
except that $\alpha$ is undefined [since in this case, the argument of
$\arg$ in Eq.$\,$(\ref{alphadef}) vanishes].  This corresponds to
the fact that for a zero singular value, the corresponding
(normalized) Takagi vector is only unique up to an overall arbitrary
phase [cf. footnote~\ref{fn}].\s

We provide one illuminating example of the above results.  Consider
the complex symmetric matrix:
\begin{eqnarray} \label{mex}
M=\left(\begin{array}{cc} 1 &  \phm i\\ i &  -1
\end{array}\right)\,.
\end{eqnarray}
The eigenvalues of $M$ are degenerate and equal to zero.  However,
there is only one linearly independent eigenvector, which is proportional
to $(1\,,\,i)$.  Thus, $M$ cannot be diagonalized by a similarity
transformation~\cite{horn}.
In contrast, all complex symmetric matrices are Takagi-diagonalizable.
The singular values of $M$ are 0 and 2 (since these
are the non-negative square roots of the eigenvalues of $M^\dagger M$),
which are \textit{not} degenerate.  Thus, all the formulae derived above
apply in this case.  One quickly determines that $\theta=\pi/4$,
$\phi=\pi/2$, $\beta=\pi/2$ and $\alpha$ is indeterminate (so one is
free to choose $\alpha=0$).  The resulting Takagi diagonalization is
$U^TMU={\rm diag}(0\,,\,2)$ with:
\begin{eqnarray}
U=\frac{1}{\sqrt{2}}\left(\begin{array}{cc} 1 & \phm i\\
      i &  \phm 1 \end{array}\right)\,\left(\begin{array}{cc} 1 &
      \phm 0\\ 0 &  -i \end{array}\right)
= \frac{1}{\sqrt{2}}\left(\begin{array}{cc} 1 & \phm 1\\
     i  &  -i \end{array}\right)\,.
\end{eqnarray}
This example clearly indicates the distinction between the
(absolute values of the) eigenvalues of $M$ and its singular values.
It also exhibits the fact that one cannot always perform a Takagi
diagonalization by using the standard
techniques for computing eigenvalues and eigenvectors.\footnote{For
\textit{real} symmetric matrices $M$, one
can always find a real orthogonal $V$ such that $V^TMV$ is diagonal.  In
this case the Takagi diagonalization is achieved by
$U=VP$, where $P$ is a diagonal matrix whose $kk$ element
is $1$ [$i$] if the corresponding \textit{eigenvalue} $m_k$ is
positive (negative).  Of course, this procedure fails for
complex symmetric matrices [such as $M$ in Eq.$\,$(\ref{mex})]
that are not diagonalizable.}\s

We end this subsection by treating the
case of degenerate (non-zero) singular values, which arises when
$bc^*=-a^*c$.  Special considerations are required since not all
the formulae derived above are applicable to this case
[cf. footnote~\ref{fn}].  The condition $bc^*=-a^*c$ implies that
$|a|=|b|$, so that $|\sigma_1|^2=|\sigma_2|^2=|b|^2+|c|^2$.  After
noting that $a/c=-b^*/c^*$, Eq.$\,$(\ref{t2th}) then yields:
\begin{eqnarray} \label{singtan}
\tan 2\theta=\left[\Re(b/c)\,c_\phi+\Im(b/c)\,s_\phi\right]^{-1}\,,
\end{eqnarray}
where $c_\phi\equiv\cos\phi$ and $s_\phi\equiv\sin\phi$.
The reality of $\tan 2\theta$
imposes no constraint on $\phi$; hence, $\phi$ is indeterminate
[a fact that is suggested by Eq.$\,$(\ref{eiphi})].
The same conclusion also follows immediately from Eq.$\,$({\ref{app:takagi}).
Namely, if $M_D=m\mathds{1}_{2\times 2}$, then $(U{\cal O})^TM(U{\cal O})
={\cal O}^TM_D{\cal O}=M_D$ for any real orthogonal matrix ${\cal O}$.
In particular, $\phi$ simply represents the freedom to choose
${\cal O}$ [see, e.g., Eq.$\,$(\ref{uspecial})].  Since $\phi$ is
indeterminate, Eq.$\,$(\ref{singtan}) implies that $\theta$ is
indeterminate as well.  In practice, it is often simplest to choose
a convenient value, say $\phi=0$, which would then fix
$\theta$ such that $\tan 2\theta=[\Re(b/c)]^{-1}$.
For pedagogical reasons, we shall keep $\phi$ as a free parameter below.\s

Naively, it appears that $\alpha$ and $\beta$ are also indeterminates.
After all, the arguments of $\arg$ in both Eqs.$\,$(\ref{alphadef}) and
(\ref{betadef}) vanish in the degenerate limit.  However, this is not
a correct conclusion, as the derivation of Eqs.$\,$(\ref{alphadef}) and
(\ref{betadef})
involve a division by $|bc^*+a^*c|$, which vanishes in the degenerate
limit.  Thus, to determine $\alpha$ and $\beta$ in the degenerate
case, one must return to Eqs.$\,$(\ref{sig1}) and (\ref{sig2}).}  A
straightforward calculation [which uses Eq.$\,$(\ref{singtan})] yields:
\begin{eqnarray} \label{sig12}
\frac{\sigma_2}{c}=-\frac{\sigma_1^*}{c^*}\,,
\end{eqnarray}
which implies
\begin{eqnarray}
\alpha+\beta=\arg c\pm\frac{\pi}{2} \,.
\end{eqnarray}
Note that separately, $\alpha$ and $\beta$ depend on the choice of $\phi$
(although $\phi$ drops out in the sum).  Explicitly,
\begin{eqnarray}
&&\hspace{-0.5in}
\sigma_1=-c\,e^{-i\phi}\left\{\sqrt{1+\bigl[c_\phi\Re\left(b/c\right)
+s_\phi\Im\left(b/c\right)\bigr]^2}
+i\bigl[s_\phi\Re\left(b/c\right)-c_\phi\Im\left(b/c\right)
\bigr]\right\}\,,\\[8pt]
&&\hspace{-0.5in} \sigma_2=\,\,\,\phm
c\,e^{i\phi}\left\{\sqrt{1+\bigl[c_\phi\Re\left(b/c\right)
+s_\phi\Im\left(b/c\right)\bigr]^2}
-i\bigl[s_\phi\Re\left(b/c\right)-c_\phi\Im\left(b/c\right)
\bigr]\right\}\,.
\end{eqnarray}
One easily verifies that Eq.$\,$(\ref{sig12}) is satisfied.  Moreover,
using Eq.$\,$(\ref{msigma}), $\alpha$
and $\beta$ are now separately determined.\s

We illustrate the above results with the classic case of
$M=\bigl(\begin{smallmatrix}0 & 1 \\ 1 & 0\end{smallmatrix}\bigr)$.
In this case
$M^\dagger M=\mathds{1}_{2\times 2}$, so $U$ cannot be deduced by
diagonalizing $M^\dagger M$.  Setting $a=b=0$ and $c=1$ in the above
formulae, it follows that $\theta=\pi/4$,
$\sigma_1=-e^{-i\phi}$ and $\sigma_2=e^{i\phi}$, which yields
$\alpha=-(\phi\pm\pi)/2$ and $\beta=\phi/2$.  Thus, Eq.$\,$(\ref{vp}) yields:
\begin{eqnarray} \label{uspecial}
U&=&\frac{1}{\sqrt{2}}\left(\begin{array}{cc} \phm 1 & e^{i\phi} \\
-e^{-i\phi} & 1
\end{array}\right)\,  \left(\begin{array}{cc}\pm  ie^{i\phi/2} & 0  \\
  0 &  e^{-i\phi/2} \end{array}\right)  =
\frac{1}{\sqrt{2}}\left(\begin{array}{cc} \pm ie^{i\phi/2} & \,\,
e^{i\phi/2} \\ \mp i e^{-i\phi/2} & \,\,e^{-i\phi/2}
\end{array}\right)\nonumber \\[10pt]
&=&\frac{1}{\sqrt{2}}\left(\begin{array}{cc} \phm i& \quad 1 \\ -i & \quad 1
\end{array}\right)\,\left(\begin{array}{cc}
\pm \cos(\phi/2) & \,\, \sin(\phi/2) \\
 \mp\sin(\phi/2) & \,\, \cos(\phi/2)\end{array}\right)\,,
\end{eqnarray}
which illustrates explicitly that in the degenerate case,
$U$ is unique only up to multiplication on
the right by an arbitrary orthogonal matrix.\s

\section*{Appendix B: The small-mixing approximation}

\setcounter{equation}{0}
\setcounter{section}{2}

The $6\times 6$ USSM neutralino mass matrix of Eq.$\,$(\ref{eq:mass_matrix})
cannot be diagonalized analytically in general.
However, simple analytical expressions for masses and mixing parameters can
be found, similarly as in the NMSSM,
by making use of approximations based on the natural assumption of small
doublet-singlet higgsino, doublet higgsino-U(1)$_X$ gaugino mixing and
kinetic gaugino mixing, i.e. for a large SUSY scale compared to the
electroweak scale.\s

In this appendix, we provide details of the neutralino mass matrix
diagonalization in the small mixing approximation, in which the weak
coupling between two off-diagonal matrix blocks can be
perturbatively treated. For mathematical clarity, we
present the solution for a general complex $(N+M)\times (N+M)$
symmetric matrix in
which the $N\times N$ and $M\times M$ submatrices are coupled weakly
so that their mixing is small:
\begin{eqnarray} \label{matrixnm}
{\mathcal M}_{N+M} =
  \left(\begin{array}{cc}
  {\cal M}_N   &   X_{NM} \\[2mm]
   X^T_{NM}    & {\cal M}_M
        \end{array}\right)
\end{eqnarray}
To obtain the corresponding physical neutralino masses, one must
perform a Takagi diagonalization of
${\cal M}_{N+M}$:\footnote{In Eq.~(\ref{takaginm}), we use
primed subscripts to indicate that the corresponding neutralino states
are continuously connected to the states of the unperturbed
block matrix, ${\rm diag}(\overline{\mathcal{M}}^D_N\,,\,
\overline{\mathcal{M}}^D_M)$, where the diagonal matrices
$\overline{\mathcal{M}}^D_N$ and $\overline{\mathcal{M}}^D_M$ are defined
in Eqs.~(\ref{dm1}) and (\ref{dm2}).
\label{fn3}}
\begin{eqnarray} \label{takaginm}
(N^{N+M})^{*}\, {\cal M}_{N+M}\, (N^{N+M})^{\dagger} = {\rm
  diag}(m_{1'}\,,\,m_{2'}\,,\,\ldots\,,\, m_{N'+M'})\,,\qquad m_{k'}\geq 0\,,
\end{eqnarray}
where $N^{N+M}$ is a unitary matrix.\footnote{When $N$ and $M$ are used in
subscripts and superscripts of matrices, they refer to the dimension of the
corresponding square matrices.  For rectangular matrices, two
subscripts will be used.}
The Takagi diagonalization of a general complex symmetric matrix is
described in Appendix~A.  The non-negative $m_{k'}$ are called the
singular values of $M$, which are defined as the non-negative square roots
of the eigenvalues of $M^\dagger M$.\s

In Eq.$\,$(\ref{matrixnm}),
${\cal M}_N$ and ${\cal M}_M$ are $N\times N$ and $M\times M$
complex symmetric submatrices
with singular values generally of the SUSY scale, $M_{\rm SUSY}$.
$X_{NM}$ is a rectangular
$N\times M$ matrix whose matrix elements are generally of the
electroweak scale.  Assuming that the electroweak scale is significantly
smaller than $M_{\rm SUSY}$, one can treat $X_{NM}$ as a perturbation
as long as there are no accidental near-degeneracies between
the singular values of ${\cal M}_N$ and ${\cal M}_M$, respectively.
(The case of such a near-degeneracy is the subject of Appendix C.)
The diagonalization of ${\cal M}_{N+M}$  can be performed using the
following steps.\s

\noindent {\bf [1]} In the first step, we separately perform
a Takagi diagonalization of ${\cal M}_N$ and
${\cal M}_M$:
\begin{eqnarray}
\overline{\cal M}^{D}_N &=& N^{N\,*} {\cal M}_N N^{N\,\dagger}
= {\rm diag} (\overline{m}_{1'}, \ldots\, \overline{m}_{N'})\,, \label{dm1}\\
\overline{\cal M}^{D}_M &=& N^{M\,*} {\cal M}_M N^{M\,\dagger}
= {\rm diag} (\overline{m}_{N'+1'}, \dots, \overline{m}_{N'+M'})\,,\label{dm2}
\end{eqnarray}
where the $\overline m_{k'}$ are real and non-negative.
The ordering of the diagonal elements above\footnote{See footnote \ref{fn3}.}
is chosen according to some convenient criterion
(e.g., see the discussion at the
end of Sect.$\,$\ref{sec:sec2}.)
Analytical expressions can be obtained for the
singular values and the Takagi vectors that comprise
the columns of the corresponding unitary matrices $N^N$ and $N^M$
for values of $N$, $M\leq 4$ \cite{ckmz}. \s

Step {\bf [1]} results in a partial Takagi diagonalization of ${\cal M}_{N+M}$:
\begin{eqnarray}
&& \overline{\cal M}_{N+M}\equiv
\left(\begin{array}{cc} N^{N\,*} & \mathds{O} \\ \mathds{O}^T &  N^{M\,*}
\end{array}\right)\,
\left(\begin{array}{cc}
      {\cal M}_N  & X_{NM}  \\
      X_{NM}^T       & {\cal M}_M
                 \end{array}\right)\,
\left(\begin{array}{cc} N^{N\,\dagger} & \mathds{O} \\ \mathds{O}^T &
  N^{M\,\dagger}\end{array}\right)
\nonumber \\[12pt]
&&\qquad\quad\,\,
=
\left(\begin{array}{cc}
\overline{\cal M}^{D}_{N} & N^{N\,*}X_{NM}N^{M\,\dagger}
\\ N^{M\,*}X_{NM}^T N^{N\,\dagger} & \overline{\cal M}^{D}_M \end{array}\right)
\,.
\label{decal}
\end{eqnarray}
where $\mathds{O}$ is an $N\times M$ matrix of zeros.  The upper left and
lower right blocks of $\overline{\cal M}_{N+M}$ are diagonal with real
non-negative entries, but the upper right and lower left off-diagonal
blocks are non-zero.\s

\noindent {\bf [2]} The ensuing $(N+M)\times (N+M)$ matrix, $\overline{\cal
M}_{N+M}$, can be subsequently block-diagonalized by performing an
$(N+M)\times (N+M)$ Takagi diagonalization of $\overline{\cal M}_{N+M}$.  Since
the elements of the off-diagonal blocks of $\overline{\cal M}_{N+M}$ are small
compared to the diagonal elements $\overline m_{k'}$, we may treat
$X_{NM}$ as a perturbation.  More precisely, $X_{NM}$ can be treated
as a perturbation if:
\begin{eqnarray} \label{pertcond}
\left|\frac{\Re(N^{N\,*}X_{NM}N^{M\,\dagger})_{i'j'}}
{\overline m_{i'}-\overline m_{j'}}\right|\ll 1\,,
\end{eqnarray}
for all choices of $i'=1',\ldots,N'$ and $j'=N'+1'\,\ldots, N'+M'$.
This condition will be an output of our computation below.\s

The perturbative block-diagonalization is accomplished by introducing an
$(N+M)\times (N+M)$ unitary matrix:
\begin{eqnarray}
\overline{\cal N}_B \simeq
\left(\begin{array}{cc}
    \mathds{1}_{N\times N}
    - \frac{1}{2} \Omega \Omega^\dagger
  &  \Omega \\[1mm]
    -\Omega^\dagger & \mathds{1}_{M\times M}
   -\frac{1}{2} \Omega^\dagger \Omega
      \end{array} \right)\,,
\end{eqnarray}
where $\Omega$ is an $N\times M$ complex matrix that vanishes when
$X_{NM}$ vanishes (and hence like $X_{NM}$ is perturbatively small).
Note that $\overline{\cal N}_B\overline{\cal N}^\dagger_B
=\mathds{1}_{(N+M)\times (N+M)}+{\cal O}(\Omega^4)$
which is sufficiently close to the identity matrix for
our purposes.  Straightforward matrix multiplication then yields:
\begin{eqnarray} \label{nbmnb}
&& \overline{\cal N}^*_B
  \left(\begin{array}{cc} \overline{\cal M}_N^D & {\cal B} \\
          {\cal B}^T & \overline{\cal M}_M^D\end{array}\right)
  \overline{\cal N}^\dagger_B
=
 \left(\begin{array}{cc}{\cal M}_N^{\prime\,D} +{\cal O}({\cal B}\Omega^3) &
 {\cal B}
+\Omega^*\overline{\cal M}_M^D -\overline{\cal M}_N^D\Omega
+{\cal O}({\cal B}\Omega^2) \\ {\cal B}^T+\overline{\cal M}_M^D\Omega^\dagger
-\Omega^T\,\overline{\cal M}_N^D +{\cal O}({\cal B}\Omega^2) &
 {\cal M}_M^{\prime\,D} +{\cal O}({\cal B}\Omega^3)\end{array}\right)\!,
 \nonumber \\[10pt] &&\phantom{line}
\end{eqnarray}
where
\begin{eqnarray} \label{calbdef}
{\cal B}&\equiv & N^{N\,*}X_{NM}N^{M\,\dagger}\,,\\
{\cal M}_N^{\prime\,D} &\equiv &
  \overline{\cal M}_N^D+ \left[\Omega^* {\cal B}^T
 +\half \Omega^* \overline{\cal M}_{M}^D\Omega^\dagger
 -\half \overline{\cal M}_N^D \Omega\Omega^\dagger
 +{\rm transp}\right]\,, \label{mnd}\\
 {\cal M}_M^{\prime\,D} &\equiv &
  \overline{\cal M}_M^D
 -\left[ \Omega^T {\cal B}
 \,-\half \Omega^T\, \overline{\cal M}_{N}^D\Omega
 \,+\half \overline{\cal M}_M^D \Omega^\dagger\Omega
 +{\rm transp}\right]\,,\label{mmd}
 \end{eqnarray}
and ``transp'' is an instruction to take the transpose of the
preceding terms inside the bracket.
For a consistent perturbative expansion, we may neglect all terms above
that are hidden inside the various order symbols in Eq.~(\ref{nbmnb}).
Hence, a successful block-diagonalization is achieved by demanding that
\begin{eqnarray} \label{calbeq}
 {\cal B}=\overline{\cal M}_N^D\Omega-\Omega^*\overline{\cal M}_M^D\,.
\end{eqnarray}
Inserting this result in Eqs.$\,$(\ref{mnd}) and (\ref{mmd}) and
eliminating ${\cal B}$, we obtain:
\begin{eqnarray}
{\cal M}_N^{\prime\,D} &= &
  \overline{\cal M}_N^D
  -\half\left[\Omega^* \overline{\cal M}_{M}^D\Omega^\dagger
        - \overline{\cal M}_N^D \Omega\Omega^\dagger
        +{\rm transp}\right]\,, \label{mnd2}\\[2mm]
 {\cal M}_M^{\prime\,D} &= &
 \overline{\cal M}_M^D
 -\half\left[\Omega^T \,\overline{\cal M}_{N}^D\Omega
        - \overline{\cal M}_M^D \Omega^\dagger\Omega
        +{\rm transp}\right]\,.\label{mmd2}
 \end{eqnarray}
%

The results above simplify somewhat when we recall that $\overline{\cal M}_N^D$
and $\overline{\cal M}_M^D$ are diagonal matrices [see Eq.$\,$(\ref{dm1}) and
(\ref{dm2})].  Taking the real and imaginary parts of the matrix elements of
Eq.$\,$(\ref{calbeq}) yields two equations for the real and imaginary parts
of $\Omega_{ij}$:
\begin{eqnarray} \label{omegab}
\Re\Omega_{i'j'}\equiv \frac{\Re{\cal B}_{i'j'}}
{\overline m_{i'}-\overline m_{j'}}\,,
\qquad\qquad
\Im\Omega_{i'j'}\equiv \frac{\Im{\cal B}_{i'j'}}
{\overline m_{i'}+\overline m_{j'}}\,,
\end{eqnarray}
with $i'=1',\ldots,N'$ and $j'=N'+1'\,\ldots,N'+M'$.
Since the $\Omega_{i'j'}$ are
the small parameters of the perturbation expansion, it follows that
$|\Re{\cal B}_{i'j'}/ (\overline m_{i'}- \overline m_{j'})| \ll 1$,
which is the
perturbativity condition previously given in Eq.$\,$(\ref{pertcond}).\s

At this stage, the result of the perturbative block diagonalization is:
\begin{eqnarray} \label{stage}
\overline{\cal N}^*_B
\left(\begin{array}{cc} \overline{\cal M}_N^D & {\cal B} \\
      {\cal B}^T &  \overline{\cal M}_M^D\end{array}\right)
\overline{\cal N}^\dagger_B
=
\left(\begin{array}{cc}{\cal M}_N^{\prime\,D} & {\cal O}(\Omega^3) \\
{\cal O}(\Omega^3) &
{\cal M}_M^{\prime\,D}\end{array}\right)\,.
\end{eqnarray}
We can neglect the ${\cal O}(\Omega^3)$ terms above.  Thus, the
remaining task is to re-diagonalize the two diagonal blocks above.
However, as long as we work self-consistently up to second order in
perturbation theory, no further re-diagonalization is necessary.
Indeed, the off-diagonal elements of ${\cal M}_N^{\prime\,D}$ and
${\cal M}_M^{\prime\,D}$ are of ${\cal O}(\Omega^2)$.  However, in the
Takagi diagonalization, the off-diagonal
terms of the diagonal blocks only effect the corresponding
diagonal elements at ${\cal O}(\Omega^4)$ which we neglect in this
analysis.  The diagonal elements of ${\cal M}_N^{\prime\,D}$ and
${\cal M}_M^{\prime\,D}$ also contain terms of ${\cal O}(\Omega^2)$,
which generate second-order shifts of the diagonal elements relative
to the $\overline m_{k'}$ obtained at step {\bf [1]}.  These are easily
obtained from the diagonal matrix elements of Eqs.$\,$(\ref{mnd2}) and
(\ref{mmd2}) after making use of Eq.$\,$(\ref{omegab}):
\begin{eqnarray}
&& m_{i'}  \simeq \overline m_{i'}+\sum_{j'=N'+1'}^{N'+M'}
\left\{\frac{[\Re{\cal B}_{i'j'}]^2}{\overline m_{i'}-\overline m_{j'}}
+ \frac{[\Im{\cal B}_{i'j'}]^2}{\overline m_{i'}+\overline m_{j'}}
+\frac{2{\rm i}\, m_{j'} \Re{\cal B}_{i'j'}\Im{\cal B}_{i'j'}}{
 \overline m_{i'}^2-\overline m_{j'}^2} \right\}\,,\label{complexmi}\\[8pt]
&& m_{j'}  \simeq \overline m_{j'}-\sum_{i'=1'}^{N'}
\left\{\frac{[\Re{\cal B}_{ij}]^2}{\overline m_{i'}-\overline m_{j'}}
- \frac{[\Im{\cal B}_{i'j'}]^2}{\overline m_{i'}+\overline m_{j'}}
+\frac{2{\rm i}\, m_{i'} \Re{\cal B}_{i'j'}\Im{\cal B}_{i'j'}}{
  \overline m_{i'}^2-\overline m_{j'}^2} \right\}\,,
\label{complexmj}
\end{eqnarray}
with $i'=1',..,N'$ and $j'=N'+1',..,N'+M'$.
Although the $\overline m_{k'}$ are real and non-negative by
construction, we see that the shifted mass parameters $m_{k'}$ are in
general complex.
Thus, to complete the perturbative Takagi diagonalization, we perform
one final step.\s

{\bf [3]} The diagonal neutralino mass matrix is given by:
\begin{eqnarray}
{\cal M}^D_{N+M}
= {\cal P}^*\overline{\cal N}^*_B
  \left(\begin{array}{cc} \overline{\cal M}_N^D & {\cal B} \\
        {\cal B}^T &  \overline{\cal M}_M^D\end{array}\right)
  \overline{\cal N}^\dagger_B {\cal P}^\dagger
  ={\rm diag} ({m}^{ph}_{1'}, \ldots, {m}^{ph}_{N'+M'})\,,
\end{eqnarray}
where ${\cal P}$ is a suitably chosen diagonal matrix of phases
\begin{eqnarray}
{\cal P}={\rm diag}(e^{-i\phi_{1'}}\,,\,\ldots\,,\,e^{-i\phi_{N'+M'}})\,,
\end{eqnarray}
such that the elements of the diagonal mass matrix ${\cal M}^D_{N+M}$
(denoted by $m^{ph}_{k'}$) are real and non-negative.
We identify the $m^{ph}_{k'}$ with the physical neutralino masses.
The unitary neutralino mixing matrix is then identified as:
\begin{eqnarray}
N^{N+M}={\cal P}\overline{\cal N}_B
\left(\begin{array}{cc}N^N & \mathds{O}\\ \mathds{O}^{\bf T} & N^M
\end{array}\right)\,.
\end{eqnarray}
\s\vskip -0.4cm

Starting from Eqs.$\,$(\ref{complexmi}) and (\ref{complexmj}), one can evaluate
${\cal P}$ to second order in the
perturbation $\Omega$.  In particular, for $\epsilon_{1,2}\ll a$, we
have $a+\epsilon_1+i\epsilon_2\simeq (a+\epsilon_1)e^{i\epsilon_2/a}$.
From this result, we easily derive the second-order expressions
for the physical neutralino masses $m^{ph}_{k'}$:
\begin{eqnarray}
&& \hspace{-0.5in} m^{ph}_{i'}  \simeq \overline m_{i'}+\sum_{j'=N'+1'}^{N'+M'}
 \left\{
 \frac{[\Re{\cal B}_{i'j'}]^2}{\overline m_{i'}-\overline m_{j'}}
+ \frac{[\Im{\cal B}_{i'j'}]^2}{\overline m_{i'}+\overline m_{j'}}
\right\}\,,
\quad [i'=1'\,,\,\ldots\,,\,N']\,, \label{physmi}\\[7pt]
&& \hspace{-0.5in} m^{ph}_{j'}  \simeq \overline m_{j'}-\sum_{i'=1'}^{N'}
 \left\{
 \frac{[\Re{\cal B}_{i'j'}]^2}{\overline m_{i'}-\overline m_{j'}}
- \frac{[\Im{\cal B}_{i'j'}]^2}{\overline m_{i'}+\overline m_{j'}}
\right\}\,,
\quad [j'=N'+1'\,,\,\ldots\,,\,N'+M']\,.\label{physmj}
\end{eqnarray}
and the phases $\phi_{k'}$:
\begin{eqnarray}
&& \hspace{-0.5in} \phi_{i'} \simeq
\, -\sum_{j'=N'+1'}^{N'+M'}\,\frac{\overline m_{j'}}{\overline m_{i'}
(\overline m_{i'}^2 -\overline m_{j'}^2)}\Re{\cal B}_{i'j'}\,
\Im{\cal B}_{i'j'}\,, \qquad [i'=1',\ldots,N']\,,\\[5pt]
&&\hspace{-0.5in} \phi_{j'} \simeq\phm
\sum_{i'=1'}^{N'}\,\frac{\overline m_{i'}}{\overline m_{j'}
(\overline m_{i'}^2 -\overline m_{j'}^2)}
\Re{\cal B}_{i'j'}\,\Im{\cal B}_{i'j'}\,, \qquad
[j'=N'+1'\,,\,\ldots\,,\,N'+M']\,,
\end{eqnarray}
\vskip 0.1cm

This completes the perturbative Takagi diagonalization of the mass
matrix for $N$-dimensional and $M$-dimensional subsystems of
Majorana fermions weakly coupled by an off-diagonal perturbation.  As
noted in Eq.$\,$(\ref{pertcond}), the perturbation theory breaks down
if any mass $\overline m_{i'}$ from the $N$-dimensional subsystem is
nearly degenerate with a corresponding mass $\overline m_{j'}$ from
the $M$-dimensional subsystem (assuming that the
corresponding residue, $\Re {\cal B}_{i'j'}$, does not
vanish).  We provide an analytic approach to
this case of near-degeneracy in Appendix C.\s

\section*{Appendix C: Degenerate mass eigenvalues}

\setcounter{equation}{0}
\setcounter{section}{3}

If the value of one of the diagonal $\overline{\cal M}^D_N$ elements,
$\overline{m}_{k'}$, is nearly equal to one of the diagonal
$\overline{\cal M}^D_M$ elements, say $\overline{m}_{\ell'}$, and
the corresponding residue $\Re {\cal B}_{k'\ell'}$ does not vanish
[cf. Eqs.$\,$(\ref{physmi}) and (\ref{physmj})], then
the techniques for degenerate states must be applied to diagonalize the
full $(N+M) \times (N+M)$ matrix.  We begin with the matrix
$\overline{\cal M}_{N+M}$ given in Eq.$\,$(\ref{decal}), which contains
off-diagonal blocks of ${\cal O}(X)$, which characterizes the small couplings
between the original MSSM matrix and the new USSM singlino/gaugino
submatrix. \s

We first interchange the first row and the $k'$th row of
$\overline{\cal M}_{N+M}$
followed by an interchange of the first column and the $k'$th
column, in order that $\overline{m}_{k'}$ occupy the $1''1''$ element of
the matrix.\footnote{To distinguish the ordering of the physical neutralino
states that arises from the manipulations performed in this
appendix from the ordering of states based on Eqs.~(\ref{dm1}) and
(\ref{dm2}), we employ double-primed subscripts here.}
Next, we interchange the second row and the $\ell'$th row
followed by an interchange of the second column and the $\ell'$th
column, in order that $\overline{m}_{\ell'}$ occupy the $2''2''$ element of
the matrix.  This sequence of interchanges has the effect of grouping
the two nearly degenerate diagonal elements next to each other,
resulting in a new matrix $\overline{\cal M}^{\prime}_{N+M}$ with the following
structure:
\begin{eqnarray} \label{dprime}
\overline{\cal M}^{\prime}_{N+M} =\left(\begin{array}{c|c}
                   \begin{array}{cc}
                   \overline{m}_{1''}    &  \delta \\
                   \delta    & \overline{m}_{2''}
                    \end{array}  & \Delta \\
                    \cline{1-2}
                   \Delta^{T}    & {\cal M}_{N+M-2}
                   \end{array}\right)\,,
\label{eq:full_matrix}
\end{eqnarray}
where the parameter $\delta$ and the submatrix $\Delta$ are of
${\cal O}(X)$.  The submatrix ${\cal M}_{N+M-2}$ is no longer
diagonal, although its new off-diagonal elements are all of ${\cal O}(X)$.
Thus, we may perform a perturbative Takagi diagonalization using
the block diagonal unitary matrix, ${\rm diag}(\mathds{1}_{2\times 2}\,,\,
N^{N+M-2})$, with
\begin{eqnarray} \label{rediag}
\overline{\cal M}_{N+M-2}=(N^{N+M-2})^{*}\, {\cal M}_{N+M-2}
\, (N^{N+M-2})^{\dagger} = {\rm diag}(\overline m'_{3''},\overline m'_{4''}\,
\ldots, \overline m'_{N''+M''})\,,
\end{eqnarray}
where the $\overline m'_{j''}$ [$j''=3'', 4'', \ldots, N''+M''$]
are slightly shifted from the original
non-degenerate $\{\overline m_{i''}\}$, $\{\overline m_{j''}\}$ by the
perturbation of ${\cal O}(X)$.\footnote{For consistency with the
second-order perturbative results of Appendix B, this diagonalization
should be carried out including all contributions quadratic in $X$.}\s

As a result of this procedure, the matrix
$\overline{\cal M}^{\,\prime}_{N+M}$ in
Eq.$\,$(\ref{dprime}) is modified by replacing the submatrix
${\cal M}_{N+M-2}$ by a diagonal matrix with perturbatively
shifted diagonal elements, $\overline{\cal M}_{N+M-2}$.
The off-diagonal blocks $\Delta$ and $\Delta^T$,
are perturbatively shifted as well, but these shifts can be neglected
as these effects are of higher order in the perturbation $X$.
We denote the resulting matrix by $\overline{\cal M}^{\prime\prime}_{N+M}$.
\s

The complex parameter $\delta$ couples the two near-degenerate states
with mass parameters $\overline m_{1''}$ and $\overline m_{2''}$.
By definition of near-degeneracy, $|\overline m_{1''}-\overline m_{2''}|\ll
\delta$, so one cannot use perturbation theory in $\delta\sim{\cal O}(X)$.
Instead, we shall perform an exact Takagi diagonalization of the $2\times 2$
block $\bigl(\begin{smallmatrix}\overline m_{1''} & \delta \\
\delta & \overline m_{2''}\end{smallmatrix}\bigr)$ of
$\overline{\cal M}^{\prime\prime}_{N+M}$,
using the results of Appendix A.2:
\begin{eqnarray} \label{dimproved}
\left(\begin{array}{cc}
   W^*             & \mathds{O} \\
   \mathds{O}^{\bf T} & \mathds{1}\end{array}\right)
\overline{\cal M}^{\prime\prime}_{N+M}
\left(\begin{array}{cc}
    W^\dagger      & \mathds{O} \\
   \mathds{O}^{\bf T} & \mathds{1}\end{array}\right)
=\left(\begin{array}{c|c}
                   \begin{array}{cc}
                   \overline{m}_{1''}^\prime    &  0 \\
                    0   & \overline{m}_{2''}^\prime
                    \end{array}  & W^*\Delta \\
                    \cline{1-2}
                   \Delta^{T}W^\dagger    & \overline{\cal M}_{N+M-2}
                   \end{array}\right)\,,
\end{eqnarray}
where the elements of the $2\times 2$ unitary matrix $W$
(which is denoted by $U^\dagger$ in Appendix A) can be determined in
terms of $\overline m_{1''}$, $\overline m_{2''}$ and $\delta$ using the formulae
of Appendix A.2.  The (non-negative) diagonal masses $\overline m'_{1''}$
and $\overline m'_{2''}$ are obtained from Eq.$\,$(\ref{mk2}):
\begin{eqnarray} \label{newmasses}
\overline m'_{1'',2''}=
\frac{1}{\sqrt{2}}\left\{ \overline{m}_{1''}^2+\overline{m}_{2''}^2
 +2|\delta|^2\mp \sqrt{(\overline m_{2''}^2-\overline m_{1''}^2)^2
      + 4|{\overline m_{1''} \delta+\overline m_{2''}\delta^*|^2}}
      \right\}^{1/2}\,.
\end{eqnarray}
Note that if $\delta$ is real, the quantity under the square root is a
perfect square, in which case Eq.$\,$(\ref{newmasses}) reduces to the
well-known expression:
\begin{eqnarray}
\overline m'_{1'',2''}=\frac{1}{2}\left[\overline m_{1''}+\overline m_{2''}\mp
\sqrt{(\overline m_{2''}-\overline m_{1''})^2 + 4\delta^2}\right]\,,\qquad
\hbox{for real}~\delta\,.
\end{eqnarray}
\s\vskip -0.4cm

If $\delta$ is very small, the trajectories of the two eigenvalues
nearly touch each other when the parameter
$M'_1$ moves through the cross-over zone. A non-zero $\delta$ value prevents the
trajectories from crossing, keeping them at a distance $\geq\delta$.  In the
$4'$-$5'$ zone, $\delta$ is of first order in the ratio $v/M_{\rm SUSY}$.
In contrast, in the $2'$-$6'$ zone, $\delta$ vanishes at first order
due to the fact that $V^6_{2'6'}\approx V^6_{6'2'}\approx 0$.
However, as discussed below Eq.$\,$(\ref{eq:v5}), these matrix elements acquire
small non-zero corrections at higher order in $v/M_{\rm SUSY}$.
Thus, we have two very different behaviors for $\delta$,
leading to the characteristically different evolution of the trajectories.
These two cases are illustrated by the dashed lines in the two panels of
Fig.$\,$\ref{fig:fig7}; on the left for $\delta\rightarrow 0$ and on the
right for moderately non-zero $\delta$ values.\s

\begin{figure}[ht!]
\begin{center}
\includegraphics[height=8.cm,width=7.2cm,angle=0]{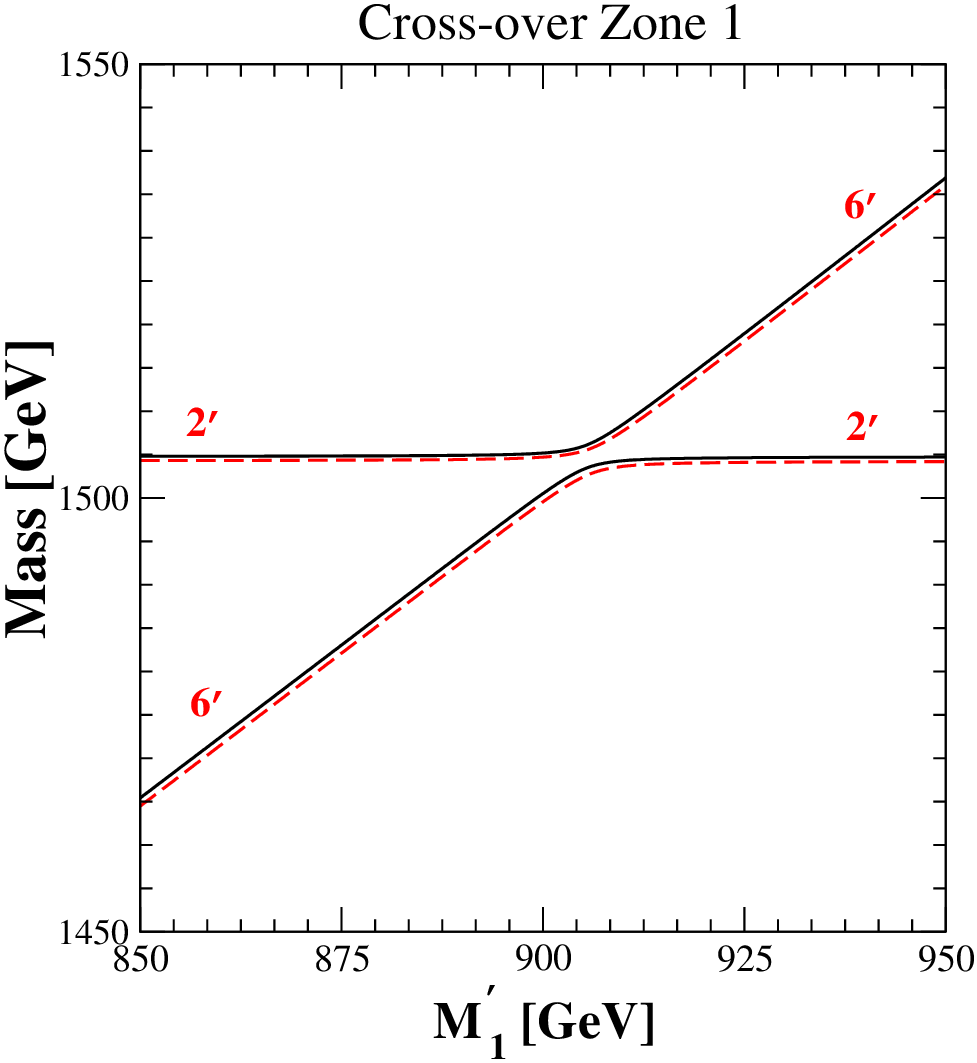}
\hskip 0.5cm
\includegraphics[height=8.cm,width=7.2cm,angle=0]{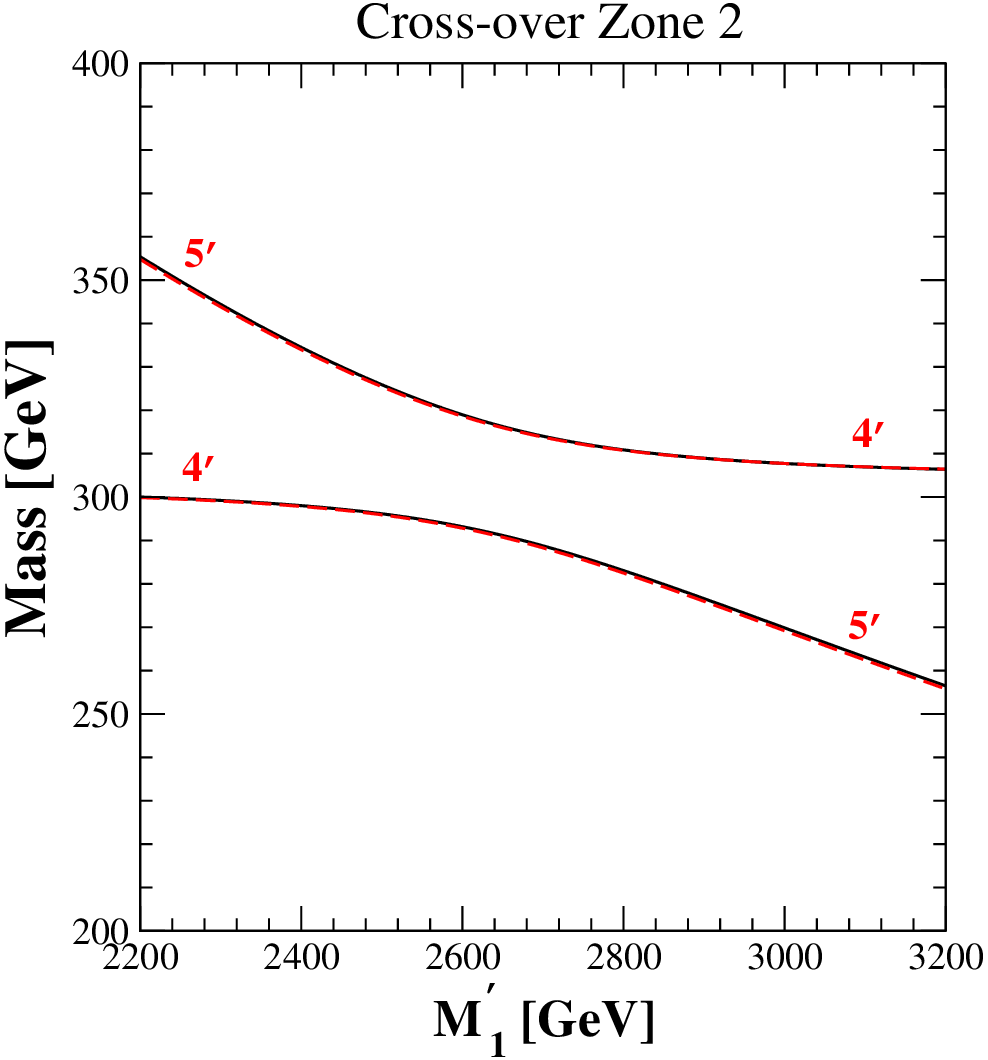}
\end{center}
\caption{\it The evolution of the neutralino masses near the cross-over
             zone 1 (left) and near the crossover zone 2 (right) when
             varying the U(1)$_X$ gaugino mass parameter $M'_1$.
             The red dashed lines represent the masses of the diagonalized
             $2\times 2$ matrix and the black solid lines after the
             subsequent approximate diagonalization of the full
             $6\times 6$ matrix [Eq.$\,$(\ref{dimproved})].}
\label{fig:fig7}
\end{figure}

We may now apply the perturbative block diagonalization technique of
Appendix B to complete the
Takagi diagonalization of Eq.$\,$(\ref{dimproved}).  The effect of this
step is to shift the diagonal masses at second order as indicated in
Eqs.$\,$(\ref{physmi}) and (\ref{physmj}). We finally arrive at the
physical neutralino masses:
\begin{eqnarray}
&& \hspace{-0.5in}
m^{ph}_{i''}  \simeq \overline m'_{i''}+\sum_{j''=3''}^{N''+M''}\left\{
 \frac{[\Re(W^*\Delta)_{i"j"}]^2}{\overline m'_{i''}-\overline m'_{j''}}
+ \frac{[\Im(W^*\Delta)_{i''j''}]^2}{\overline m'_{i''}+\overline m'_{j''}}
\right\}\,,
\quad [i''=1''\,,\,2'']\,, \label{degmi}\\[8pt]
&& \hspace{-0.5in} m^{ph}_{j''}  \simeq \overline m'_{j''}-\sum_{i''=1''}^{N''}
 \left\{
 \frac{[\Re(W^*\Delta)_{i''j''}]^2}{\overline m'_{i''}-\overline m'_{j''}}
- \frac{[\Im(W^*\Delta)_{i''j''}]^2}{\overline m'_{i''}+\overline m'_{j''}}
\right\}\,,
\quad [j''=3''\,,\,4''\,,\,\ldots\,,\,N''+M'']\,.\label{degmj}
\end{eqnarray}
Since the appearance of $\overline m'_{1''}$ and $\overline m'_{2''}$
[given by Eq.$\,$(\ref{newmasses})] takes care of the near-degeneracy
via an exact diagonalization (within the near-degenerate subspace),
the results for the physical masses given above provide a reliable
analytic description.\s

The sizes of the second-order perturbative shifts in Eqs.$\,$(\ref{degmi})
and (\ref{degmj}) vary with the parameter $M'_1$ as the
$\overline{m}_{j''}'$ $[j''=3'',..,(N''+M'')]$ depend on $M'_1$.  The effect
of these shifts can be discerned in the two cases considered
above---in the cross-over
zone $2'$-$6'$ with very small $\delta$, and in the cross-over zone
$4'$-$5'$ with moderately small $\delta$, as shown by the solid
line trajectories of Fig.$\,$\ref{fig:fig7}. \s

Thus, we have demonstrated that an analytic perturbative treatment
of the neutralino mass matrix can be carried out, and all of its
features understood, even in the case of a pair of near-degenerate
states.\s



\begin{thebibliography}{99}

\bibitem{grand_unification} For a review, see M.~Cveti$\breve{\rm c}$ and
   P.~Langacker, in {\it Perspectives on Supersymmetry}, edited by
   G.L.~Kane (World Scientific, Singapore, 1998), pp. 312-331
[hep-ph/9707451].

\bibitem{superstring} J.L.~Hewett and T.G.~Rizzo, Phys. Rept. {\bf 183} (1989)
   193; M.~Cveti$\breve{\rm c}$ and P.~Langacker, Phys. Rev. D
   {\bf 54} (1996) 3570 [hep-ph/9511378];
   Mod. Phys. Lett. A {\bf 11} (1996) 1247 [hep-ph/9602424].

\bibitem{u1ssm0}
M.~Cveti$\breve{\rm c}$, D.A.~Demir, J.R.~Espinosa,
   L.L.~Everett and P.~Langacker, Phys. Rev. D {\bf 56} (1997)
   2861 [hep-ph/9703317];
   {\bf 58} (1997) 119905(E).

\bibitem{mu_problem} D.~Suematsu and Y.~Yamagishi, Int. J. Mod. Phys. A
   {\bf 10} (1995) 4521 [hep-ph/9411239].
For the case of gauge mediated symmetry
   breaking, see P.~Langacker, N.~Polonsky and J.~Wang, Phys. Rev. D
   {\bf 60} (1999) 115005 [hep-ph/9905252].

\bibitem{nmssm} H.P.~Nilles, M.~Srednicki and D.~Wyler, Phys. Lett. B {\bf 120}
   (1983) 346; J.M.~Frere, D.R.T.~Jones and S.~Raby, Nucl. Phys. B {\bf 222}
   (1983) 11; J.P.~Derendinger and C.A.~Savoy, Nucl. Phys. B {\bf 237} (1984)
   307; M.I.~Vysotsky, K.A.~Ter-Martirosian, Sov. Phys. JETP {\bf 63}
   (1984) 307; J.R.~Ellis, J.F.~Gunion, H.E.~Haber, L.~Roszkowski, F.~Zwirner,
   Phys. Rev. D {\bf 39} (1989) 844;
   U.~Ellwanger, M.~Rausch de Traubenberg and C.A.~Savoy,
   Phys.  Lett. B {\bf 315} (1993) 331
   [hep-ph/9307322];
   Nucl. Phys. B {\bf 492} (1997) 21
   [hep-ph/9611251]; F.~Franke and H.~Fraas,
   Int. J. Mod. Phys. A {\bf 12} (1997) 479
   [hep-ph/9512366].
   For a recent summary see D.J. Miller,
   R.~Nevzorov and P.M.~Zerwas, Nucl. Phys. B {\bf 681}
   (2004) 3 [hep-ph/0304049].

\bibitem{u1prime_D} H.E.~Haber and M.~Sher, Phys. Rev. D {\bf 35} (1987) 2206;
    L.~Durand and J.L.~Lopez, Phys. Lett. B {\bf 217} (1989)
   463; M.~Drees, Int. J. Mod. Phys. A {\bf 4} (1989) 3635.

\bibitem{erler} J.~Erler, Nucl. Phys. B {\bf 586} (2000) 73 [hep-ph/0006051].

\bibitem{u1ssm3} S.~King, S.~Moretti, R.~Nevzorov, Phys. Lett. B {\bf 634}
    (2006) 278 [hep-ph/0511256]; Phys. Rev. D {\bf 73} (2006)
     035009 [hep-ph/0510419].

\bibitem{u1ssm1}
   J.~Erler, P.~Langacker and T.~Li, Phys. Rev. D {\bf 66} (2002)
   015002 [hep-ph/0205001].

\bibitem{u1ssm2} D.~Suematsu, Phys. Rev. D {\bf 57} (1998) 1738
  [hep-ph/9708413]; S.~Hesselbach, F.~Franke and H.~Fraas, Eur. Phys. J.
   C. {\bf 23} (2002) 149 [hep-ph/0107080]; F.~Franke and
  S.~Hesselbach, Phys. Lett. B {\bf 526} (2002) 370 [hep-ph/0111285];
  V.~Barger, P.~Langacker and H.S. Lee,
   Phys. Lett. B {\bf 630} (2005) 85 (2005) [hep-ph/0508027];
   V.~Barger, P.~Langacker and G.~Shaughnessy, arXiv:hep-ph/0609068.

\bibitem{Suematsu:1997tv}
  D.~Suematsu,
  Mod. Phys. Lett. A {\bf 12} (1997) 1709
  [hep-ph/9705412];
  Phys. Lett. B {\bf 416} (1998) 108
  [hep-ph/9705405];
  G.A.~Moortgat-Pick, S.~Hesselbach, F.~Franke and H.~Fraas,
  arXiv:hep-ph/9909549;
  S.~Hesselbach, F.~Franke and H.~Fraas,
  arXiv:hep-ph/0003272;
  V.~Barger, C.~Kao, P. Langacker and H.S.~Lee,
  Phys. Lett. B {\bf 600} (2004) 104
  [hep-ph/0408120];
  {\bf 614} (2005) 67
  [hep-ph/0412136].


\bibitem{ckmz} S.Y.~Choi, J.~Kalinowski, G.A.~Moortgat-Pick and P.M.~Zerwas,
   Eur. Phys. J. C {\bf 22} (2001) 563 [hep-ph/0108117]; {\bf 23} (2002) 769
[hep-ph/0202039].

\bibitem{NMSSM_CMZ} S.Y.~Choi, D.J.~Miller and P.M.~Zerwas, Nucl. Phys. B
   {\bf 711} (2005) 83 [hep-ph/0407209].

\bibitem{LHC} ATLAS Technical Proposal, CERN/LHCC/94-43, LHCC/P2 (1994);
   CMS Physics, Technical Design Report, CERN/LHCC/2006/021.

\bibitem{ILC} E.~Accomando {\it et al.}, Phys. Rept. {\bf 299} (1998) 1,
   hep-ph/9705442 J.A.~Aguilar-Saavedra {\it et al.} [ECFA/DESY LC Physics
   Working Group Collaboration], hep-ph/0106315; T.~Abe {\it et al.}
   [American LC Working Group], hep-ex/0106055-58; K.~Abe {\it et al.}
   [ACFA LC Working Group], hep-ex/0109166; E.~Accomando {\it et al.}
   [CLIC Physics Working Group Collaboration], hep-ph/9412251;
   W.~Kilian and P.M.~Zerwas,
   Proc. 2005 Snowmass ILC Workshop, hep-ph/0601217.

\bibitem{kinetic_mixing_loop} B.~Holdom, Phys. Lett. B {\bf 166} (1986) 196.

\bibitem{kinetic_mixing} F.~del Aguila, Acta. Phys. Pol. B {\bf 25} (1994)
   1317 [hep-ph/9404323]; F.~del Aguila,
   M. Cveti$\breve{\rm c}$ and P. Langacker,
   Phys. Rev. D {\bf 52} (1995) 37 [hep-ph/9501390];
   K.S.~Babu, C.~Kolda and J.~March-Russell,
   Phys. Rev. D {\bf 54} (1996) 4635 [hep-ph/9603212];
   D.~Suematsu, Phys. Rev. D {\bf 59} (1999) 055017 [hep-ph/9808409].

\bibitem{bailinlove} D.~Bailin and A.~Love, \textit{Supersymmetric
  Gauge Field Theory and String Theory} (Institute of Physics
  Publishing, Bristol, UK, 1994).

\bibitem{kinetic_mixing_string} K.R.~Dienes, C.~Kolda, J.~March-Russell,
   Nucl. Phys. B {\bf 492} (1997) 104 [hep-ph/9610479].

\bibitem{Abel:2006qt}
  S.A.~Abel, J.~Jaeckel, V.V.~Khoze and A.~Ringwald,
  arXiv:hep-ph/0608248.

\bibitem{Zprime_bound} P.~Abreu {\it et al.} [DELPHI Collaboration],
   Phys. Lett. B {\bf 485} (2000) 45; R.~Barate {\it et al.} [ALEPH
   Collaboration], Eur. Phys. J. C {\bf 12} (2000) 183;
   A.~Abulencia {\it et al.} [CDF Collaboration], hep-ph/0602045.

\bibitem{SPA} J.A.~Aguilar-Saavedra {\it et al.}, Eur. Phys. J. C {\bf 46}
  (2006) 43 [hep-ph/0511344].

\bibitem{takagi} T.~Takagi, Japan J.\ Math. {\bf 1} (1925) 83.

\bibitem{horn} R.A.~Horn and C.R.~Johnson, {\it Matrix Analysis} (Cambridge
   University Press, Cambridge, England, 1990).

\bibitem{takcompute} A.~Bunse-Gerstner and W.B.~Gragg, J. Comp. Appl. Math.
   {\bf 21} (1988) 41; W.~Xu and S.~Qiao, ``A Divide-and-Conquer Method for
   the Takagi Factorization,'' Technical Report No. CAS 05-01-SQ,
   (February 2005).

\bibitem{takcompute2} X.~Wang and S.~Qiao, in the Proceedings of
   the International Conference on Parallel and Distributed Processing
   Techniques and Applications, Vol. I, edited by H.R. Arabnia, pp. 206-212,
   Las Vegas, Nevada, USA, June 2002, pp. 206-212;
   F.T.~Luk and S.~Qiao, in {\it Advanced Signal Processing Algorithms,
   Architectures, and Implementations XI}, edited by F.T.~Luk,
   Proc. SPIE {\bf 4474} (2001) 254.

\bibitem{hahn} T.~Hahn, arXiv:physics/0607103.

\bibitem{carruthers} P.A.~Carruthers, J. Math. Phys. {\bf 9} (1968) 1835;
   {\it Spin and Isospin in Particle Physics} (Gordon and Breach,
   New York, NY, 1971);
   B.~Kayser, Phys. Rev. D {\bf 30} (1984) 1023;
   B.~Kayser, F.~Gibrat-Debu and F.~Perrier,
   World Sci. Lect. Notes Phys. {\bf 25} (1989) 1.

\bibitem{gunhab} J.F.~Gunion and H.E.~Haber, Phys. Rev. D {\bf 37} (1988) 2515;
   S.Y.~Choi, M.~Drees and B.~Gaissmaier, Phys. Rev. D {\bf 70} (2004)
   014010 10,2004 [hep-ph/0403054].

\bibitem{Leike} A.~Leike, Phys. Rept. {\bf 317} (1999) 143 [hep-ph/9805494];
   W.M.~Yao {\it et al.}  [Particle Data Group], J. Phys. G {\bf 33} (2006) 1.

\bibitem{haberkane} H.E.~Haber and G.L.~Kane, Phys. Rept. {\bf 117} (1985) 75.

\bibitem{ref:chargino} S.Y.~Choi, A.~Djouadi, M.~Guchait, J.~Kalinowski,
  H.S.~Song and P.M.~Zerwas, Eur. Phys. J. C {\bf 14}, 535 (2000)
  [hep-ph/0002033].

\bibitem{Barger:2006dh} T.~Han, P.~Langacker and B.~McElrath, Phys. Rev. D
  {\bf 70} (2004) 115006 [hep-ph/0405244];
  V.~Barger, P.~Langacker, H.S.~Lee and G.~Shaughnessy,
  Phys. Rev. D {\bf 73} (2006) 115010 [hep-ph/0603247].

\bibitem{Demir:2003ke} D.A.~Demir and L.L.~Everett, Phys. Rev. D {\bf 69}
   (2004) 015008 [hep-ph/0306240].

\bibitem{Barger:2006rd}  V.~Barger, P.~Langacker and G.~Shaughnessy,
  arXiv:hep-ph/0611112; arXiv:hep-ph/0611239.

\bibitem{radcors} For a recent review and references to the original literature,
   see e.g., M.~Carena and H.E.~Haber, Prog. Part. Nucl. Phys. {\bf 50} (2003)
   63 [hep-ph/0208209].

\bibitem{HaWy} H.E.~Haber and D.~Wyler, Nucl. Phys. B {\bf 323} (1989) 267.

\bibitem{horn2} R.A.~Horn and C.R.~Johnson, \textit{Topics in Matrix Analysis}
   (Cambridge University Press, Cambridge, England, 1991).

\bibitem{autonne} L.~Autonne, \textit{Sur les matrices hypohermitiennes et sur
   les matrices unitaire}, Annales de l'Universit\'e de Lyon, Nouvelle
   S\'erie I, Fasc. {\bf 38} (1915) 1-77.

\bibitem{dreeschoi} S.Y.~Choi and M.~Drees, unpublished. This proof was
   inspired by the diagonalization algorithm of hermitian matrices in
   W.H.~Williams, B.P.~Flannery, S.A.~Teukolsky and W.T.~Vetterling,
   \textit{Numerical Recipes in Fortran 77} (Cambridge University Press,
   Cambridge, England, 1999), section 11.4. A similar method of proof is
   outlined in Ref.~\cite{horn}, section 4.4, problem 2 (on pp.~212-213) and
   section 4.6, problem 15 (on p.~254).

\bibitem{murnaghan} F.D.~Murnaghan, \textit{The Unitary and Rotation Groups}
   (Spartan Books, Washington, DC, 1962).

\end{thebibliography}
\end{document}
